%% file: main.tex
\documentclass{article}

\usepackage[a4paper]{geometry}
\usepackage{amsmath,amssymb,mathtools}
\usepackage[sectionbib,round]{natbib}
\usepackage{amsthm}
\usepackage[utf8]{inputenc}
\usepackage[T1]{fontenc}
\usepackage{booktabs}
\usepackage{longtable}
\usepackage{amsfonts}
\usepackage{microtype}
\usepackage[font=small,skip=10pt,labelfont=bf,tableposition=top]{caption}
\usepackage[font=small,skip=0.3pt]{subcaption}
\usepackage[section]{placeins}
\usepackage{array}
\usepackage{multirow}
\usepackage{graphicx}
\graphicspath{{figures/}}
\usepackage{float}
\usepackage{pdflscape}
\usepackage{setspace}
\usepackage{xcolor}
\usepackage{hyperref}
\hypersetup{colorlinks=true,citecolor=blue,linkcolor=blue,urlcolor=blue}

\title{Forecasting Realized Volatility with Time Series Foundation Models:\\A Comparison with Econometric Benchmarks}
\author{Alessio Brini\footnote{Corresponding author.}}
\date{}


\begin{document}

\maketitle
\thispagestyle{empty}
\vspace{-1.5em}
\begin{center}
\textit{Duke University, Durham, NC (USA). E-mail: alessio.brini@duke.edu}
\end{center}

\medskip

\begin{abstract}
\noindent
Time series foundation models (TSFMs) are pretrained on large collections of series and forecast a new series without being trained on it. We compare nine TSFMs, used in this zero-shot setting, with eight econometric models, including the Heterogeneous Autoregressive (HAR) family. We forecast the realized volatility of 50 assets in equities, currencies and futures from the VOLARE dataset one, five and 22 days ahead, and compare the forecasts with pairwise and multi-model tests. The foundation models do not outperform the econometric benchmarks as a class. Only Tiny Time Mixers (TTM), the smallest of them, has a lower average QLIKE loss than the logarithmic HAR (Log-HAR) at every tested horizon, by a narrow margin. For most assets, the Model Confidence Set (MCS) contains TTM together with Log-HAR and several other models. Accuracy differs far more across the foundation models than between the best of them and Log-HAR, so which foundation model one chooses matters more than whether one uses a foundation model or an econometric one. For an investor who targets a constant volatility, TTM's edge is worth a few basis points a year before trading costs. Since TTM and Log-HAR make different errors, their equal-weight average matches the best single model and enters the MCS for nearly every asset.
\end{abstract}

\medskip


\newpage

\section{Introduction}
\label{sec:introduction}

Forecasting realized volatility is central to risk management, derivative pricing, and portfolio allocation. Since the work of \citet{andersen1998} and \citet{barndorffnielsen2002}, realized volatility constructed from high-frequency returns has become the standard model-free measure of ex-post price variation. The Heterogeneous Autoregressive (HAR) model of \citet{corsi2009} is the dominant forecasting benchmark. Its three-component structure, which aggregates past realized volatility at daily, weekly, and monthly frequencies, provides a parsimonious approximation to the long-memory dynamics that characterize volatility. Extensions incorporating jump variation \citep{andersen2007roughing}, semivariance asymmetries \citep{patton2015}, and measurement error corrections \citep{bollerslev2016harq} refine the model. Machine learning methods, including neural networks \citep{bucci2020, zhang2024intraday}, random forests \citep{luong2018}, graph-based approaches \citep{zhang2025graph, brini2025spotv2net}, and convolutional architectures \citep{morenopino2024deepvol}, have also been applied. \citet{christensen2023ml} find that machine learning beats the HAR family, with the largest gains at longer horizons, and attribute these gains to the higher persistence of the machine learning forecasts, which approximates the long memory of realized variance.

The models discussed so far, econometric and machine learning alike, are estimated on the target volatility series itself. Time series foundation models (TSFMs) are a different class of models that dispense with this step. These are large neural networks, and most of them are transformers \citep{vaswani2017}, which learn dependencies between positions in a sequence through an attention mechanism without imposing a fixed lag structure. The smallest model in our comparison is the exception, since it is built from multilayer perceptrons. TSFMs are trained on large and diverse corpora of time series from multiple domains and can produce forecasts for previously unseen series without any retraining, which is known as zero-shot forecasting. \citet{gruver2023} also demonstrate zero-shot forecasting for general-purpose large language models. Leading examples of TSFMs include Chronos \citep{ansari2024chronos}, Moirai \citep{woo2024moirai}, and Lag-Llama \citep{rasul2024lagllama}, several of which have released second-generation versions with improved architectures \citep{ansari2025chronos2, liu2025moirai2}. Surveys document the growth of this area \citep{liang2024tsfmsurvey, ye2024tsfmsurvey, miller2024tsfmsurvey}. Outside finance, \citet{carriero2024macro} apply zero-shot TSFMs to macroeconomic forecasting and find that these models are not yet a clear replacement for macroeconometric baselines. We test whether the same conclusion holds for realized volatility.

Applications of TSFMs to financial time series remain limited, and we review them in Subsec.~\ref{subsec:fm_volatility}. On realized volatility, \citet{goel2025rv} find that TimesFM~2.0 needs fine-tuning to compete with HAR. On daily excess returns, \citet{rahimikia2025revisiting} report uniformly negative zero-shot results, and fine-tuning gives only limited gains. Realized volatility is strictly positive, mean-reverting, and long-memory \citep{andersen2001exchange, andersen2003}, properties that distinguish it from returns and that may favor foundation models. A single-model study cannot settle whether these properties make a general-purpose pretrained model competitive with a benchmark designed for volatility.

Since no study has yet evaluated multiple TSFMs on realized volatility with formal statistical testing, in this paper we compare zero-shot TSFMs with established econometric benchmarks. We evaluate nine TSFMs, spanning eight distinct architectures, against eight econometric specifications across 50 assets in three asset classes (equities, foreign exchange (FX), futures) and three forecast horizons ($h = 1, 5, 22$ days), using the VOLARE (VOLatility Archive for Realized Estimates) dataset \citep{volare2026}. We study how pretrained models forecast realized volatility without any training on the target series, so every foundation model is used zero-shot, which isolates the value of general time series pretraining. The forecast target is the point-in-time realized volatility. We apply Diebold--Mariano (DM) tests \citep{diebold1995} and the Model Confidence Set (MCS) \citep{hansen2011mcs}, which controls for multiple comparisons. We supplement them with Mincer--Zarnowitz (MZ) regressions \citep{mincer1969} for forecast efficiency and Giacomini--Rossi (GR) fluctuation tests \citep{giacomini2010} for time-varying relative performance, and we check the results on pre- and post-COVID subsamples and across context windows of the foundation models. In Appendix~\ref{app:finetune} we report how fine-tuning changes the accuracy of two of them, Tiny Time Mixers (TTM) and Sundial. TTM is the most accurate foundation model in our comparison, and Sundial, with about 150 times as many weights, is about as accurate as the logarithmic HAR (Log-HAR) at the daily horizon.

We find, first, that pretrained foundation models are not uniformly more accurate than the econometric benchmarks. We measure accuracy by each asset's quasi-likelihood (QLIKE) loss relative to Log-HAR, averaged over the 50 assets so that each asset counts equally. Six of the nine TSFMs do not beat Log-HAR at any horizon. HAR, the Autoregressive Fractionally Integrated Moving Average (ARFIMA) model, the Autoregressive Moving Average (ARMA) model and the multiplicative error model (MEM) stay within a few percent of Log-HAR at the daily horizon. In our sample the HAR variants with jump, semivariance and quarticity regressors fall well behind at the daily and weekly horizons. Under mean squared error (MSE) several foundation models are more accurate than Log-HAR, but not under QLIKE. These models forecast a lower variance than Log-HAR, and QLIKE penalizes a variance forecast that is too low more than one that is too high.

Second, TTM, the smallest model in the evaluation ($<$1M parameters), is the only TSFM that beats Log-HAR at every horizon, and it does so by a narrow margin. TimesFM~2.5 and Sundial are about as accurate as Log-HAR at the daily horizon only. The MCS includes TTM for every asset, together with Log-HAR and several other models on most assets. At the daily and weekly horizons, TTM's edge disappears when we adjust the level of every model's variance forecast, so it comes mostly from the level of its forecasts. At the daily horizon, a uniform MZ recalibration also gives several other foundation models the accuracy of TTM. In an exercise with an investor who targets a constant volatility, we show that the edge is worth a few basis points of wealth a year before trading costs.

Third, accuracy differs far more across the foundation models than between the best of them and Log-HAR. At the daily horizon their loss ratios range from 0.984 to 1.535, against a gap of 1.6\% between TTM and Log-HAR, so which foundation model one chooses matters more than whether to use a foundation model or an econometric one.

Finally, TTM and Log-HAR make different errors, so their equal-weight average matches the best single model and enters the MCS for 98 to 100\% of assets across horizons. A forecaster therefore need not identify the best model for each asset in advance.

In Sec.~\ref{sec:literature} we review the related literature, and in Sec.~\ref{sec:data} we describe the VOLARE dataset. In Sec.~\ref{sec:methodology} we present the econometric and foundation model specifications and the forecast evaluation framework. In Sec.~\ref{sec:results} we report the empirical results, and in Sec.~\ref{sec:significance} we assess their statistical significance with formal forecast-comparison tests. In Sec.~\ref{sec:economic} we measure the economic value of the forecasts for an investor who targets a constant volatility, and in Sec.~\ref{sec:robustness} we run robustness checks. In Sec.~\ref{sec:combination} we study the equal-weight combination of TTM and Log-HAR and why it is more accurate than either model, and in Sec.~\ref{sec:conclusion} we conclude.

\section{Literature Review}
\label{sec:literature}

We review three strands of work. In Subsec.~\ref{subsec:rv_modeling} we cover realized volatility modeling and the HAR benchmark, in Subsec.~\ref{subsec:tsfm} we introduce TSFMs and their architectures, and in Subsec.~\ref{subsec:fm_volatility} we survey their still-limited application to volatility forecasting.

\subsection{Realized Volatility Modeling}
\label{subsec:rv_modeling}

The theory of realized volatility originates with \citet{andersen1998} and \citet{andersen2001exchange}, which show that the sum of squared intraday returns provides a consistent, nonparametric estimator of the integrated variance of asset prices. \citet{barndorffnielsen2002} develop the asymptotic distribution theory for realized variance in a stochastic volatility framework, deriving a central limit theorem and rate of convergence for the error of realized variance around integrated variance. Subsequent work identified the key stylized facts of realized volatility: approximate log-normality, long-memory dependence with a fractional integration parameter $d \approx 0.4$, and slow mean reversion \citep{andersen2003}. At ultra-high sampling frequencies, microstructure noise from bid-ask bounce and discrete price changes biases the realized variance estimator upward. \citet{barndorffnielsen2008} introduce the realized kernel as a noise-consistent alternative that remains valid in the presence of market microstructure effects. This noise is the reason we sample at 5 minutes (Sec.~\ref{sec:data}).

The HAR model of \citet{corsi2009} became the standard forecasting benchmark for realized volatility. By including daily, weekly, and monthly realized volatility components as regressors, the model approximates the long-memory decay of volatility autocorrelations. The specification is parsimonious (three regressors plus a constant) yet achieves accuracy comparable to long-memory models, although the machine learning forecasters of \citet{christensen2023ml} beat it, most clearly at longer horizons.

A large body of work has extended the HAR framework. The HAR-J model \citep{andersen2007roughing} separates realized variance into continuous and jump components using bipower variation \citep{barndorffnielsen2004bipower}. The HAR-RS model \citep{barndorffnielsen2010semivariance, patton2015} decomposes realized variance into positive and negative semivariance to capture asymmetric responses to upside and downside moves. The HARQ model \citep{bollerslev2016harq} interacts the daily realized variance regressor with realized quarticity to account for time-varying measurement error. \citet{clements2021practical} show how the accuracy of the standard HAR model depends on the choice of estimator, transformation and combination scheme.

As a complementary approach, long-memory ARFIMA models \citep{granger1980, hosking1981}, applied to realized volatility by \citet{andersen2003}, explicitly parameterize the fractional integration order. They remain competitive at longer forecast horizons, where the slow mean reversion of realized volatility becomes the dominant dynamic.

Machine learning methods have produced mixed results in realized volatility forecasting. \citet{bucci2020} reports that recurrent networks outperform ARFIMA-type benchmarks for S\&P 500 realized volatility, and \citet{luong2018} find gains from random forests. In contrast, \citet{branco2024} conclude that simple linear models are difficult to beat after correcting for multiple testing. Surveys by \citet{gunnarsson2024} and \citet{leushuis2026} cover this literature.

\subsection{Time Series Foundation Models}
\label{subsec:tsfm}

TSFMs differ from task-specific deep learning models (multilayer perceptrons, recurrent networks, convolutional architectures, graph neural networks, and transformers), which must be trained from scratch on the target series before producing any forecast. TSFMs are instead pretrained on large corpora drawn from diverse domains (weather, energy, retail, transport) and forecast immediately via zero-shot inference. This eliminates the need for domain-specific training data, hyperparameter tuning, and GPU-intensive estimation, a practical advantage when deploying to new domains where historical data may be limited \citep{dooley2023forecastpfn}.

Three architectural families dominate the first generation of TSFMs. Chronos \citep{ansari2024chronos} converts continuous values into discrete tokens via uniform binning and forecasts recursively with a T5 encoder-decoder architecture \citep{raffel2020}. Moirai \citep{woo2024moirai} handles an arbitrary number of input series through an ``any-variate'' attention mechanism and uses mixture distribution outputs for uncertainty quantification. Lag-Llama \citep{rasul2024lagllama} adapts the LLaMA decoder-only architecture for probabilistic time series forecasting using lag-based tokenization.

Improved successors and a wider range of models quickly followed these first-generation architectures. Among the models we evaluate, Chronos-Bolt, released by the Chronos team, is a faster variant of the original Chronos that produces direct quantile forecasts in a single forward pass and runs up to 250 times faster.\footnote{\url{https://huggingface.co/amazon/chronos-bolt-base}.} Moirai~2.0 \citep{liu2025moirai2} replaces the masked encoder, multi-patch inputs and mixture outputs of Moirai with a decoder-only architecture, a single patch size and a quantile loss, and outperforms larger models of the same family. Moirai-MoE \citep{liu2024moiraimoe} is a sparse Mixture of Experts (MoE) extension of Moirai\footnote{An MoE replaces a single dense network with several specialized sub-networks (``experts'') and a gating mechanism that routes each input to a small subset of them, so the model's total capacity can grow while the computation per forecast stays low.}. The other models we evaluate are TimesFM, Toto, Sundial and TTM. TimesFM \citep{das2024timesfm} is a patch-based decoder model, and Toto \citep{cohen2025toto} is pretrained on infrastructure monitoring metrics (e.g., server CPU usage, request latency). Sundial \citep{liu2025sundial} uses flow matching for generative forecasting, and TTM \citep{ekambaram2024ttm}, a lightweight model built on TSMixer, is the smallest in our evaluation. Among the models we do not evaluate, \citet{ansari2025chronos2} introduce Chronos-2, which adds multivariate support and covariate handling. The models we do not evaluate also include MOMENT \citep{goswami2024moment}, a masked-encoder architecture designed for multiple time series tasks (forecasting, classification, anomaly detection)\footnote{We do not evaluate MOMENT because it is pretrained with a masked reconstruction objective. Its forecasting head, a linear projection layer that maps patch embeddings to the forecast horizon, is not pretrained and must be trained on the target series before the model can produce any predictions \citep[Sections~3.3 and 3.4]{goswami2024moment}. This makes it incompatible with our zero-shot evaluation protocol.}, and Timer \citep{liu2024timer}.

Several standardized benchmarks now evaluate TSFMs across domains. GIFT-Eval \citep{aksu2024gifteval} provides a unified evaluation protocol across multiple datasets and forecast horizons. FEV-Bench \citep{shchur2025fevbench} emphasizes realistic tasks with covariates and principled aggregation across tasks. TSFM-Bench \citep{li2024tsfmbench} compares models across zero-shot and fine-tuned settings. In these benchmarks no single model dominates across all domains and horizons, which motivates domain-specific evaluations of the kind we undertake here. For finance, \citet{lu2025tsfmfinance} proposes such an evaluation protocol. It pairs probabilistic scoring with Value-at-Risk and Expected Shortfall backtests, and it controls for data snooping in its comparisons against strong econometric and machine learning baselines. We score only point forecasts of realized volatility, a narrower target than that protocol, and like it we control for multiple comparisons, here with the MCS. \citet{tan2024llmuseful} question whether language-model-based forecasters add value over simpler baselines at all, a skepticism that motivates the head-to-head design against econometric benchmarks that we adopt here.

\subsection{Foundation Models for Volatility Forecasting}
\label{subsec:fm_volatility}

The application of TSFMs to volatility forecasting is limited. The closest prior work is \citet{goel2025rv}, which tests TimesFM~2.0 on the realized variance of 21 global equity indices. In that study, zero-shot TimesFM does not consistently beat HAR, and fine-tuning is necessary to achieve competitive accuracy. The study forecasts one day ahead and compares TimesFM with HAR, ARFIMA and Realized GARCH, which extends the generalized autoregressive conditional heteroskedasticity (GARCH) model with a realized measure. In this work, we compare nine foundation models from eight architectures, so we can distinguish the performance of one model from that of the class. Moreover, the panel of data we use contains 50 individual assets in equities, FX and futures, rather than equity indices alone, and we forecast one day, one week and one month ahead.

\citet{rahimikia2025revisiting} evaluate TSFMs from the Chronos and TimesFM families on daily excess returns from 94 countries and find that zero-shot TSFMs consistently underperform strong machine learning ensembles such as CatBoost and LightGBM. In that study, models pretrained on financial data achieve substantial gains. Daily returns are close to white noise, while realized volatility is closer to the macroeconomic and physical series on which TSFMs were pretrained, so negative results on returns need not carry over to realized measures.

Other financial applications include stock price forecasting \citep{laniewski2025chronos, valeyre2024chronosstocks}, FX volatility modeling \citep{nguyen2025volabert}, Value-at-Risk forecasting \citep{goel2024var}, and finance-specific pretrained models such as FinCast \citep{zhu2025fincast} and Kronos \citep{shi2025kronos}. Kronos is pretrained on candlestick (open-high-low-close-volume, OHLCV) data from over 45 global exchanges, with a learned tokenizer that maps price patterns into discrete tokens. The study closest to our result is \citet{marconi2025}, which reports that small TTM models are competitive on financial forecasting tasks, including FX volatility, with the strongest gains obtained through fine-tuning rather than zero-shot use. That study does not use formal forecast-comparison tests or a multi-model panel. Our results show that a small model can beat a well-specified Log-HAR by a narrow margin on realized volatility across 50 assets in the zero-shot setting, without displacing the econometric benchmarks from the MCS.

The literature lacks a multi-model, multi-asset evaluation of realized volatility forecasts with formal statistical tests, and we design our empirical study to provide one.

\section{The VOLARE Dataset}
\label{sec:data}

We use the VOLARE dataset of \citet{volare2026}, which provides daily realized variance and related realized measures for a broad cross-section of financial assets. VOLARE is constructed from ultra-high-frequency tick data sourced from Kibot and covers 40 U.S.\ equities, 5 major currency pairs, and 5 commodity and index futures contracts. The equity sample begins on 2 January 2015 and runs through 30 January 2026, which gives 2,786 trading days per stock. The FX sample begins on 25 September 2009 and the futures sample on 28 September 2009 (up to 4,242 and 4,224 trading days, respectively). The sample spans multiple volatility regimes, including the low-volatility period of 2017 to 2019, the COVID-19 shock of 2020, and the subsequent recovery.

For each asset-day, VOLARE provides realized measures computed at multiple sampling frequencies. We use all realized measures at the 5-minute sampling frequency, the standard bias-variance compromise against microstructure noise in the realized-volatility literature. \citet{liu2015fiveminute} compare realized measures across asset classes and find 5-minute sampling hard to beat. The measures are realized variance ($RV$), bipower variation ($BPV$), positive and negative realized semivariance ($RS^{+}$, $RS^{-}$), and realized quarticity ($RQ$). All values are in decimal squared returns. For instance, a typical daily $RV$ for a U.S.\ equity is approximately $2.5 \times 10^{-4}$, corresponding to an annualized volatility of roughly 25\%. VOLARE is well suited for our multi-asset comparison, since it constructs all realized measures with a uniform methodology across assets. Cross-asset differences in model rankings then reflect forecasting performance rather than measurement inconsistencies.

The equity sample comprises the 40 VOLARE stocks with complete coverage over the full 2015 to 2026 window, a balanced-panel restriction that requires survival over the sample and trades breadth for a common evaluation period. The stocks are AAPL, ADBE, AMD, AMGN, AMZN, AXP, BA, CAT, CRM, CSCO, CVX, DIS, GE, GOOGL, GS, HD, HON, IBM, JNJ, JPM, KO, MCD, META, MMM, MRK, MSFT, NFLX, NKE, NVDA, ORCL, PG, PM, SHW, TRV, TSLA, UNH, V, VZ, WMT, and XOM, covering nine of the eleven Global Industry Classification Standard sectors, all except Utilities and Real Estate. The FX sample consists of five major currency pairs (AUDUSD, EURUSD, GBPUSD, USDCAD, USDJPY) and the futures sample covers five contracts (Corn, Crude Oil, E-mini S\&P 500, Gold, Natural Gas). Both FX and futures samples provide longer histories than the equity sample, because VOLARE's intraday coverage for these asset classes begins earlier.

We report descriptive statistics for the 5-minute realized variance across the three asset classes in Tab.~\ref{tab:descriptive}. As expected for a strictly positive quantity, realized variance is right-skewed in every asset class, and its heavy tails show that extreme volatility episodes are a persistent feature of the data. For equities, the first-order autocorrelation of daily $RV$ is high, consistent with the persistence of volatility documented by \citet{andersen2003}, and some dependence remains at the monthly lag. The FX pairs have a realized variance of about one eighth of that of the equities, with comparable persistence. Futures show the widest heterogeneity across assets. These moments describe realized \emph{variance} as distributed in VOLARE, while we forecast on the realized \emph{volatility} scale $\sigma_t = \sqrt{RV_t}$.

\begin{table}[htbp]
\centering
\singlespacing
\small
\begin{tabular}{lrrrrrrr}
\toprule
 & $N$ & Mean & Median & Skew & Kurt & $\rho_1$ & $\rho_{22}$ \\
\midrule
\multicolumn{8}{l}{\textit{Panel A: Equities (cross-sectional average, 40 stocks, 2015 to 2026)}} \\[3pt]
Average & 2,786 & 2.51 & 1.51 & 12.6 & 245 & 0.598 & 0.098 \\
\addlinespace
\multicolumn{8}{l}{\textit{Panel B: Foreign Exchange (5 pairs, 2009 to 2026)}} \\[3pt]
AUDUSD & 4,241 & 0.48 & 0.35 & 11.3 & 232 & 0.577 & 0.138 \\
EURUSD & 4,241 & 0.28 & 0.21 & 5.0 & 51 & 0.570 & 0.263 \\
GBPUSD & 4,242 & 0.32 & 0.23 & 27.5 & 1,118 & 0.484 & 0.061 \\
USDCAD & 4,240 & 0.25 & 0.18 & 4.3 & 35 & 0.654 & 0.368 \\
USDJPY & 4,239 & 0.33 & 0.21 & 14.4 & 350 & 0.325 & 0.063 \\
\addlinespace
\multicolumn{8}{l}{\textit{Panel C: Futures (5 contracts, 2009 to 2026)}} \\[3pt]
Corn (C)         & 4,185 & 2.48 & 1.54 & 12.3 & 238 & 0.178 & 0.079 \\
Crude Oil (CL)   & 4,223 & 6.87 & 2.97 & 57.6 & 3,550 & 0.195 & 0.047 \\
E-mini S\&P (ES) & 4,224 & 1.11 & 0.52 & 12.9 & 212 & 0.779 & 0.110 \\
Gold (GC)        & 4,206 & 3.63 & 0.66 & 62.4 & 3,968 & 0.000 & 0.000 \\
Natural Gas (NG) & 4,221 & 10.30 & 6.63 & 19.1 & 522 & 0.187 & 0.267 \\
\bottomrule
\end{tabular}
\caption{Descriptive statistics for daily realized variance measured at the 5-minute frequency. The equity panel reports cross-sectional averages across 40 stocks, and the FX and futures panels report individual assets. $N$ is the number of daily observations. Kurt is the excess kurtosis. $\rho_k$ denotes the sample autocorrelation at lag $k$. Mean and Median are in decimal squared returns ($\times 10^{-4}$).}
\label{tab:descriptive}
\end{table}

\section{Methodology}
\label{sec:methodology}

In this section, we describe the 17 forecasting models in our comparison, which are the eight econometric benchmarks (Subsec.~\ref{subsec:benchmarks}) and the nine TSFMs (Subsec.~\ref{subsec:tsfm_method}). We then assess the pretraining data of the TSFMs and their contamination risk (Subsec.~\ref{subsec:contamination}) and set out the forecast evaluation (Subsec.~\ref{subsec:evaluation}).

\subsection{Econometric Benchmarks}
\label{subsec:benchmarks}

We consider eight econometric specifications that span the main approaches to realized volatility forecasting: the HAR family and its extensions, which aggregate lagged realized volatility at daily, weekly, and monthly horizons; ARFIMA, which models long memory through fractional integration; an ARMA model on log realized volatility; and the MEM, which enforces positivity through a multiplicative structure. We forecast realized volatility $\sigma_t \equiv \sqrt{RV_t}$ throughout, where $RV_t$ is the realized variance stored in VOLARE. We estimate the five pure econometric models (HAR, Log-HAR, ARFIMA, ARMA, and MEM) directly on the volatility series $\sigma_t$. The augmented HAR variants (HAR-J, HAR-RS, and HARQ) instead carry variance-scale regressors (jumps, semivariances, and quarticity), so we estimate these on the variance scale $RV$ and map their forecasts to volatility. We restrict the comparison to models that forecast the realized volatility series directly. Realized GARCH \citep{hansen2012realgarch} and Realized EGARCH \citep{hansen2016realegarch} instead jointly model daily returns and a realized measure to forecast the return conditional variance. Mapping them onto our univariate realized-volatility target would require the return series and an auxiliary measurement equation, placing them on a different information set, so we leave them out of the comparison.

\subsubsection{Model specifications}
The HAR model of \citet{corsi2009} regresses volatility on its averages over the last day, week and month,
\begin{equation}
  \widehat{\sigma}_{t+1} = \beta_0 + \beta_1 \sigma_t + \beta_2 \sigma_{t-5,t} + \beta_3 \sigma_{t-22,t},
  \label{eq:har}
\end{equation}
where $\sigma_{t-k,t} \equiv k^{-1}\sum_{i=0}^{k-1} \sigma_{t-i}$ is the average volatility over the last $k$ days, with the overlapping weekly and monthly averages of the original specification. We estimate Eq.~\eqref{eq:har} by least squares with a non-negative intercept and non-negative coefficients, a sufficient condition for non-negative forecasts analogous to the GARCH constraints of \citet{nelson1992}.

Three extensions, which we specify on realized variance, add regressors built from the intraday returns $r_{t,i}$ of day $t$. HAR-J \citep{andersen2007roughing} adds the jump component $J_t = \max(RV_t - BPV_t,\, 0)$, where $BPV_t$ is the bipower variation of \citet{barndorffnielsen2004bipower},
\begin{equation}
  \widehat{RV}_{t+h} = \beta_0 + \beta_1 RV_t + \beta_2 RV_{t-5,t} + \beta_3 RV_{t-22,t} + \beta_J J_t.
  \label{eq:har_j}
\end{equation}
HAR-RS splits the daily component into the realized semivariances $RS_t^{-}=\sum_{i} r_{t,i}^2\,\mathbb{1}\{r_{t,i}\le 0\}$ and $RS_t^{+}=\sum_{i} r_{t,i}^2\,\mathbb{1}\{r_{t,i}>0\}$ of \citet{barndorffnielsen2010semivariance}, which sum to $RV_t$. The decomposition of all three components appears only in the online appendix of \citet{patton2015}, and in the main specification of that study, which we use, the weekly and monthly components remain total realized variance,
\begin{equation}
  \widehat{RV}_{t+h} = \beta_0 + \beta_1^+ RS_t^+ + \beta_1^- RS_t^- + \beta_2 RV_{t-5,t} + \beta_3 RV_{t-22,t}.
  \label{eq:har_rs}
\end{equation}
HARQ \citep{bollerslev2016harq} lets the coefficient on the daily component vary with the realized quarticity $RQ_t=\tfrac{n}{3}\sum_{i} r_{t,i}^4$, where $n$ is the number of intraday returns on day $t$, so the daily component receives less weight when its measurement error is large,
\begin{equation}
  \widehat{RV}_{t+h} = \beta_0 + (\beta_1 + \beta_1^Q \sqrt{RQ_t})\, RV_t + \beta_2 RV_{t-5,t} + \beta_3 RV_{t-22,t}.
  \label{eq:harq}
\end{equation}

Log-HAR applies the HAR specification to the logarithm of volatility,
\begin{equation}
  \widehat{\log \sigma}_{t+1} = \beta_0 + \beta_1 \log \sigma_t + \beta_2 \log \sigma_{t-5,t} + \beta_3 \log \sigma_{t-22,t}.
  \label{eq:log_har}
\end{equation}
We recover point forecasts in levels with the bias-corrected retransformation $\widehat{\sigma}_{t+1} = \exp(\widehat{\log \sigma}_{t+1} + s^2/2)$, where $s^2$ is the estimated residual variance of the log-volatility regression. The retransformation gives the conditional mean of volatility when the residuals of the log regression are Gaussian. \citet{corsi2008volatility} show that the residuals of commonly used time-series models for logarithmic realized variance exhibit non-Gaussianity, so in our setting the correction is an approximation. \citet{demetrescu2020bias} compare bias corrections for log forecasts on realized volatilities and, while finding no correction method that is uniformly best, recommend the variance-based correction we use. Log-HAR is our headline econometric benchmark, and we compute the relative loss ratios of Sec.~\ref{subsec:evaluation} against it. It is a widely used log-space specification and does not require a positivity constraint.

ARFIMA \citep{granger1980, hosking1981} models the long memory of volatility through fractional differencing,
\begin{equation}
  \Phi(L)\,(1 - L)^d\,(\log \sigma_t - \mu) = \Theta(L)\,\varepsilon_t,
  \label{eq:arfima}
\end{equation}
where $d$ is the fractional integration parameter, $\Phi(L)$ and $\Theta(L)$ are autoregressive and moving-average lag polynomials of orders $p$ and $q$, and $L$ is the lag operator. We estimate $d$ on the interval $(-0.49, 0.49)$, where the process is stationary and invertible, by the local Whittle estimator of \citet{robinson1995}. We then fit an ARMA$(p,q)$ to the fractionally differenced series $(1-L)^{\hat{d}}(\log \sigma_t - \hat{\mu})$, with $(p,q)$ selected over $\{0,1,2\}^2$ without $(0,0)$ by the Bayesian information criterion (BIC), and re-integrate its forecasts with $(1-L)^{-\hat{d}}$. The ARMA model \citep{box1970} is the short-memory counterpart of ARFIMA, so the comparison of the two isolates the value of modeling long memory. The ARMA model is $\Phi(L)(\log \sigma_t - \mu) = \Theta(L)\varepsilon_t$, with $(p,q)$ selected over $\{0,1,2\}^2$ without $(0,0)$ by BIC at each origin. ARMA and ARFIMA forecasts in levels use the retransformation of Log-HAR.

The MEM of \citet{engle2002} writes volatility as $\sigma_t = \mu_t\,\varepsilon_t$, with $\mathrm{E}[\varepsilon_t \mid \mathcal{F}_{t-1}] = 1$ and
\begin{equation}
  \mu_t = \omega + \alpha\, \sigma_{t-1} + \beta\, \mu_{t-1},
  \label{eq:mem}
\end{equation}
with $\omega > 0$, $\alpha, \beta \ge 0$ and $\alpha + \beta < 1$, so $\mu_t > 0$ and the process is stationary. We estimate $(\omega,\alpha,\beta)$ by exponential quasi-maximum likelihood, which is consistent for the conditional mean independently of the density of the innovation.

\subsubsection{Forecast construction and estimation}
We use the five pure econometric models (HAR, Log-HAR, ARFIMA, ARMA, and MEM) in iterated multistep mode, the configuration for which these specifications are designed. This mode is typically more accurate than direct projection when the model is not badly misspecified \citep{marcellino2006}. For the volatility-specific comparison of direct versus iterated multiperiod forecasts, see \citet{ghysels2019}. HAR and Log-HAR iterate by recursive plug-in, feeding each one-step forecast back as the most recent observation, and ARFIMA, ARMA, and MEM iterate natively through their recursive structure. For Log-HAR, the value fed back into the recursion is the retransformed volatility forecast, so the log-normal correction enters every step. ARFIMA and ARMA apply the retransformation once, to the $h$-step forecast of $\log \sigma$, with the one-step residual variance $s^2$.

We estimate the three augmented HAR variants (HAR-J, HAR-RS, and HARQ) directly at each horizon $h$, because we cannot project their auxiliary regressors (jumps, realized semivariances, and the quarticity interaction) forward without an auxiliary model for each one. Estimating these specifications directly at the target horizon follows \citet{patton2015} and \citet{bollerslev2016harq}. We estimate each direct regression on the 1{,}000 most recent days whose $h$-step target is observed at the forecast origin, so no future realization enters the fit. Since these regressions are linear in realized variance, they can occasionally produce a negative or implausibly large forecast. As in \citet{bollerslev2016harq}, when a forecast falls outside the range of values of the target observed in the estimation window, we replace it with the mean of the target over that window. The other econometric models produce positive forecasts by construction, and we clip the forecasts of every model to the bounds of Sec.~\ref{subsec:evaluation}.

We re-estimate every econometric model at every origin of the rolling window defined in Sec.~\ref{subsec:evaluation}, as \citet{patton2015} do with a rolling window of four years. We do not tabulate estimated coefficients or their standard errors, since the quantities we report are out-of-sample forecast losses.

\subsection{Time Series Foundation Models}
\label{subsec:tsfm_method}

We evaluate nine TSFMs in a zero-shot setting, applied directly to realized volatility series without any domain-specific training or fine-tuning. The nine models span eight distinct architectures: Chronos-Bolt (small and base checkpoints), Moirai~2.0, Moirai-MoE, Lag-Llama, TimesFM~2.5, Toto, Sundial, and TTM.\footnote{\citet{carriero2024macro} additionally evaluate TimeGPT \citep{garza2024timegpt} for macroeconomic forecasting. We exclude TimeGPT because it is a proprietary, closed-source API that does not permit inspection of model weights or training data, so its results could not be reproduced.} Where a model is offered in multiple sizes we evaluate the small checkpoint, which the results tables denote with an ``-S'' suffix (for example, Moirai-2.0-S and Moirai-MoE-S), and the surrounding text refers to each model by its architecture name. The zero-shot approach is a deployment mode of direct practical relevance, since it does not require labeled financial data or retraining and applies directly to any new asset. We fine-tune two of the models, TTM and Sundial, in a robustness check (Sec.~\ref{sec:robustness} and Appendix~\ref{app:finetune}).

\subsubsection{Architecture and point forecast}
We write a TSFM as a parametric mapping from an observed history to a forecast distribution. We denote by $\sigma_{1:T} = (\sigma_1, \ldots, \sigma_T)$ the context window of $T$ past observations (in our case, $T = 1000$ daily realized volatility values). A TSFM with pretrained parameters $\hat{\theta}$ produces a forecast of the next $H$ values:
\begin{equation}
  \widehat{\sigma}_{T+1:T+H} = f_{\hat{\theta}}(\sigma_{1:T}),
  \label{eq:tsfm}
\end{equation}
where $f_{\hat{\theta}}$ maps the input sequence to a predictive distribution over future values.\footnote{Formally, $f_{\hat{\theta}}$ outputs a distribution $\hat{p}(\sigma_{T+1:T+H} \mid \sigma_{1:T}; \hat{\theta})$ from which we take a point forecast, as described below.} The model learns $\hat{\theta}$ during a pretraining phase on large external corpora of time series spanning weather, energy, retail, transport, and macroeconomic domains. At inference time, the model receives only the context window $\sigma_{1:T}$ and produces forecasts without any task-specific parameter updates, which is the zero-shot setting.

Every TSFM implements the mapping in Eq.~\eqref{eq:tsfm} through three stages. In the \emph{tokenization} stage, the model converts $\sigma_{1:T}$ into its internal tokens with a model-specific strategy, which we summarize in Tab.~\ref{tab:tsfm_summary}. In the \emph{encoding} stage, a transformer \citep{vaswani2017} attends across positions without imposing HAR's fixed daily, weekly, and monthly structure, and in TTM multilayer perceptrons mix across time. Chronos-Bolt uses an encoder-decoder transformer, and the other transformer models are decoder-only. In the \emph{decoding} stage, a head maps representations back to forecasts. It outputs quantiles directly, parameterizes an explicit distribution (e.g., Student-$t$) that is sampled, generates trajectories via continuous normalizing flows for Sundial, or outputs a single point forecast for TTM.

We take the mean of each model's predictive distribution as its point forecast, because the QLIKE loss of Sec.~\ref{subsec:evaluation} is minimized in expectation by the conditional mean of the variance. We use the mean rather than the median, since the median of a right-skewed volatility distribution lies below its mean. How we obtain the mean depends on what each model outputs. Specifically, TimesFM~2.5 outputs a mean directly, and the sampling models (Lag-Llama, Sundial, Moirai-MoE and Toto) use the mean of their sampled trajectories. The quantile-only models (Chronos-Bolt and Moirai~2.0) use the average of their nine predicted quantiles, at the levels 0.1 to 0.9, which approximates the mean of the predictive distribution without its tails. Finally, TTM outputs a single point forecast trained on squared error.

\begin{table}[htbp]
\centering
\singlespacing
\small
\setlength{\tabcolsep}{4.5pt}
\resizebox{\textwidth}{!}{%
\begin{tabular}{lllrll@{\hspace{6pt}}l}
\toprule
Model & Architecture & Tokenization & Params & Output & Release & Reference \\
\midrule
Chronos-Bolt-S & Encoder-decoder & Patch projection & 48M & Direct quantiles & 2024 & \citet{ansari2024chronos} \\
Chronos-Bolt-B & Encoder-decoder & Patch projection & 205M & Direct quantiles & 2024 & \citet{ansari2024chronos} \\
Moirai~2.0-S & Decoder-only & Patch projection & 11M & 9 quantile levels & 2025 & \citet{liu2025moirai2} \\
Moirai-MoE-S & Decoder-only & Patch projection & 117M & Mixture distribution & 2024 & \citet{liu2024moiraimoe} \\
Lag-Llama & Decoder-only & Lag features & 2.5M & Student-$t$ & 2024 & \citet{rasul2024lagllama} \\
TimesFM~2.5 & Decoder-only & Patch + residual & 200M & Mean + 9 quantiles & 2025 & \citet{das2024timesfm} \\
Toto & Decoder-only & Patch projection & 151M & Student-$t$ mixture & 2025 & \citet{cohen2025toto} \\
Sundial & Decoder-only & Patch projection & 128M & Generative samples & 2025 & \citet{liu2025sundial} \\
TTM & MLP-Mixer & Patch + freq.\ prefix & $<$1M & Point forecast & 2025 & \citet{ekambaram2024ttm} \\
\bottomrule
\end{tabular}%
}
\caption{Summary of TSFM architectures evaluated, with the reference for each. All models are applied zero-shot with a 1{,}000-day context window, except TTM, whose r2.1 release, the one pretrained with daily series, supports at most 512 days. ``Params'' refers to the total number of pretrained parameters. The point forecast is the mean of each model's predictive distribution, which for Chronos-Bolt and Moirai~2.0 we approximate by the average of nine quantiles. TTM outputs a single point forecast. Per-model architectural detail and the exact pretrained checkpoints are given in Appendix~\ref{app:tsfm_details}. TTM uses TSMixer, a variant of the MLP-Mixer, an architecture built from multilayer perceptrons.}
\label{tab:tsfm_summary}
\end{table}

\subsubsection{Zero-shot protocol}
All TSFMs receive a rolling context window of 1{,}000 daily realized volatility observations as input, with no additional covariates. We set the context length to 1{,}000 days to match the 1{,}000-day estimation window used for the econometric models, so that both classes of model condition on the same span of history.\footnote{We run TTM at 512 days, the longest context of its r2.1 release, the release pretrained with daily series and a frequency token. All other TSFMs use 1{,}000 days.}

For multi-step horizons ($h > 1$), the TSFM produces a trajectory of $h$ individual-step forecasts $(\widehat{\sigma}_{T+1}, \ldots, \widehat{\sigma}_{T+h})$. Our primary target is point-in-time realized volatility $\sigma_{T+h}$, the value $h$ steps ahead, so the point forecast at horizon $h$ is the $h$-th element of this trajectory, $\widehat{\sigma}_{T+h}$, not the average over the trajectory. Each trajectory element is the model's point forecast at that step, obtained as described above.

\subsection{Pretraining Data and Contamination Risk}
\label{subsec:contamination}

A natural concern with zero-shot evaluation is whether TSFM performance is inflated because the models' pretraining corpora overlap with the test series. Tab.~\ref{tab:pretraining_data} summarizes each model's training data and its financial content.

\input{tables/table_pretraining_data}

Contamination of the test series remains possible. The models we evaluate were released in 2024 and 2025, and their pretraining windows can overlap our evaluation period in calendar time. Exact pretraining data cutoffs are not published for most of these models. The date on which each model's pinned weight file was last changed on Hugging Face, its weights date, gives an upper bound, because a model cannot have been trained on data observed after its weights were written. In Sec.~\ref{sec:robustness} we use this bound for a direct check, which compares each model's accuracy before and after its weights date. The papers that introduce the foundation models we test do not report realized volatility in the pretraining data of these models, but they do not document every corpus in full. Moirai~2.0, TTM and Toto give no breakdown of their corpora by domain, Sundial does not describe the contents of its own finance collection, and for Chronos-Bolt, TimesFM~2.5 and TTM the documented corpus is that of an earlier release. For these models we cannot verify everything the corpus contains, and data seen during pretraining could carry information about market volatility indirectly, even without realized variance series.

Several pieces of evidence nonetheless make contamination an unlikely full explanation of our findings. First, the pattern of results that we report in Sec.~\ref{sec:results} is the opposite of what memorization or leakage would produce. Our findings are not a broad TSFM win, since under robust loss ratios most TSFMs do not beat Log-HAR, and only TTM does so consistently. Widespread contamination would inflate many models at once, not leave the bulk of models at or below the econometric benchmark. Second, the single model that wins, TTM, has the smallest capacity in our study (fewer than 1M parameters) and therefore the least room to memorize specific series. Leakage that benefits the lowest-capacity model is hard to reconcile with a memorization account. Third, realized variance is a specific intraday-derived second moment computed at 5-minute sampling, absent from the known public corpora (e.g., Monash,\footnote{\url{https://forecastingdata.org/}} LOTSA\footnote{The Large-scale Open Time Series Archive of \citet{woo2024moirai}.}), which contain raw series rather than this derived statistic. The known corpora also hold little financial data. Where the papers allow a count, finance and economics series make up at most 0.09\% of pretraining observations, mostly competition and macroeconomic data, except for Sundial, whose own finance collection is 1.02\% of its corpus (Tab.~\ref{tab:pretraining_data}). The more parsimonious reading of the pattern is that the modest advantage we document reflects transfer of general temporal structure (mean reversion, long memory, regime persistence) rather than memorization of specific financial dynamics. We carry contamination into the conclusion as a limitation.

\subsection{Forecast Evaluation}
\label{subsec:evaluation}

We employ a walk-forward evaluation scheme with a rolling origin. Econometric models use a fixed estimation window of 1{,}000 trading days (approximately four years). At each origin, we estimate the model on the most recent 1{,}000 observations, and the window slides forward one day at a time, producing daily re-estimated forecasts. Daily re-estimation is the most demanding refresh cadence and avoids any look-ahead from stale coefficients, at no cost to the comparison because the same timing applies to every model. A window of roughly four years is standard in the realized volatility forecasting literature and is long enough to limit the sensitivity of least-squares estimates to individual volatility spikes \citep{bollerslev2016harq, clements2021practical}. Foundation models follow the same walk-forward timing but do not require an estimation step, and only the context window slides forward.

We compare all models on the common out-of-sample period in which both econometric and TSFM forecasts are available. For the equity sample (2,786 days), a 1,000-day window leaves 1,786 daily forecasts per model. The three augmented HAR variants (HAR-J, HAR-RS, HARQ) need an extra 22-day monthly lag to construct their auxiliary regressors and therefore begin 22 days later. The common out-of-sample window shared by all 17 models is thus 1,764 forecasts per asset at $h = 1$. The FX and futures samples (over 4,000 days) yield substantially more.

We evaluate three forecast horizons: $h = 1$ (one day), $h = 5$ (one week), and $h = 22$ (one month). The forecast target is the point-in-time realized volatility $h$ days ahead, $\sigma_{t+h} = \sqrt{RV_{t+h}}$, not an average over the intervening days. A multi-day average shares $(h-1)/h$ of its content with the average at the next origin, which induces serial correlation in the target and overstates its persistence relative to the underlying daily series. The average is therefore not the quantity that a forecaster with day-$t$ information wants to predict at horizon $h$. The point-in-time target is our primary target, and in Appendix Tab.~\ref{tab:avg_target} we repeat the comparison for the average of daily realized variance over the next $h$ days, where the main conclusions hold.

We report MSE on the volatility scale, computed on $\sigma_{t+h}$ and $\widehat{\sigma}_{t+h}$ directly. We focus, however, on the QLIKE loss of \citet{patton2011}, which is robust to noise in the realized-variance proxy and less sensitive to extreme volatility observations than MSE:
\begin{equation}
  L_{\text{QLIKE}}(\widehat{RV}_t, RV_t) = \frac{RV_t}{\widehat{RV}_t} - \log\frac{RV_t}{\widehat{RV}_t} - 1,
  \label{eq:qlike}
\end{equation}
where $\widehat{RV}_t$ is the variance forecast and $RV_t$ is the realized variance. QLIKE is a loss on the variance scale, and its proxy-robustness property in \citet{patton2011}, consistency of the forecast ranking under a conditionally unbiased variance proxy, holds for the variance, not its square root. We therefore evaluate QLIKE by squaring the volatility forecast back to a variance, $\widehat{RV}_t = \widehat{\sigma}_t^2$, and using realized variance $RV_t = \sigma_t^2$ as the proxy. This keeps MSE on the interpretable volatility scale while preserving QLIKE's proxy-robustness guarantee.

Since a few high-loss series can dominate pooled averages across assets, we also report average loss ratios relative to Log-HAR, for which we divide each asset's QLIKE by Log-HAR's QLIKE on that asset and average the ratios across assets. These appear in Tab.~\ref{tab:loss_ratios}.

Since the forecast that minimizes expected QLIKE is the conditional mean of the variance proxy, $\mathrm{E}_t[RV_{t+h}]$, the square of a volatility forecast misses this target even when the volatility forecast is the exact conditional mean. By Jensen's inequality the square falls below the target, since $\mathrm{E}_t[RV_{t+h}] = (\mathrm{E}_t[\sigma_{t+h}])^2 + \mathrm{Var}_t(\sigma_{t+h})$, and QLIKE penalizes a variance forecast below the realized variance more than one above it by the same amount. The three augmented HAR variants are the exception, because they forecast realized variance directly. We square each model's volatility forecast as the model issues it and do not adjust its level, because the object of our comparison is the zero-shot forecast itself. A level adjustment would use the model's past forecast errors, which a zero-shot forecast does not use.

The shortfall is not the same for every model, because the models' point forecasts target different quantities. For example, Log-HAR, ARFIMA and ARMA retransform a forecast of log volatility with a correction that gives the mean of volatility only when the residuals are Gaussian. Across models and horizons, the realized variance exceeds the squared forecast by 3 to 72\% on average. Since every model forecasts too low a variance, a model whose variance forecast lies closer to the realized level has a lower QLIKE for that reason alone. This holds even if its forecast follows the movements of volatility no more closely than that of another model. For models whose loss ratios differ by a few percent, the level of the forecasts can therefore change the ranking. To measure how much, in Appendix~\ref{app:forecast_level} we recompute the loss ratios after two level adjustments, each relative to Log-HAR adjusted the same way. The first scores Log-HAR, ARFIMA and ARMA on their log-normal expected variance, and the second rescales every model in real time by its average ratio of realized to forecast variance over past forecasts.

We winsorize each forecast to the range of realized volatility observed for that asset over the 1{,}000 trading days before the forecast origin, $[\min_{t-1000 \le s < t} \sigma_s,\ \max_{t-1000 \le s < t} \sigma_s]$ for a forecast made at origin $t$. We estimate the econometric models on the same window, so the bounds use only information available at the origin. The range is wide and leaves all but the most extreme forecasts unchanged. Since QLIKE diverges as the forecast approaches zero, an unbounded or floor-clipped forecast can distort it. Bounding to the data's own support removes that distortion without an arbitrary constant. The lower bound keeps forecasts within the range of volatility the model was estimated on, and the upper bound guards against the occasional extreme spike that the heavy-tailed predictive distributions of some TSFMs can produce. We apply the bound symmetrically to all 17 models, eight econometric and nine TSFM specifications, so that we treat every model class in the same way. The bound binds rarely and matters mainly for Toto, whose heavy-tailed predictive distribution produces most of the forecasts outside it.

Pairwise forecast comparisons use the DM test on QLIKE loss differentials, with a Newey--West heteroskedasticity-and-autocorrelation-consistent variance \citep{newey1987} using a Bartlett kernel and $\max(1, h-1)$ lags to absorb the autocorrelation that multi-step loss differentials carry by construction. To address the multiple comparison problem inherent in evaluating 17 models, we compute the MCS, which identifies the subset of models whose forecasting ability cannot be statistically distinguished from the best model at a given significance level. We use the range statistic $T_R$ of \citet{hansen2011mcs}, the largest absolute $t$-statistic of the pairwise loss differences. When the test rejects, we eliminate the model with the higher mean loss in the pair that attains $T_R$, the elimination rule that the same study pairs with this statistic. The bootstrap is a moving-block bootstrap (block length 22 days, one trading month and the longest forecast horizon, $B = 10{,}000$ replications) at the $\alpha = 0.10$ level.

Additionally, we assess forecast efficiency using the MZ regression \citep{mincer1969}:
\begin{equation}
  \sigma_t = \alpha + \beta \, \widehat{\sigma}_t + \varepsilon_t,
  \label{eq:mz}
\end{equation}
where $\widehat{\sigma}_t$ is the model's volatility forecast and $\sigma_t$ the realized volatility. We run the regression on the volatility scale. Under forecast optimality, $\alpha = 0$ and $\beta = 1$, meaning the forecast is unbiased and captures the correct scale of variation. We test the joint null $H_0\!: \alpha = 0, \beta = 1$ using a Wald test with Newey--West standard errors and report cross-asset averages and rejection rates at the 5\% level.

To examine whether relative forecast performance is stable over time, we apply the GR fluctuation test \citep{giacomini2010}. For each model paired against the Log-HAR benchmark, we compute a rolling DM statistic $S_t$ over a window of size $m = \lfloor 0.3 \times P \rfloor$, where $P$ is the number of out-of-sample loss differentials. The path of $S_t$ over time traces the relative QLIKE performance. The window fraction follows that study. The test statistic is $\sup_t |S_t|$, compared against the critical values that the study tabulates for the limit of the statistic. The limit is the supremum of $|B(\tau) - B(\tau - \mu)|/\sqrt{\mu}$, with $B$ a standard Brownian motion and $\mu$ the window as a fraction of the sample. A rejection indicates that the two models do not have equal predictive ability at every point in the sample, with one model significantly more accurate over some subperiod.

\section{Empirical Results}
\label{sec:results}

We first examine aggregate forecast accuracy across loss functions and horizons, and then analyze cross-asset and cross-market heterogeneity.\footnote{Replication code is available at \url{https://github.com/Alessiobrini/tsfm-rv}.}

\subsection{Aggregate Forecast Accuracy}
\label{subsec:main_results}

\begin{figure}[!tb]
  \centering
  \includegraphics{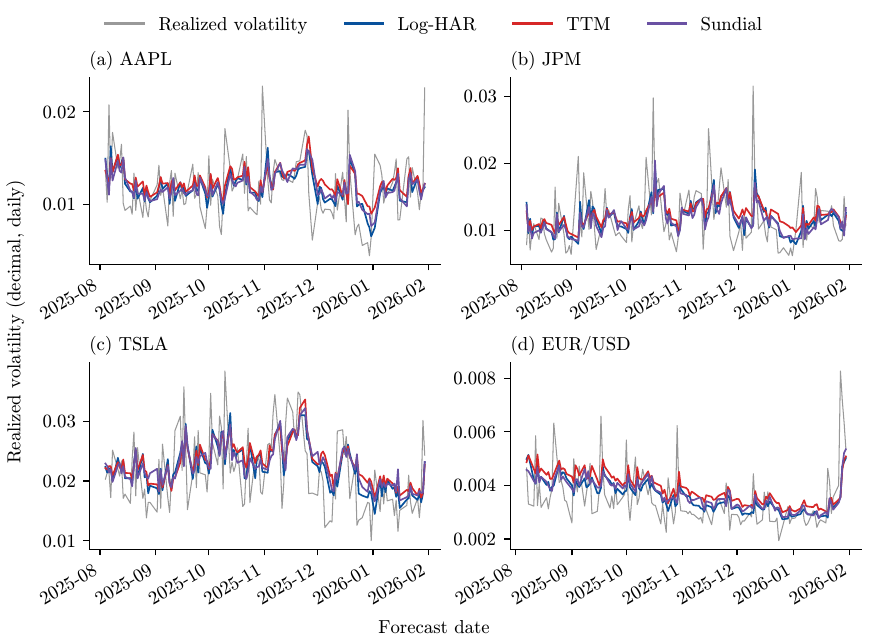}
  \caption{Daily realized volatility and the one-day-ahead ($h = 1$) forecasts of Log-HAR, TTM and Sundial for AAPL (a), JPM (b), TSLA (c) and EUR/USD (d), over the last 125 days of each asset's evaluation sample, from August 2025 to 30 January 2026. Realized volatility is the square root of the daily realized variance from 5-minute returns, in decimal units.}
  \label{fig:forecast_vs_actual}
\end{figure}

In Tab.~\ref{tab:main_results} we report the cross-sectional average of per-asset losses across the 40 equities, and in Tab.~\ref{tab:pooled50} we pool the same per-asset losses across all 50 assets (equities, FX, and futures). In Fig.~\ref{fig:forecast_vs_actual} we plot the forecasts against realized volatility for three equities and one currency pair.

\input{tables/table_equity_metrics}

\input{tables/table_pooled50}

On the pooled cross-sectional average across all 50 assets (Tab.~\ref{tab:pooled50}), TTM achieves the lowest pooled QLIKE at every horizon. Log-HAR, HAR and several foundation models follow within a narrow band, with the gaps among the leaders rarely exceeding 0.02 to 0.03 in QLIKE. Since the pooled mean gives the most weight to the assets with the largest losses, we rank the models by the average loss ratio to Log-HAR of Subsec.~\ref{subsec:evaluation} (Tab.~\ref{tab:loss_ratios}), which weights every asset equally. Under this measure, TTM is the only foundation model that beats Log-HAR at every horizon, with an average loss 1.1 to 1.6\% below that of Log-HAR. TTM, the smallest model in the evaluation, has fewer than one million parameters.

\input{tables/table_loss_ratios}

No other foundation model matches TTM's consistency. At the daily horizon TimesFM~2.5 and Sundial are about as accurate as Log-HAR, and both fall behind at the two longer horizons. Moirai~2.0, Moirai-MoE-S, Chronos-Bolt, and Toto have loss ratios above one at all three horizons, so they lose to Log-HAR on the typical asset. The competitive econometric benchmarks, by contrast, cluster near Log-HAR. HAR matches it at $h = 1$ before deteriorating at longer horizons, and ARFIMA, ARMA, and the MEM are about as accurate as Log-HAR at the daily horizon and within about 13\% of it at the monthly horizon (Tab.~\ref{tab:loss_ratios}).

The pooled means and the loss ratios agree on which models are at or below Log-HAR. At the daily horizon these are HAR, TimesFM~2.5, Sundial and TTM, and at the two longer horizons TTM alone. The loss function matters more than the aggregation. Chronos-Bolt and TimesFM~2.5 have a lower pooled MSE than Log-HAR at every horizon (Tab.~\ref{tab:pooled50}), while their QLIKE loss ratios are above one beyond the daily horizon. QLIKE penalizes a variance forecast below the realized variance more than one above it, and these models forecast a lower variance than Log-HAR on average (Appendix Tab.~\ref{tab:forecast_level}). Under either level adjustment of Subsec.~\ref{subsec:evaluation}, TTM no longer beats Log-HAR at the daily and weekly horizons, while at the monthly horizon it remains the most accurate model (Appendix~\ref{app:forecast_level}). The ranking by QLIKE loss ratio does not depend on how the per-asset ratios are averaged. Without the adjustment and under a scale-symmetric geometric mean, TTM remains the only model below one at every horizon, so the arithmetic average is not what produces the result.

In Tab.~\ref{tab:loss_ratios}, the three augmented HAR variants are less accurate than HAR at the daily horizon and move close to it as the horizon lengthens, and at $h = 22$ HAR-J and HAR-RS carry lower loss ratios than HAR. Most of the daily gap comes from the estimation method. The augmented variants are linear regressions on realized variance, and a HAR estimated in the same way falls behind the HAR of Tab.~\ref{tab:loss_ratios} by a similar margin.\footnote{A HAR estimated as a linear regression on realized variance, with the same replacement of out-of-range forecasts as the augmented variants, has loss ratios to Log-HAR of 1.187, 1.137, and 1.049 at $h = 1$, 5, and 22.} Relative to this HAR, HAR-J and HAR-RS are within three percent at every horizon, while HARQ falls behind it at every horizon, by about seven percent at the daily horizon and less at the longer ones. \citet{bollerslev2016harq} find HARQ ahead of HAR for individual stocks, and our sample does not reproduce that gain at the short horizons. The quarticity regressor is itself measured from 5-minute returns, so it adds measurement noise that a direct regression can pass into the forecast, and we read HARQ's weaker short-horizon result as specific to this sample and sampling frequency. The long-memory and multiplicative benchmarks, by contrast, stay close to Log-HAR at the daily horizon and within about 13 percent of it at the longer horizons.

At $h = 1$ and $h = 5$, Toto ranks eighth of the nine foundation models on both pooled QLIKE and the loss ratio. Its largest losses come from the E-mini S\&P 500 futures contract (ES), where its QLIKE is the highest of any asset at every horizon, and at $h = 5$ this contract alone accounts for 15.7\% of Toto's pooled QLIKE.

\subsection{Cross-Asset and Cross-Market Heterogeneity}
\label{subsec:cross_asset}

In Tab.~\ref{tab:fx_futures_results} we extend the analysis to the FX and futures markets.

\input{tables/table_fx_futures_metrics}

The FX sample covers five major currency pairs with lower volatility levels and moderate persistence. On FX, Log-HAR is at or near the top at every horizon, and the foundation-model advantage, where it exists, is small. On the cross-sectional average (Tab.~\ref{tab:fx_futures_results}, Panel~A), the spread among the leading models is narrow. At $h = 1$, TTM and the leading econometric and foundation models cluster within about 0.005 of each other on QLIKE. At $h = 5$, Log-HAR ties the best foundation models on QLIKE, so no foundation model beats the best econometric specification. At $h = 22$, Log-HAR has the lowest QLIKE, with HARQ, HAR-RS and TTM close behind, and Moirai-MoE-S has the highest QLIKE of the 17 models.

On the futures (Tab.~\ref{tab:fx_futures_results}, Panel~B), Sundial and TTM tie for the lowest QLIKE at $h = 1$, with HAR, TimesFM~2.5 and Log-HAR just behind. At $h = 5$, Log-HAR has the lowest QLIKE, just ahead of HAR and TTM, and at $h = 22$ Log-HAR leads by a wider margin, with HAR, HAR-J, HAR-RS and TTM following. Toto's QLIKE spikes on the ES contract, consistent with the outlier noted above.

The first of three patterns is that TTM is the most consistent foundation model across classes. It is at or near the lowest QLIKE on equities, FX, and futures at most horizons, and it is the only foundation model to beat Log-HAR under the equal-weighted loss ratio at every horizon (Tab.~\ref{tab:loss_ratios}). Second, Log-HAR is the best or near-best specification on FX and futures across horizons, and on equities it is second to TTM at $h = 5$ and within 0.02 of the lowest QLIKE at the other two horizons. This result is consistent with the practical value of the log transformation, which keeps forecasts positive and aligns them with QLIKE. Third, dispersion within the foundation-model class is wide. Apart from TTM, only TimesFM~2.5 and Sundial are about as accurate as Log-HAR, and only at the daily horizon, while Moirai~2.0, Moirai-MoE-S, Chronos-Bolt, and Toto lose to Log-HAR on the typical asset. Realized volatility is persistent but stationary and mean-reverting, which makes it closer to the series these models encounter in pretraining than macroeconomic series are. Some foundation models are nonetheless relatively less accurate on the more persistent assets (Appendix~\ref{subsec:drivers}).

\begin{figure}[H]
  \centering
  \includegraphics{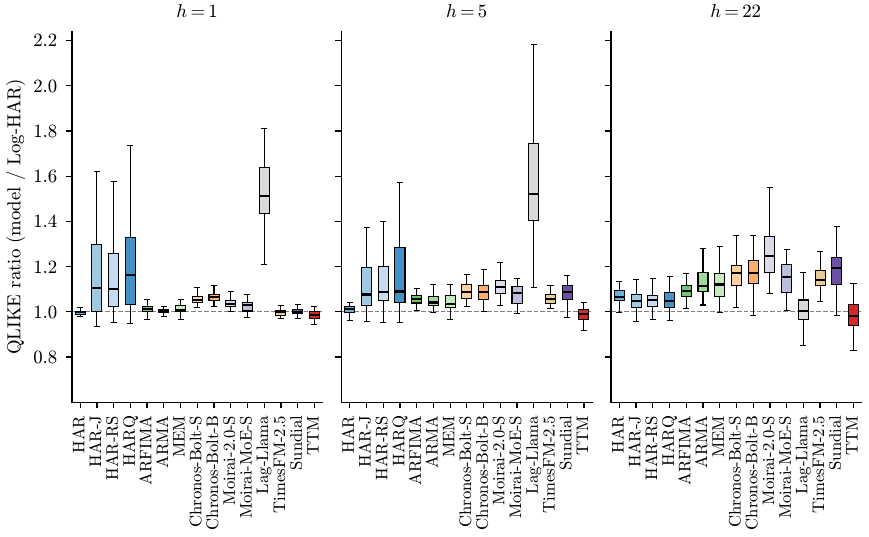}
  \caption{Distribution across the 50 assets of the per-asset QLIKE ratio of each model to Log-HAR at $h = 1$ (left), $h = 5$ (center) and $h = 22$ (right). A ratio below one means that the model has a lower QLIKE than Log-HAR on that asset. Boxes span the interquartile range, whiskers extend to 1.5 times the interquartile range, and outliers are not shown. The dashed line marks parity with Log-HAR. The plot shows every econometric benchmark and every foundation model except Toto, whose per-asset QLIKE ratio reaches 6.8 at the weekly horizon and would compress the scale.}
  \label{fig:qlike_boxplot}
\end{figure}

In Fig.~\ref{fig:qlike_boxplot} we plot the distribution of the QLIKE ratios to Log-HAR across all 50 assets. Values below one indicate the model outperforms Log-HAR. At the daily horizon the competitive econometric benchmarks (HAR, ARFIMA, ARMA, MEM) cluster at or just above parity, and at longer horizons they move above it. The augmented HARs lie above parity at the short horizons and close to it at the monthly horizon. TTM is the only model whose distribution sits predominantly below one at all three horizons.

\section{Statistical Significance and Model Confidence Sets}
\label{sec:significance}

The rankings in Sec.~\ref{sec:results} show differences across models, but mean loss comparisons can be misleading when distributions are skewed and sample sizes vary across assets. We apply the MCS \citep{hansen2011mcs} to identify the subset of models that cannot be distinguished from the best, and pairwise DM tests \citep{diebold1995} to quantify directional win rates. We also run MZ regressions \citep{mincer1969} to assess forecast efficiency and GR fluctuation tests \citep{giacomini2010} to detect time variation in relative performance. We report MCS inclusion rates and pairwise DM win rates for all 50 assets in Tab.~\ref{tab:mcs}.

\input{tables/table_mcs_dm}

TTM enters the MCS for every asset at every horizon, and Log-HAR for nearly every asset (Tab.~\ref{tab:mcs}). TimesFM~2.5 and Sundial also enter for nearly every asset at every horizon, and at the daily horizon so do ARMA, HAR and HAR-J, and ARFIMA for most. Chronos-Bolt and Moirai~2.0 enter for few assets at the daily horizon and more often at the longer horizons. Lag-Llama enters for half of the assets at the daily horizon and for most of them at the monthly horizon, in line with its low pooled QLIKE there. The MCS therefore never excludes TTM, but at the 10\% level it also contains Log-HAR, TimesFM~2.5, Sundial and several econometric models on most assets. The set is wide because the QLIKE differences on a single asset are noisy and the MCS controls for all 136 pairs of models at once. At the daily horizon HAR-J has a loss ratio of 1.181, yet a DM test finds it significantly worse than Log-HAR on only 11 of 50 assets, and the MCS keeps it on 8 of those 11. The set for each asset is in Appendix Fig.~\ref{fig:mcs_heatmap}.

We compute the DM test pairwise. For each of the 17 models in Tab.~\ref{tab:dm_summary}, we test it against each of the remaining 16 on each of the 50 assets. In Tab.~\ref{tab:dm_summary} we report the fraction of those tests in which each model has a significantly lower QLIKE at the 5\% level.\footnote{These win rates are unadjusted for multiple comparisons. When we apply a Benjamini--Hochberg false-discovery-rate correction \citep{benjamini1995} at 5\% within each model's 800 tests, every win rate shrinks, most at the monthly horizon where pairwise differences are weak, but the ranking stays the same. TTM retains a higher adjusted win rate than Log-HAR at $h = 1$ and $h = 5$, and the small unadjusted gap at $h = 22$ becomes a tie.} TTM wins the highest fraction at the short and medium horizons. Log-HAR is the most consistent econometric model and wins slightly more comparisons than TTM at $h = 22$. Win rates compress across all models at $h = 22$, consistent with forecast differences narrowing as horizons lengthen. The Chronos-Bolt checkpoints and Moirai-MoE-S win a small share of comparisons beyond the daily horizon, so their loss-ratio shortfall carries over to pairwise losses. At the weekly horizon TimesFM~2.5 wins as many comparisons as ARFIMA and the MEM.

We report the average MZ regression results across 50 assets for all three horizons in Tab.~\ref{tab:mz_all}. Efficiency varies with the horizon. At the daily horizon Log-HAR, HAR, Sundial, ARFIMA and ARMA reject joint efficiency for few assets, while TTM rejects it for most assets, TimesFM~2.5 for fewer than half, and the other foundation models for nearly all. TTM's average intercept and slope across assets are nonetheless close to zero and one. As the horizon lengthens to $h = 22$, the econometric benchmarks undershoot more sharply, with slopes well below one and MZ $R^2$ falling toward zero. TTM, Moirai~2.0 and TimesFM~2.5 have slopes closer to one than any econometric benchmark, consistent with the higher persistence that \citet{christensen2023ml} find in flexible models at longer horizons. The MZ regressions describe the calibration and efficiency of each forecast and do not by themselves show where its accuracy comes from.

\input{tables/mz_regression_all}

\begin{figure}[htbp]
  \centering
  \includegraphics{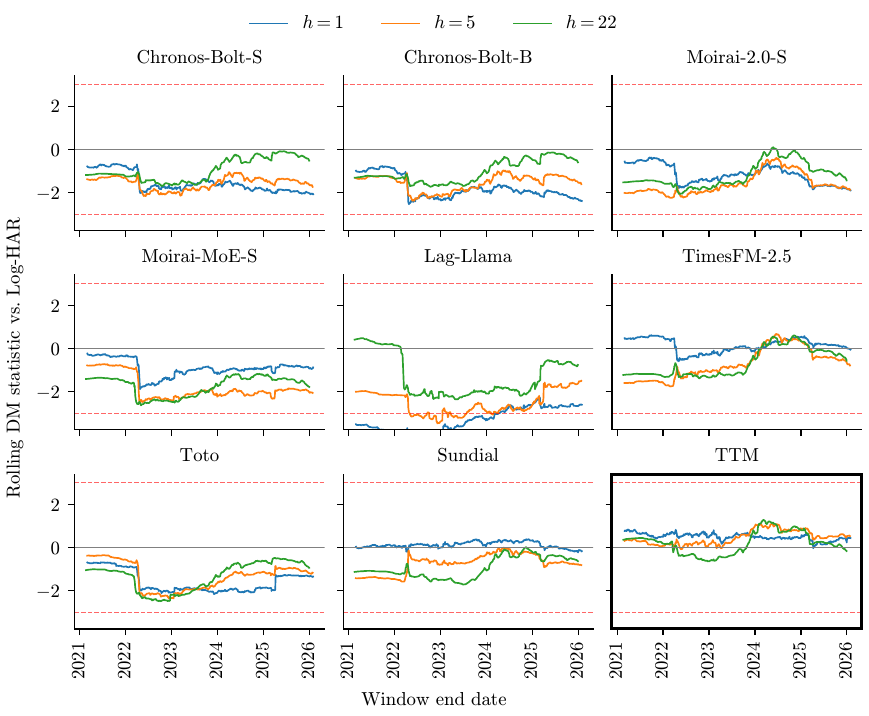}
  \caption{GR fluctuation test of each foundation model against Log-HAR, one panel per model, with the TTM panel outlined. For each asset we compute the DM statistic of the QLIKE loss of Log-HAR minus that of the model over a rolling window of 30\% of the asset's forecast dates, and date it at the end of the window. Its variance is a Newey--West estimate with $\max(1, h-1)$ lags. Each line is the average of this statistic across the 50 assets on the dates at which every asset has a complete window, February 2021 to January 2026, for $h = 1$, 5 and 22, and the solid gray line marks zero. Positive values mean that the model has the lower QLIKE. The dashed lines mark the 5\% critical values of \citet{giacomini2010} for the statistic of a single asset in this window. The vertical axis spans the 1st to 99th percentiles of all plotted values, widened where needed to show the critical values.}
  \label{fig:gr_fluctuation}
\end{figure}

In Fig.~\ref{fig:gr_fluctuation} we plot the rolling DM statistic (QLIKE loss) of each foundation model against Log-HAR across the three forecast horizons, averaged across the 50 assets. The average covers each date from February 2021 to January 2026, the dates on which every asset has a complete rolling window. A positive rolling DM indicates that the comparison model produces lower QLIKE than Log-HAR during that window, so the foundation model outperforms the benchmark, and a negative value indicates that Log-HAR is the more accurate. The $\pm 3.01$ bands are the 5\% critical values for the statistic of a single asset. Since the average of the 50 statistics has a variance of at most one, a value beyond the bands still rejects equal predictive ability at the 5\% level, and a value inside them does not show equal accuracy.

In the GR tests, the relative performance of most foundation models against Log-HAR keeps the same sign on most dates. TTM's rolling paths are the most stable. Its rolling statistic is above zero on nearly every date at the daily and weekly horizons and on about two thirds of the dates at the monthly horizon. Its advantage over Log-HAR therefore has a stable sign, and its path stays well inside the 5\% bands. At the daily horizon Sundial and TimesFM~2.5 are above parity on most dates (83\% and 70\%), and the other foundation models are below parity on most dates at every horizon, consistent with their loss ratios. Most rolling statistics nonetheless remain within the 5\% bands. Only Lag-Llama breaches the lower band over sustained windows, at the daily and weekly horizons, and in those periods Log-HAR is significantly more accurate. The unconditional DM and MCS results therefore describe the sign of each comparison on most dates, while the size and significance of each difference vary over the sample.

The four tests point to the same conclusion as the loss ratios. The foundation-model class is not uniformly strong, and only TTM has the lowest loss ratio at every horizon and is never excluded from the MCS, though a competitive set of econometric benchmarks stays close behind it. The wide differences in accuracy across foundation-model architectures indicate that the choice of model within the class matters more than membership in the foundation-model class.

\section{Economic Value of the Forecasts}
\label{sec:economic}

The losses and tests in Secs.~\ref{sec:results} and~\ref{sec:significance} give a statistical ranking of the forecasts, but they do not measure what a more accurate forecast is worth to an investor. We measure that value with the realized utility of \citet{bollerslev2018risk}, which values the volatility forecast of one asset and does not need forecasts of returns or correlations. It therefore fits our setting, in which we forecast the volatility of each asset separately. An investor with mean-variance preferences and relative risk aversion $\gamma$ holds one asset with a constant Sharpe ratio ($SR$) and a risk-free account. With a forecast $F_t$ of the daily variance over the holding period, the investor holds the fraction $x_t = (SR/\gamma)/\sqrt{252 F_t}$ of wealth in the asset, which targets an annual volatility of $SR/\gamma$. The realized utility of this position, per unit of wealth and per year, is
\begin{equation}
{U_t = \frac{SR^2}{\gamma}\left(\sqrt{\frac{RV^{(h)}_t}{F_t}} - \frac{1}{2}\,\frac{RV^{(h)}_t}{F_t}\right),}
\label{eq:utility}
\end{equation}
where $RV^{(h)}_t$ is the mean daily realized variance over the $h$ days of the holding period. As \citet{bollerslev2018risk} do, we set $SR = 0.4$ and $\gamma = 2$, so the investor targets a volatility of 20\% a year and a perfect forecast earns 4\% of wealth a year. The utility depends on the forecast only through the ratio $RV^{(h)}_t/F_t$, so before trading costs the ranking of the models does not depend on $SR$ and $\gamma$. Like QLIKE, the utility is highest when the variance forecast equals the realized variance, and it falls more for a forecast that is too low than for one that is too high. The level of the forecasts documented in Appendix~\ref{app:forecast_level} therefore matters here as well.

The investor rebalances every day, so at $h = 5$ and $h = 22$ the holding periods of consecutive forecasts overlap. At $h = 1$, $F_t$ is the square of the point forecast of the main comparison and $RV^{(1)}_t$ is the realized variance of the next day. At $h = 5$ and $h = 22$, $F_t$ is the square of the forecast of the average volatility over the next $h$ days, the averaged target of Appendix Tab.~\ref{tab:avg_target}, and for the three augmented HAR variants it is their forecast of the average variance. In that case $RV^{(h)}_t$ is the mean daily realized variance over those days, computed in the same way for every model.

With full costs, every daily change in position, $|x_t - x_{t-1}|$, pays the asset's proportional bid-ask spread. With gradual trading, the position moves 15\% of the way toward $x_t$ each day and pays the same spread on its changes. This is the rate of \citet{bollerslev2018risk}, which applies the partial adjustment of \citet{garleanu2013dynamic} to slow down trading. The spread of each asset is the median over the last nine months of the sample, May 2025 to January 2026. For the stocks it is the median daily quoted spread, which we obtain from the Trade and Quote (TAQ) intraday indicators of Wharton Research Data Services, between 0.7 and 9.9 basis points. For the futures it is the minimum price increment of the contract divided by the median price. For the currencies it is one pip divided by the median price, between 0.7 and 1.5 basis points, in line with the effective spreads that \citet{karnaukh2015understanding} report for the major currency pairs.

We average each model's utility over the forecast origins of each asset and then over the 50 assets. We test the difference from Log-HAR with a DM test on a daily series that weights the assets equally. We also report in Tab.~\ref{tab:economic_value} the equal-weight average of the TTM and Log-HAR forecasts, which we study in Sec.~\ref{sec:combination}, and the infeasible perfect forecast.

\input{tables/table_economic_value}

Before trading costs, the gains from a more accurate forecast are small next to the value of a perfect one, which earns 46 basis points a year more than Log-HAR at $h = 1$. TTM adds about 2 basis points a year at $h = 1$ and $h = 5$, both significant, and its gain at $h = 22$ is not significant. Toto and Lag-Llama lose at every horizon, and every other foundation model earns less than Log-HAR at $h = 5$ and $h = 22$. The average of TTM and Log-HAR gains less than TTM at the daily and weekly horizons but on more assets, with a significant gain for 42 of the 50 assets at $h = 1$, against 28 for TTM.

Trading costs lower the utility of every forecast, but less for a forecast that changes less from one day to the next. The position $x_t$ is proportional to $1/\sqrt{F_t}$, so a forecast that changes less moves the position less, and every change in position pays the spread. At $h = 1$, TTM's daily changes in position add up to about 24 times wealth a year, against about 29 for Log-HAR. With full costs, TTM's gain over Log-HAR therefore rises to about 21 basis points, and to about 7 with gradual trading. Chronos-Bolt and MEM, which earn less than Log-HAR before costs, gain under full costs at $h = 1$ for the same reason. With full costs, TTM's gain comes mostly from the stocks, and for the futures it is negative at every horizon, significantly at $h = 5$ and $h = 22$. A perfect forecast of the next day's variance turns over about 73 times a year, and with full costs its utility falls from 4\% to about 1.6\% of wealth a year. Daily rebalancing on a noisy target is therefore costly, and gradual trading reduces that cost.

For an investor who targets a constant volatility, the accuracy of TTM over Log-HAR is worth a few basis points of wealth a year before costs. After costs, TTM gains up to about 21 basis points at the daily horizon, primarily because its forecasts change less from one day to the next.

\section{Robustness Checks}
\label{sec:robustness}

The results in Secs.~\ref{sec:results} and~\ref{sec:significance} establish model rankings and their statistical significance. We test whether these rankings are stable, in this order, across the COVID-19 shock, across the weights date of each model, under the averaged target, under the winsorization bounds and across context lengths. We then feed the foundation models log volatility and compare them with supervised machine learning models fitted to each asset and with fine-tuned versions of TTM and Sundial. Finally, we ask how much of the headline advantage an affine recalibration of every forecast removes.

We split the full sample at 1 March 2020, the onset of the COVID-19 volatility spike in U.S.\ markets, by the date of the forecast target. In Tab.~\ref{tab:subsample} we report forecast accuracy separately for the pre-COVID and post-COVID periods across all 50 assets. The model ranking is stable across the COVID structural break, apart from the augmented HARs and the position of TTM at the monthly horizon before March 2020. After March 2020 the ordering of the main analysis carries over, and TTM has the lowest QLIKE at every horizon. Before March 2020 the augmented HARs have the lowest QLIKE at every horizon. TTM is then close to HAR and Log-HAR at the daily and weekly horizons and behind Log-HAR at the monthly horizon. The foundation models that lose to Log-HAR on the full sample also lose within each sub-period, except Lag-Llama at the monthly horizon before March 2020. The augmented HARs change the most across the split, since after March 2020 they are among the least accurate models at the daily and weekly horizons.

The losses are also more dispersed across models after March 2020. Part of this is a matter of scale, since squared errors grow with the level of volatility and Log-HAR's MSE doubles or more at every horizon. Relative to Log-HAR, the standard deviation of the QLIKE ratios across the 17 models doubles at the daily horizon, nearly triples at the weekly horizon and is unchanged at the monthly horizon. At the daily and weekly horizons the wider spread comes from five models whose ratios rise after March 2020, the three augmented HARs, Lag-Llama and Toto. The ratios of the other twelve models are as close together as before the split or closer. The high-volatility period therefore pulls a few models away from the rest rather than spreading all models apart.

For the augmented HARs, the rise of their ratios after March 2020 fits the measurement noise discussed in Sec.~\ref{subsec:main_results}. Their additional regressors, the jump component, the realized semivariances and the realized quarticity, are realized measures built from 5-minute returns, each with its own measurement error. For realized variance, the measurement error has a variance proportional to the integrated quarticity, so on days when the integrated quarticity is high its signal is relatively weak \citep{barndorffnielsen2002, bollerslev2016harq}. The integrated quarticity is the integral of the fourth power of volatility, so these days cluster in periods of high volatility.

\input{tables/table_subsample}

The COVID split divides the sample by market regime, while a check for contamination needs a split by what each model could have seen during pretraining. If pretraining on market data that overlaps our sample helped a foundation model, its loss relative to models that cannot be contaminated would rise once the forecast targets fall outside its pretraining data. We split the sample by the date of the forecast target in two ways, reported in Tab.~\ref{tab:contamination_split}. The first split is at 1 January 2024 for every model. The second splits each foundation model at the date its pinned weight file was last changed on Hugging Face. This date cannot be earlier than the release of the weights and is therefore an upper bound on the end of its pretraining data. We estimate HAR, ARFIMA, ARMA and MEM only on the 1{,}000 days before each origin, so they cannot be contaminated, and we use the average of their losses as a control for changes in market conditions across the split. We test the change with a regression of the cross-asset daily mean of each foundation model's QLIKE minus the control average on a dummy for the later period, with Newey--West standard errors. Taking the mean across assets accounts for the shocks that all assets share.

TTM's loss relative to the control does not change significantly after either split at any horizon. Across the nine foundation models at their own weights dates, no test of the 27 is significant at the 5\% level, and the largest positive $t$-statistic is 1.61. At the 2024 split, Chronos-Bolt and Moirai~2.0 lose ground significantly at the daily horizon, but the change does not line up with their weights dates. For Chronos-Bolt the loss is concentrated in 2024, before its weights date, and eases in 2025 (Appendix Tab.~\ref{tab:contamination_years}).

One limit of the check is that the weights date is an upper bound, so the earlier period can include data a model never saw, which makes a contamination effect harder to detect. Another is that the period after the weights date is short for the latest releases, 83 trading days for TimesFM~2.5. Within these limits, the split gives no evidence that overlap between pretraining data and the evaluation period drives our results.

\input{tables/table_contamination_split}

Our headline results use the point-in-time realized volatility as the target. When we re-run the full pipeline under an alternative, $h$-day-average target, TTM remains the only model below one at every horizon and the main result does not change. We report the details in Appendix~\ref{sec:appendix_tables} (Tab.~\ref{tab:avg_target}).

Every result in the paper uses winsorization bounds taken from the 1{,}000 days before each origin (Sec.~\ref{subsec:evaluation}). To show how much this choice matters, we recompute every loss ratio with bounds taken from each asset's full sample, which includes the evaluation period. The rankings do not change, because the bound rarely binds. Toto, whose heavy-tailed predictive distribution produces most of the forecasts outside the bounds, is the only model whose loss ratio moves by more than 0.004. The window bounds clip 0.1\% to 0.2\% of its forecasts, and its loss ratio to Log-HAR is 1.169, 1.241 and 1.132 with them and 1.215, 1.277 and 1.158 with full-sample bounds.

We also examine how foundation-model accuracy varies with the TSFM context length. Context effects are model-specific and horizon-dependent (Appendix Tab.~\ref{tab:context_sensitivity}). Since TTM runs at 512 days while the other foundation models use 1{,}000, we test whether TTM's result comes from its shorter memory, in two ways reported in Appendix Tab.~\ref{tab:equal_context}. With every foundation model and Log-HAR at 512 days, TTM still beats Log-HAR at $h = 1$ and $h = 5$, and no other foundation model matches TTM at those horizons. At $h = 22$, where Log-HAR gains slightly from the shorter window, TTM and Lag-Llama are about as accurate as Log-HAR. To give TTM a longer memory, we use its earlier r2 release, which offers contexts of 512, 1{,}024 and 1{,}536 days but is pretrained for minute to hourly data. The r2.1 release we use adds daily series and allows at most 512 days. Within r2, a context longer than 512 days lowers the loss ratio at the weekly and monthly horizons and raises it at the daily horizon, and every r2 checkpoint does worse than our r2.1 checkpoint at any context. TTM's result therefore points to the daily pretraining of its r2.1 release rather than to its shorter context.

For a practitioner who deploys one foundation model, the best context length differs by model and horizon, and a longer context does not guarantee higher accuracy. A practitioner should therefore choose the context length on a validation period that ends before forecasting starts. We apply one context rule to every model instead, since we compare the models on a common basis rather than select the best foundation model. If we chose each model's context on the evaluation sample, we would also select on the test data.

Log-HAR models the logarithm of volatility, and it is more accurate than HAR on levels at $h = 5$ and $h = 22$ (Tab.~\ref{tab:loss_ratios}). To test whether the foundation models gain from the logarithm as well, we feed each of them log volatility over the same context and retransform its forecast $\hat{m}_{t+h}$ of $\log \sigma_{t+h}$ to volatility once, as ARFIMA and ARMA do (Subsec.~\ref{subsec:benchmarks}), with $\exp(\hat{m}_{t+h} + \hat{v}_t/2)$. We call the resulting models the log versions of the foundation models, and name each one, after Log-HAR, by adding the prefix Log- to the model's name, as in Log-TTM and Log-Sundial. For Log-HAR, $\hat{v}_t$ is the residual variance of its one-step regression over the 1{,}000-day estimation window. A foundation model has no regression residuals, so we take the variance of its own one-step errors in log volatility over the 1{,}000 forecasts before the origin. Until 1{,}000 such forecasts exist, at the start of the evaluation period, we use all earlier ones, and we set $\hat{v}_t = 0$ for the first 22 forecasts. As for every other forecast, we clip the log forecast to the bounds of Sec.~\ref{subsec:evaluation}, here on the logarithmic scale. At $h = 1$ the log input lowers the loss ratio of seven of the nine foundation models (Tab.~\ref{tab:log_input}). Log-Sundial has a lower loss ratio than TTM and beats Log-HAR on 47 of the 50 assets. At $h = 5$ and $h = 22$ no log version beats Log-HAR on average, and Log-TTM does worse than TTM at all three horizons.

Toto and Lag-Llama do not suit the log input. Toto's predictive mean occasionally takes extreme values on the logarithmic scale. With volatility as input, we set Lag-Llama's negative sample values to zero. On log volatility, which is negative, we keep them. Without the zero floor its sampled paths drift upward at the monthly horizon, and many of its forecasts there are clipped at the upper bound.

\input{tables/table_log_input}

The two inputs differ mainly in the level of the forecasts and little in how closely the forecasts follow volatility. For seven of the nine foundation models, the standard deviation of the error of the log forecast changes by at most 5\% between the two inputs, and for TTM, Chronos-Bolt, TimesFM~2.5 and Sundial by at most 1.5\%. Toto and Lag-Llama, the two models that do not suit the log input, change by more. The level matters because every model forecasts too low a variance on average (Appendix~\ref{app:forecast_level}), and QLIKE penalizes a forecast that is too low more heavily than one that is too high (Subsec.~\ref{subsec:evaluation}). A higher forecast therefore tends to have a lower loss. At the daily horizon the retransformation gives seven of the nine models a higher forecast than they issue on volatility, and it gives TTM, whose forecasts of volatility are among the highest, a lower one. At $h = 5$ and $h = 22$ it gives most models a lower forecast than the same model issues on volatility. The term $\hat{v}_t/2$ raises a forecast by about 4\%, and we estimate it from one-step errors, while the errors of a log forecast 5 and 22 days ahead have about 1.5 and 2 times their variance. The retransformation therefore raises the exponential of the log forecast at the longer horizons by less than the log-normal mean of volatility would require. This pattern is consistent with a gain from the log input that is concentrated at the daily horizon.

To isolate the effect of the level, we rescale every forecast in real time, Log-HAR's included. We multiply each forecast by the average ratio of realized to forecast variance over the model's last 250 forecasts whose targets are observed by the origin (Appendix~\ref{app:forecast_level}). After the rescaling, TTM and Log-TTM have loss ratios within 0.003 of each other at $h = 1$ and $h = 5$, and at $h = 22$ Log-TTM is more accurate. Without the rescaling, the log input leaves the comparison with Log-HAR unchanged at $h = 5$ and $h = 22$, where no log version beats Log-HAR, and TTM remains the only foundation model below Log-HAR at every horizon. At $h = 1$ the log input improves most foundation models, and a forecaster who uses one of them one day ahead may prefer its log version. We keep volatility as the input of the main comparison, because the retransformation of a log forecast requires an error variance estimated from realized outcomes, which a zero-shot forecast does not use. We repeat the DM tests and the MCS with the log versions, and test each log version against the same model with volatility as input, in Appendix~\ref{app:log_input}.

The foundation models learn from many series during pretraining, while each econometric benchmark is a linear model in a few lags of the asset's own volatility. A gap between the two classes could therefore come from pretraining or from the flexibility of a neural network. To distinguish the two, we add three supervised machine learning models fitted to each asset's own history, of the kind that \citet{christensen2023ml} compare with the HAR family.

Gradient-boosted trees \citep[XGBoost,][]{chen2016xgboost} and a feedforward network take the regressors of Log-HAR, the logarithms of the average volatility over the last 1, 5 and 22 days. A long short-term memory network \citep[LSTM,][]{hochreiter1997long} takes the last 22 values of log volatility, from which we build those averages. Like the foundation models, each learner receives only the asset's past volatility. We fit each learner on the 1{,}000 days before the origin with targets observed by the origin. It forecasts log volatility $h$ days ahead directly and retransforms its forecast to volatility once, as ARFIMA and ARMA do. Since fitting the networks at every origin is costly, we refit them every 22 trading days. XGBoost would be cheap enough to refit every day, but we keep it on the same schedule so that we refit the three learners in the same way. The learners have hyperparameters, which the econometric models and the zero-shot foundation models do not, so they need a validation sample. Once a year, every 252 trading days, we fit each setting of a small grid on the earlier days of the 1{,}000-day window and score it by QLIKE on the last 250 days, which it was not fitted on. We keep the best setting until the next tuning and refit it every 22 trading days on all 1{,}000 days, the validation days included (Appendix~\ref{app:ml}).

\input{tables/table_ml_baselines}

The three learners are less accurate than Log-HAR at every horizon and on almost every asset (Tab.~\ref{tab:ml_baselines}). The LSTM comes closest at $h = 1$, where it ranks behind TTM and four other foundation models. At $h = 5$ and $h = 22$ every learner has a loss ratio of about 1.2 or more. Part of this gap at the longer horizons comes from our design, since a direct forecast loses to the iterated forecast of Log-HAR there (Appendix~\ref{app:ml}). The 22-day schedule costs little, since Log-HAR refitted every 22 days has a loss at most 2.5\% above that of Log-HAR refitted at every origin (Tab.~\ref{tab:ml_baselines}), which is small next to the gap of the learners. The remaining gap does not come from the level of the forecasts. On average, the forecasts of the three learners are at least as close to the realized level as those of Log-HAR, but they follow the day-to-day movements of volatility less closely, so their errors in log volatility are more dispersed. The three learners therefore stay behind Log-HAR after the real-time rescaling of Appendix~\ref{app:forecast_level}, which corrects only the level (Appendix Tab.~\ref{tab:ml_diag}), while the errors of TTM are about as dispersed as those of Log-HAR. Within our design, a network fitted to the 1{,}000 days of a single asset does not reach the accuracy of TTM, which forecasts each series from patterns learned on many other series during pretraining. Since TTM is itself a small neural network, this points to its pretraining on many series, rather than the flexibility of a network fitted to one asset, as the source of its advantage.

With the HAR lags as the only inputs, \citet{christensen2023ml} find that neural networks forecast realized variance one day ahead with an MSE about 5\% below that of HAR, while gradient boosting does worse than HAR. Our boosted trees agree with the second finding, and our results differ for the networks. Our feedforward network uses the architectures, training settings and ensemble of networks reported in that study. Under the loss of that study, the MSE of realized variance, our two networks are 15 to 23\% less accurate than Log-HAR at the daily horizon and about as accurate at the monthly horizon (Appendix~\ref{app:ml}).

However, \citet{christensen2023ml} estimate the networks once, on a training sample of 2{,}964 days that ends at a fixed date, with a further 424 days for validation, and do not update them over the evaluation period. Our networks are refitted on a rolling window of 1{,}000 days. With a training sample of 1{,}000 days, the networks of that study remain between 5\% below and 1\% above Log-HAR in MSE, so a shorter sample alone does not explain the difference. Longer training does not explain it either. In our base design the LSTM stops after 20 epochs without improvement on the validation days. With the early stopping of \citet{christensen2023ml}, after 100 epochs without improvement and at most 500, it trains 1.6 to 2.8 times as many epochs. Its loss ratios change by at most 0.025, slightly lower at $h = 5$ and slightly higher at $h = 1$ and $h = 22$ (Tab.~\ref{tab:ml_baselines}). The two studies also differ in the assets and years they cover and in the target on which the networks are trained, realized variance there and log volatility here.

We apply every foundation model zero-shot in our comparison. When we fine-tune TTM and Sundial on the realized volatility of U.S.\ stocks up to 2021, from a corpus separate from VOLARE, their accuracy does not improve consistently from 2022 to 2026. The result holds when we train all of TTM's weights or low-rank adapters on Sundial, and when we train only the layers after each model's pretrained backbone (Appendix~\ref{app:finetune}).

A model's low loss could come from forecasts whose level and scale happen to match realized volatility rather than from how closely they follow its movements. We recalibrate every forecast with a recursive MZ regression to measure how much of each model's advantage a rescaling estimated from the model's own past forecasts can create or remove. It does not split a model's accuracy uniquely into calibration and information, because an affine correction removes only linear departures and its estimated coefficients carry sampling error. We apply the correction symmetrically to all models, econometric and TSFM alike, because the unbiasedness of least squares is an in-sample property that need not carry over to the out-of-sample forecasts produced under a rolling window. Specifically, at each origin $t$ we estimate $\hat{\alpha}_t$ and $\hat{\beta}_t$ on an expanding window of the past forecasts whose targets are observed by $t$, those issued at origins up to $t-h$. We then form the corrected forecast $\widehat{\sigma}_t^{\text{MZ}} = \hat{\alpha}_t + \hat{\beta}_t \widehat{\sigma}_t$ at each of the three horizons. A forecaster can apply this correction in real time only by using historical information on past forecasts and realizations for each model and asset, and when this history is short the estimates of $\alpha_t$ and $\beta_t$ are noisy. We compare the original and corrected QLIKE at $h = 1, 5, 22$ in Tab.~\ref{tab:mz_bias_corrected}.\footnote{We begin scoring the recursively corrected forecast only after a 252-day (one trading year) expanding-window warm-up, so the first correction uses only forecasts whose targets fall within that year. We score the uncorrected forecasts of the ``Orig.'' column on this same post-warm-up window. Its QLIKE therefore differs slightly from the full-window pooled means in Tab.~\ref{tab:pooled50} (for example, TTM at $h = 1$ is 0.192 here versus 0.190 there), and the two are not in conflict.}

Only at the daily horizon does a corrected forecast beat the uncorrected TTM, and there the correction reorders the leading models. The econometric benchmarks are already close to affine-efficient, so it leaves them almost unchanged, and at $h = 1$ HAR and Log-HAR each move by at most about $0.002$ in QLIKE. Several foundation models improve once we remove the affine bias. At $h = 1$, the QLIKE of TimesFM~2.5, Chronos-Bolt (base) and Moirai~2.0 falls to about that of the uncorrected TTM. Their losses therefore come mostly from a level and scale that an affine correction repairs. TTM moves the opposite way at $h = 1$, dropping to eighth on the corrected QLIKE, where the eight best models lie within $0.01$ of one another. At the daily horizon an affine rescaling gives several econometric and foundation models the accuracy that TTM reaches without one, so TTM's raw advantage there rests largely on the level and scale of its forecasts. Good calibration is useful in itself, and TTM achieves it without the history of past forecasts that the correction requires.

At the longer horizons the correction raises the loss of most models. At $h = 5$ it lowers the QLIKE of the foundation models with the largest level errors, such as Chronos-Bolt and Moirai~2.0, but no corrected forecast reaches the uncorrected TTM. At $h = 22$ it raises the QLIKE of every model, by about half for Log-HAR. We estimate the coefficients at these horizons from forecasts whose $h$-day horizons overlap, so the estimation error of the coefficients outweighs the bias they remove. We therefore read the recalibration at the daily horizon only.

\input{tables/mz_bias_corrected}

\section{Combining TTM and Log-HAR}
\label{sec:combination}

TTM and Log-HAR draw on different information, so combining them may improve on either alone, as \citet{carriero2024macro} find for macroeconomic series, where the errors of TSFMs and econometric models are nearly orthogonal at short horizons. We form two combinations of the TTM and Log-HAR volatility forecasts \citep{bates1969, timmermann2006}. The first is an equal-weight average, and the second a recursive Bates--Granger combination whose variance-minimizing weight is estimated from the errors of the forecasts whose targets are observed by the forecast origin (expanding window, clipped to $[0,1]$, with an equal-weight warm-up). Tab.~\ref{tab:combination} reports the results. Both two-way combinations beat Log-HAR on the average loss ratio at every horizon. The equal-weight combination is also at or below TTM's loss ratio at all three horizons. The Bates--Granger combination, however, does worse than TTM at the monthly horizon.

By DM test the equal-weight combination has a significantly lower QLIKE than Log-HAR on 35 of 50 assets at $h = 1$, but it significantly beats TTM on at most 12. We also add a third member, ARMA, which is the most accurate of ARFIMA, ARMA and MEM at $h = 1$ (Tab.~\ref{tab:loss_ratios}). The three-way combination beats Log-HAR at the daily horizon, is at parity at the weekly horizon and falls behind at the monthly horizon. It adds nothing beyond the two-way combination at any horizon.

The combination gains most in MCS inclusion. When we add the combinations to the candidate set and recompute the MCS across all 50 assets, the equal-weight two-way combination enters the set for 98 to 100\% of assets across horizons. It enters at least as often as any single model at every horizon, including TTM and Log-HAR.

\input{tables/table_combination}

To explain why the equal-weight average beats both of its members, we measure how much the two forecasts differ and how their errors are related. From the error variance of a combination in \citet{bates1969}, the average of two forecasts has a lower MSE than the better of the two when the correlation $\rho$ of their errors is below $(3-k^2)/(2k)$. Here $k \ge 1$ is the ratio of the larger to the smaller root mean squared error (RMSE). For two forecasts of nearly equal accuracy $k$ is close to one, so the threshold is also close to one and even highly correlated errors leave room for a gain. This threshold holds for squared error only, and for QLIKE we use an exact decomposition instead. The QLIKE loss in Eq.~\eqref{eq:qlike} depends on the forecast only through the ratio $RV_t/\widehat{RV}_t$ and has the form of the Itakura--Saito divergence, the member of the Bregman family of losses generated by the convex function $\phi(x) = -\log x$ \citep[Table~2]{wood2023diversity}. Squared error is the member of the same family generated by $\phi(x) = x^2$. For any loss in this family, \citet[Theorem~11]{wood2023diversity} give an exact decomposition, the Bregman ambiguity decomposition, of the loss of a combination whose members are averaged on the scale of the derivative $\phi'$. This average is the arithmetic mean for squared error and the harmonic mean for QLIKE. For the variance forecasts of TTM and Log-HAR, $\widehat{RV}^{T}_t$ and $\widehat{RV}^{L}_t$, and their harmonic mean $\widehat{RV}^{H}_t = 2/(1/\widehat{RV}^{T}_t + 1/\widehat{RV}^{L}_t)$, this decomposition gives at every date and for any realized variance
\begin{equation}
{\begin{aligned}
\tfrac{1}{2}\left[L_{\text{QLIKE}}(\widehat{RV}^{T}_t, RV_t) + L_{\text{QLIKE}}(\widehat{RV}^{L}_t, RV_t)\right] - L_{\text{QLIKE}}(\widehat{RV}^{H}_t, RV_t) &= D_t, \\
D_t = \tfrac{1}{2}\left[\log\frac{\widehat{RV}^{T}_t}{\widehat{RV}^{H}_t} + \log\frac{\widehat{RV}^{L}_t}{\widehat{RV}^{H}_t}\right] &\ge 0.
\end{aligned}}
\label{eq:diversity}
\end{equation}
The diversity term $D_t$ depends only on the ratio of the two forecasts and is zero when they coincide. Averaged over dates, Eq.~\eqref{eq:diversity} implies that the harmonic mean beats the better of the two models when the mean of $D_t$ exceeds half the difference between their mean QLIKE losses, which is the QLIKE counterpart of the squared-error threshold. Our combination squares the average of the two volatility forecasts, which lies above $\widehat{RV}^{H}_t$, so its gain over the average of its members is $D_t$ plus a level term. The harmonic mean isolates the part of the gain that does not come from the level.

\input{tables/table_combination_diagnostics}

The errors of TTM and Log-HAR are highly correlated and almost equal in size (Tab.~\ref{tab:combination_diagnostics}, Panel~A), so the squared-error condition holds for most assets at the daily horizon and for few at the monthly horizon. Under QLIKE, the diversity term accounts for close to 90\% of the gain of the combination over the average of its members at every horizon. The harmonic mean, which has no level gain, also beats both models on average (Panel~B). The combination therefore gains because TTM and Log-HAR make different errors, and it beats both because their accuracy differs by less than this gain for most assets at the daily horizon. The gain does not come from opposite biases that cancel in the average, since both models forecast too low a variance on average at every horizon and in each tercile of the previous day's volatility. The level of the combination therefore lies between theirs. When we first correct each model's level with its own past errors, as in Appendix~\ref{app:forecast_level}, Log-HAR gains more than TTM, and the combination still beats both at the daily horizon but no longer at the weekly one (Panel~C).

The model with the lower loss also changes over time (Tab.~\ref{tab:combination_diagnostics}, Panel~E). TTM has the lower QLIKE in a little more than half of the 126-day blocks, and the leading model changes about seven times per asset, so a forecaster who selects one model on past performance often holds the weaker one. At the daily horizon, TTM's advantage over Log-HAR comes from calm periods, and after days of high volatility TTM alone does no better than Log-HAR while the combination keeps its gain. A forecast encompasses another if the optimal combination puts all the weight on it \citep{harvey1998encompassing}. We test encompassing on the derivative of QLIKE with respect to the combination weight, as \citet{giacomini2005quantile} do for quantile forecasts under the tick loss. At $h = 1$ the test rejects in both directions for 24 of the 50 assets, so each model carries information that the other lacks (Panel~D). The tests lose power at the longer horizons, where overlapping errors leave fewer independent observations. The recursive Bates--Granger weight stays near one half at the daily and weekly horizons and moves toward TTM at the monthly horizon, where it does worse than equal weights. This is the forecast combination puzzle, which \citet{smith2009puzzle} and \citet{claeskens2016puzzle} attribute to the estimation error in the weights.

TTM is the best partner for Log-HAR at the daily and weekly horizons, but it is not the only model that improves Log-HAR when we average the two (Appendix Tab.~\ref{tab:combination_pairs}). Several weaker models lower Log-HAR's loss because their errors are less correlated with Log-HAR's than TTM's errors are. For HAR-J and Toto at the daily and weekly horizons, and for HAR-RS at the weekly horizon, this gain comes from the level of the averaged volatility and disappears under the harmonic mean, which changes the TTM pair little. Under the harmonic mean, TTM is the best partner at $h = 1$ and $h = 5$, and Lag-Llama at $h = 22$. At the daily and weekly horizons TTM also combines with HAR slightly better than with Log-HAR, and at the monthly horizon about as well with HAR-J as with Log-HAR. At the weekly horizon, TTM is the only foundation model that is about as accurate as the HAR models and still makes different errors. At the daily horizon, where TimesFM~2.5 and Sundial are about as accurate as well, TTM's pair with Log-HAR gains the most. The combination result is therefore a property of TTM and the HAR family, and averaging TTM with one of the HAR models gives a small but consistent gain, largest at the daily horizon.

A forecaster cannot know in advance which foundation model will work, and in our results this choice matters more than the choice between a foundation model and an econometric one. By averaging Log-HAR with TTM, the one small foundation model that beats it, that forecaster obtains accuracy matching the best single model while improving on the benchmark for most assets.

\section{Conclusion}
\label{sec:conclusion}

In this work, we evaluate nine zero-shot TSFMs for realized volatility forecasting across 50 assets spanning U.S.\ equities, FX, and futures, and compare them to eight econometric benchmarks with formal pairwise and multi-model forecast-comparison tests.

Our first conclusion is that the foundation models do not outperform the econometric benchmarks as a class. Six of the nine never beat Log-HAR on the average QLIKE loss ratio, and HAR, ARFIMA, ARMA and the MEM stay within a few percent of Log-HAR at the daily horizon. Under MSE several foundation models are more accurate than Log-HAR, but not under QLIKE. Second, TTM, the smallest model in the evaluation, beats Log-HAR at every horizon by a narrow margin. For most assets the MCS contains it together with Log-HAR and several other models. At the daily and weekly horizons its edge comes mostly from the level of its forecasts, since it disappears when we adjust the level of every model's forecasts. Before trading costs, the edge is worth a few basis points of wealth a year to an investor who targets a constant volatility. Third, accuracy differs far more across the foundation models than between the best of them and Log-HAR, so the choice of foundation model matters more than the choice between foundation and econometric models.

For practitioners, a general-purpose foundation model is not a drop-in improvement over HAR for realized volatility, and most of the models we test do not beat Log-HAR on a typical asset. TTM is the exception, as it beats Log-HAR by a narrow margin at all three horizons and runs on a CPU, which makes it a reasonable alternative to consider. At the daily horizon, an affine correction estimated on past forecasts brings several underperforming TSFMs close to TTM, which reaches its accuracy without such a history. Since the edge is thin and even TTM is not best on every asset, a simple equal-weight average of TTM and Log-HAR matches the best single model. It enters the MCS for 98 to 100\% of assets across horizons, so a forecaster need not identify the best model for each asset in advance. Since accuracy differs widely across architectures, a practitioner cannot treat TSFMs as a uniform class and should evaluate several of them before deployment.

A first limitation is that our main comparison is zero-shot. Fine-tuning TTM and Sundial in two ways on a separate corpus of U.S.\ stock volatilities does not improve their accuracy consistently. Our fine-tuning follows a generic recipe, since each model keeps the volatility input of the zero-shot runs and we train it on a standard loss, the squared error for TTM and its own flow-matching loss for Sundial. Both losses differ from the QLIKE loss of our evaluation. A study of fine-tuning designed for realized volatility, with a loss aligned with QLIKE and a choice of input such as log volatility, remains a natural next step, since even the best zero-shot model beats the benchmark only by a narrow margin.

A second limitation is that we evaluate only point forecasts, specifically the mean of each model's predictive distribution, or the closest summary the model outputs. Density evaluation under proper scoring rules such as the continuous ranked probability score would extend the comparison. A third limitation concerns the pretraining corpora. The papers that introduce the foundation models we test do not report realized volatility or another intraday-derived statistic in the pretraining data of these models. Finance and economics series make up at most 0.09\% of the documented corpora, or 1.02\% for Sundial. For the models whose corpus is not documented in full, we cannot fully rule out indirect exposure to related financial series. Future work could pretrain a foundation model from scratch on point-in-time vintages of financial data, so that every forecast uses only data available at its origin, which removes the question of pretraining contamination by construction. Realized covariance forecasting and portfolio construction are a further extension.

\section*{Statements and declarations}

\paragraph{Conflict of interest.} The author declares no conflict of interest.

\paragraph{Funding.} This research received no specific grant from any funding agency in the public, commercial, or not-for-profit sectors.

\paragraph{Data availability.} The realized measures analyzed in this study are derived from the VOLARE dataset, which is constructed from proprietary high-frequency tick data licensed from Kibot. The underlying tick data are not redistributable by the author, and the VOLARE-derived realized measures are available from the VOLARE project subject to its terms of use. The code that reproduces all results in this paper is openly available at \url{https://github.com/Alessiobrini/tsfm-rv}.

\bibliography{references}

\appendix

\renewcommand{\thesection}{\Alph{section}}
\setcounter{section}{0}

\section{Foundation Model Details}
\label{app:tsfm_details}

In this appendix, we give the architecture and the exact pretrained checkpoint for each of the nine TSFMs summarized in Tab.~\ref{tab:tsfm_summary}. We apply all of them zero-shot in the main comparison and fine-tune TTM and Sundial in Appendix~\ref{app:finetune}. We load each checkpoint below from Hugging Face.

Chronos-Bolt \citep{ansari2024chronos} is a faster, non-autoregressive variant of Chronos. The T5 encoder-decoder produces all forecast quantiles in a single forward pass rather than generating tokens one at a time, a 250$\times$ speedup at competitive accuracy.\footnote{Chronos-Bolt is a later release by the Chronos authors without a separate publication. The non-autoregressive architecture and the 250$\times$ speedup are documented in the model card (\url{https://huggingface.co/amazon/chronos-bolt-base}), whereas the cited paper \citep{ansari2024chronos} describes the original autoregressive Chronos.} The original Chronos tokenizes values into 4,094 uniform bins. It is pretrained on about 890K series and 84B observations, of which M4, whose series mix finance with other domains, is at most 0.03\%. We evaluate the small (48M) and base (205M) checkpoints\footnote{\texttt{amazon/chronos-bolt-small} and \texttt{amazon/chronos-bolt-base} on Hugging Face.} and take the average of the nine predicted quantiles as the point forecast, because Chronos-Bolt outputs quantiles only.

Moirai~2.0 \citep{liu2025moirai2} is a patch-based, decoder-only model that uses multi-token prediction (generating several future patches per pass) and outputs nine quantile levels per step. We use the small checkpoint\footnote{\texttt{Salesforce/moirai-2.0-R-small} on Hugging Face.} and take the average of the nine quantiles as the point forecast.

Moirai-MoE \citep{liu2024moiraimoe} extends Moirai with a sparse MoE layer, so only a fraction of parameters are active per input. The small checkpoint has 117M total parameters, roughly 11M active per forward pass. Like Moirai~2.0 it uses patch tokenization, and its output layer predicts the parameters of a mixture distribution. We use this checkpoint,\footnote{\texttt{Salesforce/moirai-moe-1.0-R-small} on Hugging Face.} sample 20 trajectories, and report the sample mean as the point forecast.

Lag-Llama \citep{rasul2024lagllama} is a decoder-only model based on LLaMA that tokenizes each step with lag features (the current value and a fixed set of lagged values), where the models above use patches, and parameterizes a Student-$t$ distribution per step. We run the pretrained checkpoint\footnote{\texttt{time-series-foundation-models/Lag-Llama} on Hugging Face.} with 8 layers, a 144-dimensional embedding split into 4 attention heads of 36 dimensions and mean scaling of the context, sample 100 trajectories, and report the sample mean as the point forecast. Since volatility is positive, we set the negative values of the sampled trajectories to zero, an option of the Lag-Llama predictor.

TimesFM~2.5 \citep{das2024timesfm} is a decoder-only model with patch-plus-residual tokenization (200M parameters) that outputs a mean forecast and nine quantiles per step. The TimesFM paper describes the pretraining corpus of version 1.0, 381B points made mostly of Wikipedia pageviews, with Google Trends, synthetic series and public datasets. M4 is about 0.006\% of this corpus. The model card of version 2.5 lists GIFT-Eval Pretrain, Wikimedia pageviews to November 2023, Google Trends to the end of 2022 and synthetic and augmented data, without their sizes. We use the TimesFM~2.5 checkpoint\footnote{\texttt{google/timesfm-2.5-200m-pytorch} on Hugging Face.} and take the conditional mean as the point forecast.

Toto \citep{cohen2025toto} is a decoder-only model (151M parameters), pretrained primarily on observability metrics (CPU usage, request latency), the training domain furthest from finance in our study. It is pretrained on about 2.36 trillion points, 43\% of them Datadog observability metrics and about a third synthetic series, with the rest from the GIFT-Eval Pretrain collection and the Chronos datasets, among them the exchange\_rate series, and its paper gives no share by domain. Its decoding head parameterizes a Student-$t$ mixture, and we take the mean of 20 sampled trajectories as the point forecast. We use the Toto checkpoint.\footnote{\texttt{Datadog/Toto-Open-Base-1.0} on Hugging Face.}

Sundial \citep{liu2025sundial} is a generative model (128M parameters) that uses flow matching \citep{lipman2023flow}. It learns a continuous transformation from noise to the distribution of forecast trajectories and produces full distributional forecasts without quantile bins. We use the base checkpoint,\footnote{\texttt{thuml/sundial-base-128m} on Hugging Face.} sample 20 trajectories, the same number as for Moirai-MoE, and report the sample mean as the point forecast.

TTM \citep{ekambaram2024ttm} is a lightweight model built on TSMixer, a multi-layer perceptron that mixes across time and channels without self-attention. It uses frequency prefix tuning (a learned frequency embedding prepended to the input), and the r2.1 release adds daily-frequency support for context lengths of 52 to 512 days. TTM is the smallest model in our evaluation, with 855{,}100 parameters, and runs on CPU. The TTM paper documents a corpus from Monash and LibCity, including the NN5 and Bitcoin datasets, without a breakdown by domain, and the r2.1 release we use is trained on about 1B samples. We use the r2.1 checkpoint\footnote{\texttt{ibm-granite/granite-timeseries-ttm-r2}, branch \texttt{512-96-ft-r2.1} on Hugging Face, whose 855{,}100 parameters we counted from the loaded model. The suffix ``ft'' denotes frequency tuning, not fine-tuning, and the model is fully zero-shot.} with a 512-day context window and read the point forecast directly from the model output.

\FloatBarrier

\section{Supplementary Tables}
\label{sec:appendix_tables}

In this appendix, we report cross-sectional median metrics as a robustness check against outlier assets (Tab.~\ref{tab:main_results_median}), and we discuss the context-length sensitivity (Tab.~\ref{tab:context_sensitivity}), the comparison at an equal context (Tab.~\ref{tab:equal_context}) and the averaged target (Tab.~\ref{tab:avg_target}).

\input{tables/table_equity_metrics_median}

In Tab.~\ref{tab:context_sensitivity} we report how the accuracy of each foundation model varies with its context length. We compare 128, 256, and 512 days with the 1{,}000-day default and hold the rest of the evaluation fixed (point-in-time target, mean forecast, volatility scale). TTM's r2.1 release supports at most 512 days, so for TTM 512 is the default, and its 128-day and 256-day columns use the 90-day and 180-day branches of the release. Context effects are model-specific and horizon-dependent. Among the models that can use the 1{,}000-day window, TimesFM~2.5 and Moirai-MoE-S are most accurate with it at the daily and weekly horizons, and Sundial is most accurate around 512 days. Moirai~2.0 is most accurate at 128 to 256 days, and Chronos-Bolt has no single best context across horizons. The default 1{,}000-day context, matched to the econometric estimation window, is therefore at or near the best setting for several foundation models but is not universally optimal. TTM's QLIKE is lowest at its 512-day default at every horizon, so the headline conclusion that TTM is the only foundation model to beat Log-HAR under the equal-weighted loss ratio at every horizon is unaffected by the context choice.

\input{tables/table_context_sensitivity}

In Tab.~\ref{tab:equal_context} we place every foundation model and Log-HAR at 512 days, and we vary TTM's context within its earlier r2 release. Log-HAR estimated on 512 days is as accurate as on 1{,}000 days at $h = 1$ and $h = 5$ and slightly more accurate at $h = 22$, so the shorter memory does not handicap the benchmark.

\input{tables/table_equal_context}

In Tab.~\ref{tab:avg_target} we replace the point-in-time target of the main results with the average of daily realized variance over the next $h$ days \citep{patton2015}, against which we score the square of each model's forecast of average volatility. The three augmented HAR variants forecast the average variance directly, and we score that forecast. TTM is again the only foundation model to beat Log-HAR at every horizon, with average QLIKE loss ratios close to those of the point-in-time target. At the daily horizon TimesFM~2.5, Sundial and HAR are also below one, as on the point-in-time target, and no other model is below one at the longer horizons. The choice of forecast target therefore does not drive the main result.

\input{tables/table_avg_target}

\FloatBarrier

\section{Additional Results}
\label{app:additional}

In this appendix, we relate forecast performance to the characteristics of each asset (Appendix~\ref{subsec:drivers}) and show the MCS inclusion of each model by asset (Appendix~\ref{subsec:mcs_heatmap}). We measure the level of each forecast (Appendix~\ref{app:forecast_level}), split the loss ratios by calendar year (Appendix~\ref{app:years}) and report every equal-weight pair of models (Appendix~\ref{app:pairs}). We then repeat the DM tests and the MCS with the log versions of the foundation models and test each log version against the original model (Appendix~\ref{app:log_input}). Finally, we describe the supervised machine learning baselines (Appendix~\ref{app:ml}) and the fine-tuned TTM and Sundial (Appendix~\ref{app:finetune}).

\subsection{Drivers of Forecast Performance}
\label{subsec:drivers}

Since foundation-model rankings vary across asset classes (Subsec.~\ref{subsec:cross_asset}), we relate relative forecast performance to the characteristics of each asset. \citet{carriero2024macro} find that foundation models struggle with highly persistent macroeconomic series. We test whether a similar pattern holds for realized volatility by examining the relationship between each asset's first-order autocorrelation $\rho_1$ and the foundation model's QLIKE ratio relative to HAR.

\begin{figure}[!tb]
  \centering
  \includegraphics{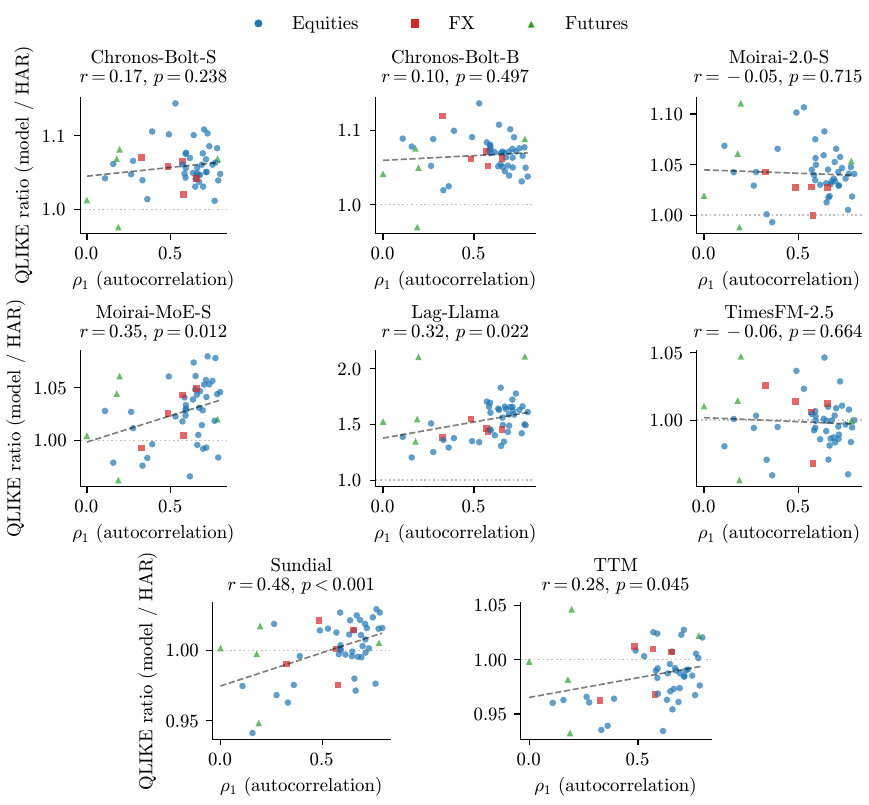}
  \caption{QLIKE loss ratio of each foundation model to HAR at $h = 1$ against the first-order autocorrelation $\rho_1$ of the asset's daily realized variance from 5-minute returns over its full sample, one panel per foundation model. Toto is omitted, as in Fig.~\ref{fig:qlike_boxplot}. Each point is one of the 50 assets. A ratio below one means that the model has a lower QLIKE than HAR, and the dotted line marks a ratio of one. The dashed line is the least-squares fit, and $r$ and $p$ in each panel title are the Pearson correlation and its $p$-value. HAR rather than Log-HAR is the denominator so that the ratios can be compared directly with prior studies of realized volatility, in which HAR is the standard baseline \citep{bollerslev2016harq}.}
  \label{fig:persistence_drivers}
\end{figure}

In Fig.~\ref{fig:persistence_drivers} we plot the QLIKE ratio of each foundation model to HAR at $h = 1$ against $\rho_1$ for the 50 assets. For four of the eight models plotted (Toto is omitted), namely Sundial, Moirai-MoE-S, Lag-Llama and TTM, the correlation is positive and significant at the 5\% level. These models are relatively less accurate on more persistent assets, as \citet{carriero2024macro} find for macroeconomic series, and for the other four the correlation is small and insignificant. The effect is moderate, since TTM beats HAR on 68\% of the more persistent half of the assets and on 72\% of the less persistent half, the largest shares of any foundation model in both groups.

Persistence therefore explains part of the variation across assets for some models, but less than the model itself does. Realized volatility is stationary and mean-reverting, with a positive, right-skewed distribution, the kind of series these models see in pretraining.

\FloatBarrier

\subsection{MCS Inclusion Heatmap}
\label{subsec:mcs_heatmap}

\begin{figure}[htbp]
  \centering
  \includegraphics{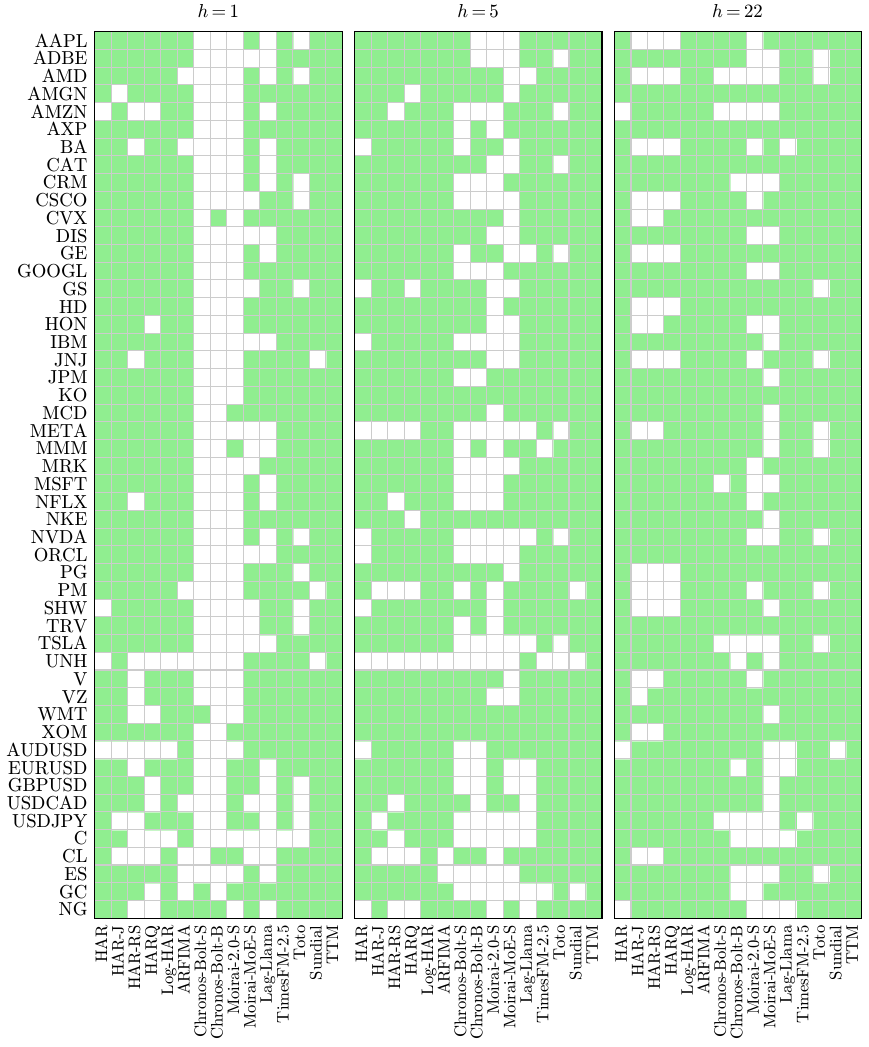}
  \caption{MCS inclusion by asset and model at $h = 1$ (left), $h = 5$ (center) and $h = 22$ (right). Rows are the 50 assets, grouped into equities, FX and futures and sorted alphabetically within each group, and columns are 15 of the 17 models. A green cell marks a model that is in the MCS of that asset at the 10\% level, computed on QLIKE losses with the range statistic and its elimination rule (Sec.~\ref{subsec:evaluation}). ARMA and MEM are omitted for space, and their inclusion rates are in Tab.~\ref{tab:mcs}.}
  \label{fig:mcs_heatmap}
\end{figure}

In Fig.~\ref{fig:mcs_heatmap}, we show for each asset and horizon which models the MCS of Tab.~\ref{tab:mcs} contains.

\FloatBarrier

\subsection{Forecast Level}
\label{app:forecast_level}

For each model and horizon, we report the average across assets of the ratio of realized variance to the squared volatility forecast, $RV_{t+h}/\widehat{\sigma}^{2}_{t+h}$, which measures how far the variance forecast falls below the realized variance on average. A ratio of one means the variance forecast is correct on average. We then recompute the QLIKE loss ratios under two level adjustments, each relative to Log-HAR adjusted the same way, so Log-HAR has a ratio of one under both. The first applies to the three log models. For Log-HAR, ARFIMA and ARMA we replace the squared volatility forecast with the log-normal expected variance, $\exp(2\hat{m}_{t+h} + 2\hat{v}_{t+h})$, where $\hat{m}_{t+h}$ is the forecast of $\log \sigma_{t+h}$ and $\hat{v}_{t+h}$ its $h$-step forecast-error variance. The second applies to all 17 models. At each origin we multiply a model's variance forecast by the average ratio of realized to forecast variance over its last 250 forecasts whose targets are observed before the origin. We leave the forecast unchanged until 22 such forecasts are available. This adjustment uses only information available at the origin, and we keep it out of the main comparison because it adjusts each model with realized outcomes.

The results are in Tab.~\ref{tab:forecast_level}. Every model forecasts too low a variance on average, with a realized variance 3\% to 72\% above the squared forecast, more at the longer horizons. Among the models with loss ratios near one, TTM's forecast is closest to the realized level at the daily horizon. Under either adjustment TTM no longer beats Log-HAR at $h = 1$ and $h = 5$, and it remains the most accurate model at $h = 22$. After the real-time rescaling, TimesFM~2.5, Moirai~2.0, Chronos-Bolt and Sundial beat Log-HAR at the daily horizon.

\input{tables/table_forecast_level}

\FloatBarrier

\subsection{Loss Ratios by Calendar Year}
\label{app:years}

\input{tables/table_contamination_years}

In Tab.~\ref{tab:contamination_years}, we report the loss ratio of every model to Log-HAR by calendar year of the forecast target, the context for the contamination check of Sec.~\ref{sec:robustness}. TTM's ratio after its weights date is just above the range of its earlier years at $h = 1$ in 2025, within that range otherwise, and below it in January 2026 at $h = 1$ and $h = 5$.

\FloatBarrier

\subsection{Equal-Weight Pairs}
\label{app:pairs}

In Tab.~\ref{tab:combination_pairs}, we report the equal-weight average of Log-HAR with every other model, of TTM with every econometric model, and of HAR with every other model, for the comparison of partners in Sec.~\ref{sec:combination}. For each pair we report the partner alone, the square of the average of the two volatility forecasts and the harmonic mean of the two variance forecasts, which removes the level gain of the average. We also report the correlation of the two models' errors.

\input{tables/table_combination_pairs}

\FloatBarrier

\subsection{Log Versions of the Foundation Models}
\label{app:log_input}

We repeat the DM tests and the MCS of Sec.~\ref{sec:significance} with the log version of each foundation model, defined in Sec.~\ref{sec:robustness}, in place of the model itself, with the results in Tab.~\ref{tab:log_input_mcs_dm}.

\input{tables/table_log_input_mcs_dm}

\input{tables/table_log_vs_vol_dm}

To test whether each log version differs significantly from the same model with volatility as input, we apply the DM test of Sec.~\ref{sec:significance} to each pair, asset by asset, on the dates of Tab.~\ref{tab:log_input}. The results are in Tab.~\ref{tab:log_vs_vol_dm}. At $h = 1$ the log input lowers the loss significantly on most assets for Chronos-Bolt, Moirai~2.0 and Lag-Llama, and on about half of them for Sundial. For TTM and Toto it raises the loss significantly on about a third of the assets at every horizon. At $h = 5$ and $h = 22$ the significant gains remain mainly for Chronos-Bolt and Moirai~2.0, and Log-Lag-Llama has a significantly higher loss than Lag-Llama on almost every asset at $h = 22$.

\FloatBarrier

\subsection{Supervised Machine Learning Baselines}
\label{app:ml}

The three learners of Sec.~\ref{sec:robustness} forecast $\log \sigma_{t+h}$ directly at each horizon. XGBoost and the feedforward network take the regressors of Log-HAR in Eq.~\eqref{eq:log_har}. The LSTM takes the last 22 values of $\log \sigma$ up to the origin, standardized with the mean and standard deviation of the training window.

At each origin the training sample holds the origins of the previous 1{,}000 days whose targets are observed by the origin. Every 252 trading days, about once a year, we tune each learner on a small grid. We fit each setting on the origins whose targets end before the last 250 days of the window. We then score it by QLIKE on the $251 - h$ forecasts whose origins fall in those 250 days and whose targets are observed by the end of the window. We keep the best setting. Every 22 trading days we refit the kept setting on all 1{,}000 days of the window, the validation days included, and the forecasts until the next refit come from that fit. We retransform the forecast $\hat{m}_{t+h}$ of $\log \sigma_{t+h}$ to volatility as $\exp(\hat{m}_{t+h} + \hat{v}_t/2)$, where $\hat{v}_t$ is the variance of the errors on the 250 validation days. A flexible model fits its training sample closely, so its residuals there understate its forecast errors, and the validation errors do not.

XGBoost uses trees of depth 2, 3 or 4, a learning rate of 0.03 or 0.1, a minimum child weight of 1 or 10, and 80\% of the rows for each tree. Early stopping after 50 rounds without improvement on the validation days sets the number of trees, at most 2{,}000. The feedforward network follows \citet{christensen2023ml}. We take the four architectures of that study: one hidden layer of 2 neurons, two layers of 4 and 2, three layers of 8, 4 and 2, and four layers of 16, 8, 4 and 2. All four use leaky rectified linear activations, Glorot normal initialization and the Adam optimizer with a learning rate of 0.001. Each architecture trains for at most 500 epochs with early stopping at a patience of 100 epochs on the validation MSE and with dropout of probability 0.2 on the hidden layers. That study reports a dropout of 0.8, which we read as the probability of keeping a neuron. The forecast is the average of the 10 best of 100 networks with different seeds, ranked by their validation error.

The LSTM has one layer of 16 or 32 units and dropout of probability 0 or 0.2 before a linear output layer, and it uses the Adam optimizer with a learning rate of 0.001. It trains for at most 200 epochs with early stopping at a patience of 20 epochs, and its forecast is the average of three networks with different seeds. At a refit, each network trains for the number of epochs chosen at the last tuning.

The first of three checks is that the rolling procedure does not cost accuracy at the daily horizon. On AAPL, JPM, MSFT, XOM, EURUSD, USDJPY, the E-mini S\&P 500 and crude oil futures, a linear regression on the regressors of Log-HAR, fitted and tuned like the learners, has a loss ratio to Log-HAR of 1.005 at $h = 1$. Its ratios of 1.057 at $h = 5$ and 1.140 at $h = 22$ measure the cost of a direct forecast against the iterated forecast of Log-HAR. On the same eight assets the three learners have ratios of 1.06 to 1.18 at $h = 1$ and 1.28 to 1.54 at $h = 22$, so they remain behind the linear regression. Second, with the early stopping of \citet{christensen2023ml}, at a patience of 100 epochs and at most 500 epochs, the LSTM trains for 1.6 to 2.8 times as many epochs, and its loss ratios are in Tab.~\ref{tab:ml_baselines}. Third, we score the learners by the MSE of realized variance, the loss of \citet{christensen2023ml}, as the average across assets of the per-asset ratio to the MSE of Log-HAR, on dates common to every model. The two networks have an MSE 15 to 23\% above that of Log-HAR at $h = 1$ and within about 1\% of it at $h = 22$, and XGBoost is further behind at every horizon. The level and the tracking error of the forecasts, which we use in Sec.~\ref{sec:robustness} to explain the gap of the learners, are in Tab.~\ref{tab:ml_diag}.

\input{tables/table_ml_diag}

\FloatBarrier

\subsection{Fine-Tuned TTM and Sundial}
\label{app:finetune}

To measure what fine-tuning adds to the zero-shot comparison, we fine-tune two of the foundation models on a corpus of daily realized variance for U.S.\ stocks that is separate from VOLARE and evaluate them as in the main comparison. We choose TTM, the most accurate foundation model in our comparison, and Sundial, which is about as accurate as Log-HAR at the daily horizon. Sundial has 128 million weights, about 150 times as many as TTM.

The corpus holds the five-minute realized variance, computed from TAQ, of the 3{,}000 largest U.S.\ common stocks by market capitalization in each year from 2015 to 2025. We rank the stocks by their market capitalization from the Center for Research in Security Prices on the last trading day of May and keep the list from July to the following June. A stock enters the corpus only in the years in which it ranks among the 3{,}000 largest, and a stock that is later delisted stays in for those years, so the corpus is free of survivorship bias. Over the eleven years the corpus holds 5{,}143 stocks. We leave out the stock-days with fewer than 80\% of the five-minute intervals observed or a median price below \$5. The 40 stocks of our sample are in the corpus, while the FX pairs and futures are not.

We keep the input and the context of the zero-shot runs, volatility over 512 days for TTM and 1{,}000 days for Sundial. Each training example is one context followed by the next 96 days, and we take one example every five days. Training examples end by 31 December 2020 and validation examples in 2021, at least 22 trading days before 3 January 2022, the first forecast origin we score. A stock therefore enters the training examples only with at least 608 days of data before 2021 for TTM, whose context is 512 days, and 1{,}096 days for Sundial, whose context is 1{,}000 days. The training examples cover 2{,}309 stocks for TTM and 1{,}664 for Sundial. No target used in training or validation falls in the period from 2022 to 2026 on which we compare the fine-tuned and zero-shot models.

The two models differ in size, so we fine-tune them in the way suited to each. TTM has 855{,}100 weights, few enough to update all of them at a small cost, and we train them to minimize the squared error of the 96-day forecast. Sundial has 128 million weights. For a model of this size we train low-rank adapters \citep{hu2022lora}, small matrices added to the attention projections that change the model's output while its pretrained weights stay fixed. With rank 16 the adapters hold 1.2 million weights, about 1\% of the model, and we train them on the flow-matching loss with which Sundial is pretrained. Both models use the AdamW optimizer with a learning rate of $10^{-4}$ and a 500-step warmup. We compute the validation loss every 1{,}000 steps, stop after six validations without improvement and keep the weights with the lowest validation loss. The fine-tuned models then forecast with the same context and clipping as the zero-shot models.

We also fine-tune each model in a second way, which keeps the pretrained backbone fixed and trains only the layers that turn its output into the forecast, with the same data, loss and recipe. For TTM these are the decoder and the forecast head, 302{,}162 weights or 35\% of the model, and a frozen backbone is the default for few-shot fine-tuning in granite-tsfm, IBM's library for TTM. For Sundial they are the flow-matching head, 12.5 million weights or 10\% of the model, ten times as many as its adapters. This head-only fine-tuning is close to the linear probing of \citet{goel2025rv}, who keep the transformer layers of TimesFM fixed.

\input{tables/table_finetune}

Fine-tuning does not improve either model consistently from 2022 to 2026, in either way (Tab.~\ref{tab:finetune}). With all its weights trained, TTM loses accuracy at $h = 1$ and $h = 5$ and gains at $h = 22$, where the gain comes from the stocks, which are in the corpus. On the FX pairs and futures it loses accuracy at the two shorter horizons. With adapters, Sundial is about unchanged at $h = 1$ and $h = 22$ and less accurate at $h = 5$. With the head only, TTM is slightly less accurate than zero-shot at $h = 1$ and $h = 5$ and about unchanged at $h = 22$. Sundial with the head only is less accurate at $h = 1$ and more accurate at $h = 22$, where it beats its zero-shot version on 46 of the 50 assets. Its loss ratio there is the lowest in Tab.~\ref{tab:finetune}, just below that of TTM with all its weights trained and below that of zero-shot TTM.

The changes come mostly from the level of the forecasts. Fine-tuning all the weights raises the forecasts of TTM at $h = 22$, the adapters lower those of Sundial at $h = 5$, and the head-only fine-tuning raises those of Sundial at every horizon, most at $h = 1$. The standard deviation of the errors in log volatility changes by at most 2.5\%.

The fine-tuned weights are also no better than the pretrained ones in 2021, the year after the training data and before the evaluation period. We score both on a fixed random sample of the 2021 validation examples, 8{,}192 for TTM and 4{,}096 for Sundial, whose loss costs far more to compute. The two losses are on different scales, so we compare each model only with itself. With all its weights trained, the validation loss of the fine-tuned TTM is 53\% higher than that of the pretrained TTM, and with the head only it is 94\% higher. For Sundial it is 5\% higher with adapters and about unchanged, 0.06\% higher, with the head only, whose gain at $h = 22$ comes from a higher forecast level that the flow-matching loss on 2021 does not measure. In our early stopping, we compare only the weights saved during training, from step 1{,}000 on. A selection on the 2021 validation year that also considered the pretrained weights would have kept them for three of the four fine-tunings and would have been indifferent for the fourth. For those three, fine-tuning already lowers accuracy in the year before the evaluation period, so their loss of accuracy from 2022 to 2026 is not specific to the evaluation period. In our setting, fine-tuning on a large corpus of stock volatilities does not improve on the zero-shot forecasts consistently across horizons.

\FloatBarrier

\end{document}

%% file: tables/table_pretraining_data.tex
\begin{table}[htbp]
\centering
\singlespacing
\caption{TSFM pretraining corpora and financial data exposure, as documented in each model's paper. Fin.\ Share is the share of pretraining observations in datasets that the paper labels finance or economics. For Moirai-MoE it is computed from the observation counts of its corpus, the Large-scale Open Time Series Archive (LOTSA) of \citet{woo2024moirai}. Not disclosed means that no paper documents this for the checkpoint we use, or that the paper gives no breakdown by domain. None means that the paper lists its datasets and labels no dataset as finance or economics, and None named means that the paper names no financial dataset. RV? indicates whether the paper reports realized volatility or another intraday-derived statistic in the pretraining data. $^\dagger$The checkpoint we use is a later release than the one its paper documents, and Appendix~\ref{app:tsfm_details} gives the corpus of the earlier release.}
\label{tab:pretraining_data}
\small
\setlength{\tabcolsep}{3.5pt}
\begin{tabular}{lllll}
\toprule
Model & Corpus Size & Fin.\ Share & Financial Datasets & RV? \\
\midrule
Chronos-Bolt$^\dagger$ & Not disclosed & Not disclosed & Not disclosed & Not disclosed \\
Moirai 2.0   & 36M series, 295B obs & Not disclosed & None named & No \\
Moirai-MoE   & 27.6B obs (LOTSA)    & 0.09\% & LOTSA Econ/Fin & No \\
Lag-Llama    & 7{,}965 series, 352M tokens & 0\% & None & No \\
TimesFM 2.5$^\dagger$ & Not disclosed & Not disclosed & Not disclosed & Not disclosed \\
Toto & 2.36T obs & Not disclosed & exchange\_rate & No \\
Sundial      & 1{,}032B obs & 1.02\% & Own collection & Not disclosed \\
TTM$^\dagger$ & Not disclosed & Not disclosed & Not disclosed & Not disclosed \\
\bottomrule
\end{tabular}
\end{table}

%% file: tables/table_equity_metrics.tex
\begin{table}[htbp]
\centering
\singlespacing
\caption{Forecast accuracy for 40 U.S.\ equities (VOLARE). Values are cross-sectional averages of per-asset loss functions. Bold indicates the best value in each column. MSE is on the volatility scale and QLIKE on the variance scale. $^{\ast}$ marks models in the MCS at the 10\% level for a majority of the 40 equities at that horizon.}
\label{tab:main_results}
\footnotesize
\setlength{\tabcolsep}{4pt}
\begin{tabular}{lrrrrrr}
\toprule
& \multicolumn{3}{c}{MSE ($\times 10^{-6}$)} & \multicolumn{3}{c}{QLIKE} \\
\cmidrule(lr){2-4}\cmidrule(lr){5-7}
Model & $h=1$ & $h=5$ & $h=22$ & $h=1$ & $h=5$ & $h=22$ \\
\midrule
Log-HAR & 30.653 & 46.322 & 64.104 & 0.198$^{\ast}$ & 0.300$^{\ast}$ & 0.538$^{\ast}$ \\
HAR & 30.349 & 47.029 & 73.709 & 0.198$^{\ast}$ & 0.302$^{\ast}$ & 0.580$^{\ast}$ \\
HAR-J & 35.266 & 55.560 & 74.580 & 0.231$^{\ast}$ & 0.336$^{\ast}$ & 0.563$^{\ast}$ \\
HAR-RS & 37.014 & 56.473 & 74.661 & 0.239$^{\ast}$ & 0.342$^{\ast}$ & 0.565$^{\ast}$ \\
HARQ & 36.927 & 57.682 & 74.326 & 0.249$^{\ast}$ & 0.359$^{\ast}$ & 0.573$^{\ast}$ \\
ARFIMA & 30.965 & 47.022 & 62.076 & 0.200$^{\ast}$ & 0.317$^{\ast}$ & 0.590$^{\ast}$ \\
ARMA & 31.672 & 47.245 & 64.188 & 0.198$^{\ast}$ & 0.315$^{\ast}$ & 0.613$^{\ast}$ \\
MEM & 30.704 & 46.968 & 66.236 & 0.200$^{\ast}$ & 0.314$^{\ast}$ & 0.620$^{\ast}$ \\
Chronos-Bolt-S & 30.129 & 45.901 & 63.696 & 0.211 & 0.329 & 0.639$^{\ast}$ \\
Chronos-Bolt-B & 30.233 & 45.855 & 63.209 & 0.212 & 0.326$^{\ast}$ & 0.639$^{\ast}$ \\
Moirai-2.0-S & 30.131 & 46.416 & 63.326 & 0.207 & 0.336 & 0.678$^{\ast}$ \\
Moirai-MoE-S & 32.743 & 50.932 & 72.700 & 0.203$^{\ast}$ & 0.320 & 0.610$^{\ast}$ \\
Lag-Llama & 47.306 & 65.137 & 69.560 & 0.296$^{\ast}$ & 0.466$^{\ast}$ & 0.532$^{\ast}$ \\
TimesFM-2.5 & \textbf{29.780} & 45.853 & 61.855 & 0.198$^{\ast}$ & 0.318$^{\ast}$ & 0.622$^{\ast}$ \\
Toto & 37.149 & 50.941 & 74.315 & 0.225$^{\ast}$ & 0.333$^{\ast}$ & 0.604$^{\ast}$ \\
Sundial & 31.023 & 48.006 & 63.692 & 0.198$^{\ast}$ & 0.329$^{\ast}$ & 0.652$^{\ast}$ \\
TTM & 30.490 & \textbf{45.410} & \textbf{60.052} & \textbf{0.194}$^{\ast}$ & \textbf{0.294}$^{\ast}$ & \textbf{0.521}$^{\ast}$ \\
\bottomrule
\end{tabular}
\end{table}

%% file: tables/table_pooled50.tex
\begin{table}[htbp]
\centering
\singlespacing
\caption{Pooled forecast accuracy across all 50 assets (40 equities, 5 FX, 5 futures). Each cell is the simple average of the per-asset loss over the 50 assets. MSE is on the volatility scale and QLIKE is on the variance scale. The pooled mean gives the most weight to the assets with the largest losses and is reported for comparison with the equal-weight loss ratios of Tab.~\ref{tab:loss_ratios}. Bold marks the lowest MSE and lowest QLIKE in each horizon column.}
\label{tab:pooled50}
\footnotesize
\setlength{\tabcolsep}{4pt}
\begin{tabular}{lrrrrrr}
\toprule
& \multicolumn{3}{c}{MSE ($\times 10^{-6}$)} & \multicolumn{3}{c}{QLIKE} \\
\cmidrule(lr){2-4}\cmidrule(lr){5-7}
Model & $h=1$ & $h=5$ & $h=22$ & $h=1$ & $h=5$ & $h=22$ \\
\midrule
Log-HAR & 31.548 & 45.981 & 64.200 & 0.194 & 0.289 & 0.507 \\
HAR & 31.629 & 47.750 & 78.840 & 0.194 & 0.292 & 0.544 \\
HAR-J & 37.983 & 64.274 & 78.762 & 0.224 & 0.324 & 0.532 \\
HAR-RS & 39.706 & 65.074 & 77.114 & 0.231 & 0.329 & 0.533 \\
HARQ & 39.306 & 66.404 & 78.844 & 0.239 & 0.343 & 0.548 \\
ARFIMA & 32.014 & 46.745 & 62.789 & 0.196 & 0.306 & 0.562 \\
ARMA & 47.244 & 61.642 & 78.362 & 0.195 & 0.304 & 0.579 \\
MEM & 31.879 & 46.726 & 65.900 & 0.196 & 0.302 & 0.580 \\
Chronos-Bolt-S & 31.098 & 45.466 & 62.026 & 0.205 & 0.317 & 0.605 \\
Chronos-Bolt-B & 31.133 & 45.393 & 61.430 & 0.207 & 0.315 & 0.603 \\
Moirai-2.0-S & 31.180 & 46.260 & 62.871 & 0.202 & 0.323 & 0.645 \\
Moirai-MoE-S & 34.902 & 51.700 & 72.541 & 0.198 & 0.311 & 0.580 \\
Lag-Llama & 47.564 & 65.110 & 69.413 & 0.292 & 0.452 & 0.513 \\
TimesFM-2.5 & \textbf{30.798} & 45.661 & 61.135 & 0.193 & 0.307 & 0.587 \\
Toto & 37.636 & 50.292 & 74.286 & 0.225 & 0.373 & 0.575 \\
Sundial & 31.776 & 47.440 & 62.807 & 0.193 & 0.314 & 0.608 \\
TTM & 31.486 & \textbf{45.214} & \textbf{59.712} & \textbf{0.190} & \textbf{0.285} & \textbf{0.499} \\
\bottomrule
\end{tabular}
\end{table}

%% file: tables/table_loss_ratios.tex
\begin{table}[htbp]
\centering
\singlespacing
\caption{Average QLIKE loss ratios to Log-HAR across all 50 assets. Each cell is the mean across assets of the per-asset QLIKE ratio to Log-HAR. Log-HAR's own ratio is 1.000 by construction. A ratio below 1 means a lower QLIKE than Log-HAR on average, and bold marks the lowest ratio in each column.}
\label{tab:loss_ratios}
\small
\begin{tabular}{lrrr}
\toprule
Model & $h=1$ & $h=5$ & $h=22$ \\
\midrule
Log-HAR & 1.000 & 1.000 & 1.000 \\
HAR & 0.999 & 1.008 & 1.068 \\
HAR-J & 1.181 & 1.130 & 1.049 \\
HAR-RS & 1.216 & 1.147 & 1.050 \\
HARQ & 1.264 & 1.194 & 1.078 \\
ARFIMA & 1.016 & 1.060 & 1.104 \\
ARMA & 1.006 & 1.051 & 1.133 \\
MEM & 1.016 & 1.043 & 1.134 \\
Chronos-Bolt-S & 1.057 & 1.092 & 1.180 \\
Chronos-Bolt-B & 1.065 & 1.088 & 1.180 \\
Moirai-2.0-S & 1.039 & 1.113 & 1.260 \\
Moirai-MoE-S & 1.025 & 1.077 & 1.147 \\
Lag-Llama & 1.535 & 1.588 & 1.024 \\
TimesFM-2.5 & 0.997 & 1.060 & 1.149 \\
Toto & 1.169 & 1.241 & 1.132 \\
Sundial & 0.999 & 1.087 & 1.183 \\
TTM & \textbf{0.984} & \textbf{0.989} & \textbf{0.988} \\
\bottomrule
\end{tabular}
\end{table}

%% file: tables/table_fx_futures_metrics.tex
\begin{table}[htbp]
\centering
\singlespacing
\caption{Forecast accuracy for FX and futures (VOLARE), cross-sectional averages. Bold marks the best value per column within each panel. $^{\ast}$ marks models in the MCS at the 10\% level for a majority of the panel's assets at that horizon.}
\label{tab:fx_futures_results}
\footnotesize
\begin{tabular}{lrrrrrr}
\toprule
& \multicolumn{3}{c}{MSE} & \multicolumn{3}{c}{QLIKE} \\
\cmidrule(lr){2-4}\cmidrule(lr){5-7}
Model & $h=1$ & $h=5$ & $h=22$ & $h=1$ & $h=5$ & $h=22$ \\
\midrule
\multicolumn{7}{l}{\textbf{Panel A: FX (5 pairs; MSE $\times 10^{-8}$)}} \\[2pt]
\midrule
Log-HAR & 266.985 & 348.524 & 447.050 & 0.165$^{\ast}$ & \textbf{0.216}$^{\ast}$ & \textbf{0.312}$^{\ast}$ \\
HAR & 268.881 & 358.031 & 460.485 & 0.165$^{\ast}$ & 0.219$^{\ast}$ & 0.319$^{\ast}$ \\
HAR-J & 299.113 & 418.109 & 494.611 & 0.172$^{\ast}$ & 0.228$^{\ast}$ & 0.319$^{\ast}$ \\
HAR-RS & 309.222 & 415.289 & 493.557 & 0.179 & 0.229$^{\ast}$ & 0.314$^{\ast}$ \\
HARQ & 302.624 & 393.058 & 490.572 & 0.176 & 0.222$^{\ast}$ & 0.313$^{\ast}$ \\
ARFIMA & 271.434 & 348.369 & \textbf{430.306} & 0.168$^{\ast}$ & 0.223$^{\ast}$ & 0.335$^{\ast}$ \\
ARMA & 305.595 & 386.453 & 486.877 & 0.168$^{\ast}$ & 0.224$^{\ast}$ & 0.343$^{\ast}$ \\
MEM & 280.534 & 357.124 & 453.949 & 0.173 & 0.222$^{\ast}$ & 0.327$^{\ast}$ \\
Chronos-Bolt-S & 266.185 & 347.446 & 433.703 & 0.174 & 0.233 & 0.353$^{\ast}$ \\
Chronos-Bolt-B & 268.222 & 350.072 & 445.213 & 0.178 & 0.238 & 0.363$^{\ast}$ \\
Moirai-2.0-S & \textbf{261.744} & 341.991 & 434.657 & 0.170$^{\ast}$ & 0.229$^{\ast}$ & 0.367$^{\ast}$ \\
Moirai-MoE-S & 291.881 & 434.719 & 663.698 & 0.168$^{\ast}$ & 0.240 & 0.375 \\
Lag-Llama & 381.653 & 480.341 & 513.910 & 0.239 & 0.299 & 0.333$^{\ast}$ \\
TimesFM-2.5 & 261.853 & \textbf{341.291} & 432.029 & 0.166$^{\ast}$ & 0.225$^{\ast}$ & 0.349$^{\ast}$ \\
Toto & 367.462 & 468.881 & 595.400 & 0.190 & 0.235$^{\ast}$ & 0.328$^{\ast}$ \\
Sundial & 265.889 & 342.767 & 438.739 & 0.165$^{\ast}$ & 0.217$^{\ast}$ & 0.330$^{\ast}$ \\
TTM & 271.672 & 353.449 & 430.348 & \textbf{0.163}$^{\ast}$ & 0.218$^{\ast}$ & 0.315$^{\ast}$ \\
\midrule
\multicolumn{7}{l}{\textbf{Panel B: Futures (5 contracts; MSE $\times 10^{-6}$)}} \\[2pt]
\midrule
Log-HAR & 67.592 & 85.749 & 124.701 & 0.187$^{\ast}$ & \textbf{0.276}$^{\ast}$ & \textbf{0.461}$^{\ast}$ \\
HAR & 70.809 & 97.680 & 194.128 & 0.185$^{\ast}$ & 0.278$^{\ast}$ & 0.484$^{\ast}$ \\
HAR-J & 94.707 & 194.075 & 186.033 & 0.222$^{\ast}$ & 0.328$^{\ast}$ & 0.495$^{\ast}$ \\
HAR-RS & 97.851 & 194.805 & 168.915 & 0.221 & 0.329 & 0.496$^{\ast}$ \\
HARQ & 94.617 & 198.652 & 188.928 & 0.223 & 0.335$^{\ast}$ & 0.581$^{\ast}$ \\
ARFIMA & 69.703 & 87.788 & 126.977 & 0.196 & 0.305$^{\ast}$ & 0.571$^{\ast}$ \\
ARMA & 216.009 & 234.592 & 265.245 & 0.192$^{\ast}$ & 0.296$^{\ast}$ & 0.538$^{\ast}$ \\
MEM & 70.353 & 87.944 & 124.567 & 0.190$^{\ast}$ & 0.285$^{\ast}$ & 0.514$^{\ast}$ \\
Chronos-Bolt-S & 67.289 & 83.984 & 106.360 & 0.193 & 0.307 & 0.582$^{\ast}$ \\
Chronos-Bolt-B & \textbf{66.791} & \textbf{83.591} & \textbf{104.183} & 0.194 & 0.306 & 0.556 \\
Moirai-2.0-S & 68.130 & 87.853 & 117.753 & 0.194 & 0.316 & 0.659 \\
Moirai-MoE-S & 84.153 & 105.198 & 137.178 & 0.189$^{\ast}$ & 0.304 & 0.545 \\
Lag-Llama & 93.377 & 125.201 & 132.504 & 0.312 & 0.491 & 0.546$^{\ast}$ \\
TimesFM-2.5 & 67.123 & 86.376 & 112.189 & 0.186$^{\ast}$ & 0.297$^{\ast}$ & 0.551$^{\ast}$ \\
Toto & 75.489 & 90.708 & 142.389 & 0.265$^{\ast}$ & 0.831$^{\ast}$ & 0.582$^{\ast}$ \\
Sundial & 66.911 & 86.922 & 114.141 & \textbf{0.184}$^{\ast}$ & 0.295$^{\ast}$ & 0.541$^{\ast}$ \\
TTM & 68.222 & 85.329 & 112.397 & \textbf{0.184}$^{\ast}$ & 0.283$^{\ast}$ & 0.504$^{\ast}$ \\
\bottomrule
\end{tabular}
\end{table}

%% file: tables/table_mcs_dm.tex
\begin{table}[htbp]
\centering
\singlespacing
\caption{MCS inclusion rates and DM pairwise win rates for 50 assets (40 equities, 5 FX, 5 futures). MCS columns report the percentage of assets for which the model is included in the MCS at the 10\% significance level (QLIKE loss, range statistic $T_R$ and its elimination rule, moving-block bootstrap with 22-day blocks and $B = 10{,}000$). DM columns report the percentage of pairwise comparisons (across 50 assets $\times$ 16 opponents = 800 tests) in which the row model achieves significantly lower QLIKE at the 5\% level.}
\label{tab:mcs}
\label{tab:dm_summary}
\small
\begin{tabular}{lrrrrrr}
\toprule
& \multicolumn{3}{c}{MCS inclusion (\%)} & \multicolumn{3}{c}{DM win rate (\%)} \\
\cmidrule(lr){2-4}\cmidrule(lr){5-7}
Model & $h=1$ & $h=5$ & $h=22$ & $h=1$ & $h=5$ & $h=22$ \\
\midrule
Log-HAR & 94.0 & 98.0 & 100.0 & 32.9 & 47.8 & 27.4 \\
HAR & 92.0 & 80.0 & 94.0 & 34.2 & 27.4 & 9.9 \\
HAR-J & 92.0 & 90.0 & 66.0 & 13.5 & 7.8 & 4.9 \\
HAR-RS & 72.0 & 82.0 & 68.0 & 9.1 & 6.2 & 4.8 \\
HARQ & 78.0 & 84.0 & 80.0 & 8.5 & 8.0 & 5.2 \\
ARFIMA & 84.0 & 94.0 & 100.0 & 22.6 & 14.4 & 13.8 \\
ARMA & 94.0 & 74.0 & 92.0 & 30.8 & 16.2 & 4.9 \\
MEM & 78.0 & 82.0 & 86.0 & 22.4 & 14.6 & 4.6 \\
Chronos-Bolt-S & 6.0 & 46.0 & 90.0 & 11.1 & 4.2 & 2.4 \\
Chronos-Bolt-B & 6.0 & 52.0 & 80.0 & 9.2 & 3.9 & 1.4 \\
Moirai-2.0-S & 18.0 & 32.0 & 54.0 & 17.8 & 4.5 & 0.0 \\
Moirai-MoE-S & 70.0 & 42.0 & 46.0 & 14.4 & 3.4 & 0.0 \\
Lag-Llama & 50.0 & 76.0 & 90.0 & 0.0 & 0.0 & 3.9 \\
TimesFM-2.5 & 98.0 & 94.0 & 98.0 & 38.5 & 14.5 & 9.8 \\
Toto & 74.0 & 84.0 & 80.0 & 6.8 & 4.5 & 1.8 \\
Sundial & 94.0 & 94.0 & 98.0 & 33.6 & 12.4 & 6.1 \\
TTM & 100.0 & 100.0 & 100.0 & 45.4 & 57.1 & 23.4 \\
\bottomrule
\end{tabular}
\end{table}

%% file: tables/mz_regression_all.tex
{\centering
\begin{table}[htbp]
\centering
\singlespacing
\caption{MZ forecast-efficiency regressions, $\sigma_t=\alpha+\beta\widehat{\sigma}_t+\varepsilon_t$, cross-sectional averages across 50 assets. Under efficiency $\alpha=0,\beta=1$. \% Rej. is the percentage of assets on which a Wald test with Newey--West standard errors rejects the joint null hypothesis $\alpha=0,\beta=1$ at the 5\% level.}\label{tab:mz_all}
\footnotesize
\setlength{\tabcolsep}{3pt}
\resizebox{\textwidth}{!}{%
\begin{tabular}{lrrrrrrrrrrrr}
\toprule
& \multicolumn{4}{c}{$h=1$} & \multicolumn{4}{c}{$h=5$} & \multicolumn{4}{c}{$h=22$} \\
\cmidrule(lr){2-5}\cmidrule(lr){6-9}\cmidrule(lr){10-13}
Model & $\hat{\alpha}$ & $\hat{\beta}$ & $R^2$ & \% Rej. & $\hat{\alpha}$ & $\hat{\beta}$ & $R^2$ & \% Rej. & $\hat{\alpha}$ & $\hat{\beta}$ & $R^2$ & \% Rej. \\
\midrule
Log-HAR & -0.0003 & 1.036 & 0.519 & 2\% & 0.0011 & 0.919 & 0.285 & 4\% & 0.0065 & 0.502 & 0.059 & 84\% \\
HAR & 0.0008 & 0.943 & 0.525 & 4\% & 0.0028 & 0.798 & 0.292 & 32\% & 0.0087 & 0.352 & 0.044 & 84\% \\
HAR-J & 0.0008 & 0.885 & 0.456 & 98\% & 0.0033 & 0.695 & 0.201 & 98\% & 0.0096 & 0.259 & 0.029 & 96\% \\
HAR-RS & 0.0013 & 0.853 & 0.441 & 98\% & 0.0036 & 0.681 & 0.194 & 96\% & 0.0097 & 0.252 & 0.028 & 96\% \\
HARQ & 0.0014 & 0.850 & 0.438 & 100\% & 0.0040 & 0.651 & 0.186 & 100\% & 0.0098 & 0.245 & 0.027 & 96\% \\
ARFIMA & -0.0003 & 1.023 & 0.513 & 6\% & 0.0008 & 0.959 & 0.271 & 14\% & 0.0060 & 0.580 & 0.065 & 62\% \\
ARMA & 0.0003 & 0.990 & 0.494 & 10\% & 0.0015 & 0.925 & 0.268 & 34\% & 0.0068 & 0.519 & 0.044 & 48\% \\
MEM & 0.0002 & 0.999 & 0.521 & 14\% & 0.0006 & 0.973 & 0.281 & 20\% & 0.0104 & 0.245 & 0.033 & 74\% \\
Chronos-Bolt-S & 0.0006 & 0.992 & 0.532 & 100\% & 0.0019 & 0.922 & 0.303 & 100\% & 0.0065 & 0.563 & 0.070 & 68\% \\
Chronos-Bolt-B & 0.0011 & 0.958 & 0.532 & 100\% & 0.0020 & 0.906 & 0.301 & 100\% & 0.0064 & 0.564 & 0.067 & 74\% \\
Moirai-2.0-S & -0.0002 & 1.062 & 0.535 & 100\% & -0.0001 & 1.092 & 0.300 & 100\% & 0.0054 & 0.679 & 0.068 & 78\% \\
Moirai-MoE-S & 0.0018 & 0.833 & 0.507 & 100\% & 0.0045 & 0.655 & 0.265 & 100\% & 0.0093 & 0.298 & 0.046 & 100\% \\
Lag-Llama & -0.0019 & 1.109 & 0.268 & 94\% & 0.0066 & 0.481 & 0.048 & 100\% & 0.0091 & 0.309 & 0.027 & 98\% \\
TimesFM-2.5 & -0.0003 & 1.042 & 0.535 & 42\% & -0.0003 & 1.075 & 0.296 & 50\% & 0.0056 & 0.624 & 0.065 & 34\% \\
Toto & 0.0031 & 0.748 & 0.469 & 100\% & 0.0041 & 0.672 & 0.263 & 88\% & 0.0094 & 0.296 & 0.046 & 100\% \\
Sundial & -0.0007 & 1.061 & 0.516 & 4\% & 0.0002 & 1.008 & 0.258 & 12\% & 0.0071 & 0.494 & 0.051 & 56\% \\
TTM & -0.0004 & 1.010 & 0.521 & 88\% & 0.0003 & 0.969 & 0.295 & 16\% & 0.0046 & 0.670 & 0.078 & 60\% \\
\bottomrule
\end{tabular}%
}
\end{table}
}

%% file: tables/table_economic_value.tex
\begin{table}[htbp]
\centering
\singlespacing
\caption{Realized utility of the volatility forecasts. The investor holds the fraction $x_t = (SR/\gamma)/\sqrt{252 F_t}$ of wealth in the asset, with $SR = 0.4$ and $\gamma = 2$, and rebalances every day. At $h = 1$, $F_t$ is the square of the point forecast of the main comparison, and at $h = 5$ and $h = 22$ it is the square of the forecast of average volatility over the next $h$ days. Panel~A: realized utility of Eq.~\eqref{eq:utility}, in percent of wealth per year, averaged over the forecast origins of each asset and then over the 50 assets. Zero: no trading costs. Full: each change in position costs the asset's bid-ask spread. Gradual (Grad.): the position moves 15\% of the way to its target each day and pays the same spread. Panel~B: difference from Log-HAR in basis points per year. An asterisk marks a DM test that rejects equal utility at the 5\% level, on a daily series that weights the 50 assets equally, with Newey--West standard errors and $\max(22, 2h)$ lags. TTM and Log-HAR is the equal-weight average of their volatility forecasts. The perfect forecast sets $F_t$ equal to the realized variance.}
\label{tab:economic_value}
\footnotesize
\setlength{\tabcolsep}{3pt}
\begin{tabular}{lrrrrrrrrr}
\toprule
& \multicolumn{3}{c}{$h=1$} & \multicolumn{3}{c}{$h=5$} & \multicolumn{3}{c}{$h=22$} \\
\cmidrule(lr){2-4}\cmidrule(lr){5-7}\cmidrule(lr){8-10}
Model & Zero & Full & Grad. & Zero & Full & Grad. & Zero & Full & Grad. \\
\midrule
\multicolumn{10}{l}{\textbf{Panel A: realized utility (\% per year)}} \\[2pt]
HAR & 3.55 & 2.60 & 3.22 & 3.64 & 3.03 & 3.38 & 3.44 & 3.06 & 3.22 \\
Log-HAR & 3.54 & 2.57 & 3.21 & 3.64 & 3.07 & 3.38 & 3.47 & 3.14 & 3.25 \\
HAR-J & 3.48 & 2.55 & 3.27 & 3.40 & 2.67 & 3.32 & 3.34 & 2.96 & 3.22 \\
HAR-RS & 3.47 & 2.54 & 3.26 & 3.41 & 2.70 & 3.32 & 3.35 & 3.00 & 3.22 \\
HARQ & 3.44 & 2.16 & 3.20 & 3.45 & 2.40 & 3.31 & 3.35 & 2.74 & 3.21 \\
ARFIMA & 3.54 & 2.64 & 3.21 & 3.62 & 3.02 & 3.34 & 3.37 & 3.02 & 3.14 \\
ARMA & 3.54 & 2.63 & 3.21 & 3.62 & 3.01 & 3.34 & 3.36 & 2.96 & 3.13 \\
MEM & 3.54 & 2.74 & 3.22 & 3.63 & 2.97 & 3.37 & 3.37 & 2.96 & 3.17 \\
Chronos-Bolt-S & 3.50 & 2.80 & 3.18 & 3.59 & 3.07 & 3.32 & 3.32 & 2.96 & 3.08 \\
Chronos-Bolt-B & 3.50 & 2.75 & 3.17 & 3.59 & 3.04 & 3.31 & 3.31 & 2.93 & 3.06 \\
Moirai-2.0-S & 3.51 & 2.68 & 3.18 & 3.58 & 3.02 & 3.30 & 3.25 & 2.90 & 3.00 \\
Moirai-MoE-S & 3.55 & 2.48 & 3.26 & 3.64 & 2.87 & 3.39 & 3.46 & 2.93 & 3.24 \\
Lag-Llama & 3.32 & 2.41 & 3.09 & 3.33 & 2.99 & 3.15 & 3.29 & 3.17 & 3.19 \\
TimesFM-2.5 & 3.54 & 2.69 & 3.22 & 3.62 & 3.05 & 3.35 & 3.35 & 2.98 & 3.10 \\
Toto & 3.46 & 2.12 & 3.22 & 3.16 & 2.15 & 3.34 & 1.94 & 0.86 & 2.87 \\
Sundial & 3.55 & 2.65 & 3.23 & 3.62 & 3.01 & 3.35 & 3.32 & 2.87 & 3.09 \\
TTM & 3.57 & 2.78 & 3.28 & 3.66 & 3.14 & 3.42 & 3.47 & 3.16 & 3.27 \\
TTM and Log-HAR & 3.56 & 2.70 & 3.25 & 3.65 & 3.12 & 3.40 & 3.48 & 3.16 & 3.27 \\
Perfect forecast & 4.00 & 1.62 & 3.21 & 4.00 & 3.40 & 3.63 & 4.00 & 3.84 & 3.87 \\
\midrule
\multicolumn{10}{l}{\textbf{Panel B: difference to Log-HAR (basis points per year)}} \\[2pt]
HAR & 0.2 & 2.7$^{*}$ & 1.4 & $-$0.1 & $-$4.0$^{*}$ & 0.4 & $-$3.1$^{*}$ & $-$8.0$^{*}$ & $-$3.6$^{*}$ \\
HAR-J & $-$6.1 & $-$2.2 & 6.1$^{*}$ & $-$23.9 & $-$40.6 & $-$5.9 & $-$12.6 & $-$17.3 & $-$3.8 \\
HAR-RS & $-$7.5 & $-$2.9 & 5.2$^{*}$ & $-$22.8 & $-$37.4 & $-$5.6 & $-$11.4 & $-$13.8 & $-$3.2 \\
HARQ & $-$10.6 & $-$41.5$^{*}$ & $-$1.3 & $-$19.4 & $-$66.7$^{*}$ & $-$6.3 & $-$11.7 & $-$39.9$^{*}$ & $-$4.4 \\
ARFIMA & $-$0.4 & 6.4$^{*}$ & $-$0.4 & $-$2.4$^{*}$ & $-$4.9$^{*}$ & $-$3.6$^{*}$ & $-$9.6 & $-$11.5 & $-$11.0 \\
ARMA & $-$0.2 & 5.9$^{*}$ & $-$0.3 & $-$2.4$^{*}$ & $-$6.4$^{*}$ & $-$3.7$^{*}$ & $-$10.5$^{*}$ & $-$17.3$^{*}$ & $-$12.2$^{*}$ \\
MEM & $-$0.4 & 17.1$^{*}$ & 0.8 & $-$1.2 & $-$9.8$^{*}$ & $-$1.0 & $-$9.5$^{*}$ & $-$17.5$^{*}$ & $-$8.5 \\
Chronos-Bolt-S & $-$4.0$^{*}$ & 22.5$^{*}$ & $-$2.4$^{*}$ & $-$5.1$^{*}$ & $-$0.1 & $-$5.9$^{*}$ & $-$15.0$^{*}$ & $-$17.5$^{*}$ & $-$17.7 \\
Chronos-Bolt-B & $-$4.6$^{*}$ & 18.0$^{*}$ & $-$3.8$^{*}$ & $-$5.3$^{*}$ & $-$3.1 & $-$6.9$^{*}$ & $-$16.0 & $-$20.5$^{*}$ & $-$19.4 \\
Moirai-2.0-S & $-$3.2$^{*}$ & 11.0$^{*}$ & $-$3.0$^{*}$ & $-$5.9$^{*}$ & $-$5.5$^{*}$ & $-$7.7$^{*}$ & $-$21.3 & $-$24.1$^{*}$ & $-$25.0 \\
Moirai-MoE-S & 0.6 & $-$9.5$^{*}$ & 5.3$^{*}$ & $-$0.3 & $-$20.4$^{*}$ & 1.0 & $-$1.1 & $-$21.2$^{*}$ & $-$1.6 \\
Lag-Llama & $-$22.2$^{*}$ & $-$16.6 & $-$12.0 & $-$30.9 & $-$8.3 & $-$22.2 & $-$17.7 & 3.1 & $-$6.7 \\
TimesFM-2.5 & $-$0.3 & 11.6$^{*}$ & 1.1 & $-$2.5$^{*}$ & $-$2.2 & $-$3.2$^{*}$ & $-$12.2 & $-$15.7 & $-$15.1 \\
Toto & $-$8.5$^{*}$ & $-$45.5$^{*}$ & 1.1 & $-$48.5$^{*}$ & $-$91.7$^{*}$ & $-$4.1$^{*}$ & $-$152.3 & $-$228.1$^{*}$ & $-$38.2 \\
Sundial & 0.2 & 7.7$^{*}$ & 1.7$^{*}$ & $-$2.3 & $-$6.4$^{*}$ & $-$2.8 & $-$14.6 & $-$27.1$^{*}$ & $-$16.6 \\
TTM & 2.1$^{*}$ & 20.7$^{*}$ & 6.8$^{*}$ & 1.7$^{*}$ & 6.7$^{*}$ & 4.4$^{*}$ & 0.5 & 2.1 & 1.2 \\
TTM and Log-HAR & 1.6$^{*}$ & 12.3$^{*}$ & 4.0$^{*}$ & 1.4$^{*}$ & 4.6$^{*}$ & 2.7$^{*}$ & 0.9 & 2.2$^{*}$ & 1.2$^{*}$ \\
Perfect forecast & 45.5$^{*}$ & $-$95.4$^{*}$ & 0.3 & 35.9$^{*}$ & 32.6$^{*}$ & 25.6$^{*}$ & 53.2 & 70.4$^{*}$ & 62.1 \\
\bottomrule
\end{tabular}
\end{table}

%% file: tables/table_subsample.tex
\begin{table}[htbp]
\centering
\singlespacing
\caption{Sub-sample forecast accuracy for forecast targets before and from 1 March 2020, averaged across the 50 assets, on the dates common to the 17 models of each asset. MSE ($\times 10^{-6}$) is on the volatility scale and QLIKE on the variance scale. Bold marks the lowest MSE and lowest QLIKE in each horizon column within each panel, following the unrounded values.}
\label{tab:subsample}
\small
\begin{tabular}{lrrrrrr}
\toprule
& \multicolumn{3}{c}{MSE ($\times 10^{-6}$)} & \multicolumn{3}{c}{QLIKE} \\
\cmidrule(lr){2-4}\cmidrule(lr){5-7}
Model & $h=1$ & $h=5$ & $h=22$ & $h=1$ & $h=5$ & $h=22$ \\
\midrule
\multicolumn{7}{l}{\textbf{Panel A: before March 2020}} \\[2pt]
\midrule
Log-HAR & 18.925 & \textbf{25.967} & 30.833 & 0.205 & 0.305 & 0.353 \\
HAR & 18.605 & 25.971 & \textbf{30.240} & 0.203 & 0.310 & 0.364 \\
HAR-J & 19.437 & 28.271 & 32.596 & \textbf{0.199} & 0.298 & \textbf{0.344} \\
HAR-RS & 20.493 & 28.319 & 32.839 & 0.207 & 0.299 & 0.345 \\
HARQ & 19.828 & 27.910 & 32.628 & 0.199 & \textbf{0.297} & 0.345 \\
ARFIMA & 19.077 & 26.361 & 30.520 & 0.215 & 0.336 & 0.395 \\
ARMA & 20.945 & 28.031 & 31.814 & 0.207 & 0.325 & 0.378 \\
MEM & 19.286 & 26.941 & 31.061 & 0.208 & 0.314 & 0.352 \\
Chronos-Bolt-S & 18.988 & 26.670 & 30.630 & 0.218 & 0.354 & 0.415 \\
Chronos-Bolt-B & 18.644 & 26.593 & 30.569 & 0.219 & 0.354 & 0.416 \\
Moirai-2.0-S & 18.524 & 26.603 & 31.084 & 0.213 & 0.352 & 0.435 \\
Moirai-MoE-S & 19.728 & 28.810 & 34.195 & 0.207 & 0.337 & 0.414 \\
Lag-Llama & 26.299 & 33.612 & 35.728 & 0.267 & 0.330 & 0.349 \\
TimesFM-2.5 & \textbf{18.418} & 26.397 & 30.404 & 0.204 & 0.335 & 0.400 \\
Toto & 20.807 & 29.995 & 37.224 & 0.226 & 0.333 & 0.380 \\
Sundial & 18.678 & 26.707 & 30.917 & 0.206 & 0.333 & 0.404 \\
TTM & 19.243 & 26.431 & 30.571 & 0.203 & 0.303 & 0.360 \\
\midrule
\multicolumn{7}{l}{\textbf{Panel B: March 2020 to January 2026}} \\[2pt]
\midrule
Log-HAR & 38.001 & 54.807 & 77.539 & 0.192 & 0.289 & 0.548 \\
HAR & 38.434 & 57.955 & 101.103 & 0.192 & 0.291 & 0.591 \\
HAR-J & 47.518 & 85.024 & 99.280 & 0.231 & 0.334 & 0.580 \\
HAR-RS & 49.520 & 86.111 & 95.638 & 0.236 & 0.339 & 0.580 \\
HARQ & 49.092 & 88.304 & 99.631 & 0.248 & 0.357 & 0.606 \\
ARFIMA & 38.596 & 55.613 & 75.921 & 0.192 & 0.302 & 0.608 \\
ARMA & 69.539 & 86.325 & 106.707 & 0.193 & 0.304 & 0.633 \\
MEM & 38.515 & 55.646 & 79.375 & 0.194 & 0.302 & 0.638 \\
Chronos-Bolt-S & 37.439 & 53.898 & 73.458 & 0.203 & 0.312 & 0.657 \\
Chronos-Bolt-B & 37.506 & 53.790 & 72.553 & 0.205 & 0.310 & 0.652 \\
Moirai-2.0-S & 37.727 & 55.204 & 75.264 & 0.200 & 0.320 & 0.704 \\
Moirai-MoE-S & 43.181 & 62.354 & 87.073 & 0.197 & 0.307 & 0.625 \\
Lag-Llama & 56.721 & 77.811 & 82.282 & 0.301 & 0.487 & 0.556 \\
TimesFM-2.5 & \textbf{37.240} & 54.450 & 73.063 & 0.192 & 0.304 & 0.639 \\
Toto & 45.299 & 59.320 & 89.494 & 0.223 & 0.432 & 0.626 \\
Sundial & 38.313 & 56.501 & 75.034 & 0.191 & 0.314 & 0.661 \\
TTM & 37.944 & \textbf{53.765} & \textbf{71.104} & \textbf{0.188} & \textbf{0.284} & \textbf{0.535} \\
\bottomrule
\end{tabular}
\end{table}

%% file: tables/table_contamination_split.tex
\begin{table}[htbp]
\centering
\singlespacing
\caption{Contamination check by sample split. Mean across the 50 assets of the per-asset QLIKE loss ratio to Log-HAR before and after the split, by the date of the forecast target. Panel~A: every model split at 1 January 2024. Panel~B: each foundation model split at the date its pinned weight file was last changed on Hugging Face, which cannot be earlier than the release of the weights and is therefore an upper bound on the end of its pretraining data, with that date and the number of trading days after it (median across assets, at $h = 1$). $t$ is the Newey--West $t$-statistic of a dummy for the later period in a regression of the cross-asset daily mean of the model's QLIKE minus the average QLIKE of HAR, ARFIMA, ARMA and MEM. A positive $t$ means the model loses ground after the split relative to these four models, which are estimated only on the 1{,}000 days before each origin. A dash (--) marks these four control models, for which we do not run the test.}
\label{tab:contamination_split}
\small
\setlength{\tabcolsep}{3.5pt}
\resizebox{\textwidth}{!}{%
\begin{tabular}{llrrrrrrrrrr}
\toprule
& & & \multicolumn{3}{c}{$h=1$} & \multicolumn{3}{c}{$h=5$} & \multicolumn{3}{c}{$h=22$} \\
\cmidrule(lr){4-6}\cmidrule(lr){7-9}\cmidrule(lr){10-12}
Model & Weights & Days & Before & After & $t$ & Before & After & $t$ & Before & After & $t$ \\
\midrule
\multicolumn{12}{l}{\textbf{Panel A: split at 1 January 2024}} \\[2pt]
HAR &  &  & 1.000 & 0.996 & -- & 1.005 & 1.015 & -- & 1.076 & 1.029 & -- \\
ARFIMA &  &  & 1.011 & 1.027 & -- & 1.064 & 1.047 & -- & 1.115 & 1.038 & -- \\
ARMA &  &  & 1.005 & 1.007 & -- & 1.051 & 1.050 & -- & 1.146 & 1.068 & -- \\
MEM &  &  & 1.017 & 1.013 & -- & 1.043 & 1.047 & -- & 1.145 & 1.077 & -- \\
HAR-J &  &  & 1.265 & 1.003 & -- & 1.169 & 1.028 & -- & 1.037 & 1.127 & -- \\
HAR-RS &  &  & 1.307 & 1.021 & -- & 1.190 & 1.038 & -- & 1.038 & 1.129 & -- \\
HARQ &  &  & 1.374 & 1.035 & -- & 1.264 & 1.023 & -- & 1.071 & 1.118 & -- \\
Chronos-Bolt-S &  &  & 1.043 & 1.083 & 3.44 & 1.086 & 1.103 & 0.77 & 1.201 & 1.072 & -0.94 \\
Chronos-Bolt-B &  &  & 1.049 & 1.097 & 2.93 & 1.081 & 1.102 & 0.68 & 1.200 & 1.077 & -0.82 \\
Moirai-2.0-S &  &  & 1.026 & 1.066 & 2.89 & 1.110 & 1.118 & 0.90 & 1.279 & 1.164 & -0.96 \\
Moirai-MoE-S &  &  & 1.022 & 1.036 & 1.72 & 1.065 & 1.106 & 0.66 & 1.145 & 1.151 & 0.69 \\
Lag-Llama &  &  & 1.674 & 1.245 & -1.48 & 1.789 & 1.083 & -1.86 & 1.021 & 1.043 & 0.27 \\
TimesFM-2.5 &  &  & 0.993 & 1.005 & 1.10 & 1.068 & 1.041 & -0.35 & 1.165 & 1.059 & -0.75 \\
Toto &  &  & 1.200 & 1.101 & -1.10 & 1.282 & 1.075 & -1.07 & 1.139 & 1.099 & -0.16 \\
Sundial &  &  & 0.998 & 1.003 & 0.50 & 1.105 & 1.039 & -0.79 & 1.209 & 1.048 & -1.09 \\
TTM &  &  & 0.981 & 0.990 & 0.11 & 0.988 & 0.993 & -0.40 & 0.980 & 1.020 & 0.94 \\
\midrule
\multicolumn{12}{l}{\textbf{Panel B: split at the date of each model's weight file}} \\[2pt]
Chronos-Bolt-S & 2024-11-13 & 302 & 1.052 & 1.073 & 1.61 & 1.096 & 1.074 & -0.51 & 1.202 & 1.009 & -1.47 \\
Chronos-Bolt-B & 2024-11-13 & 302 & 1.060 & 1.083 & 1.16 & 1.092 & 1.072 & -0.45 & 1.202 & 1.010 & -1.33 \\
Moirai-2.0-S & 2025-08-06 & 122 & 1.038 & 1.056 & 0.74 & 1.114 & 1.083 & -0.63 & 1.265 & 1.085 & -1.49 \\
Moirai-MoE-S & 2024-11-01 & 310 & 1.022 & 1.042 & 1.40 & 1.073 & 1.090 & -0.13 & 1.150 & 1.136 & -0.28 \\
Lag-Llama & 2024-02-07 & 496 & 1.658 & 1.257 & -1.37 & 1.780 & 1.077 & -1.88 & 1.022 & 1.038 & 0.21 \\
TimesFM-2.5 & 2025-10-01 & 83 & 0.995 & 1.030 & 1.51 & 1.059 & 1.078 & 0.87 & 1.149 & 1.091 & -0.11 \\
Toto & 2025-05-06 & 185 & 1.176 & 1.113 & -1.22 & 1.249 & 1.095 & -0.97 & 1.136 & 1.102 & 0.20 \\
Sundial & 2025-05-14 & 179 & 0.998 & 1.006 & 0.08 & 1.092 & 1.021 & -1.03 & 1.193 & 0.989 & -1.09 \\
TTM & 2025-01-31 & 250 & 0.980 & 1.005 & -0.20 & 0.988 & 0.997 & -1.26 & 0.983 & 1.030 & 0.61 \\
\bottomrule
\end{tabular}%
}
\end{table}

%% file: tables/table_log_input.tex
\begin{table}[htbp]
\centering
\singlespacing
\caption{Foundation models applied to log volatility. Mean across the 50 assets of the per-asset QLIKE loss ratio to Log-HAR for each foundation model with volatility as input, as in the main comparison (Vol.), and for its log version, which receives log volatility over the same context (Log), with the percentage of the 50 assets on which the log version has a ratio below one (\% $<1$). We clip the log forecast $\hat{m}_{t+h}$ to the logarithm of the bounds of Sec.~\ref{subsec:evaluation} and retransform it to volatility as $\exp(\hat{m}_{t+h} + \hat{v}_t/2)$. Here $\hat{v}_t$ is the variance of the model's one-step errors in log volatility over the 1{,}000 forecasts before the origin, or over all earlier forecasts during the first 1{,}000 days of the evaluation period, with $\hat{v}_t = 0$ for the first 22 forecasts. Every model is scored on the dates common to every model and both inputs.}
\label{tab:log_input}
\small
\setlength{\tabcolsep}{4pt}
\begin{tabular}{lrrrrrrrrr}
\toprule
& \multicolumn{3}{c}{$h=1$} & \multicolumn{3}{c}{$h=5$} & \multicolumn{3}{c}{$h=22$} \\
\cmidrule(lr){2-4}\cmidrule(lr){5-7}\cmidrule(lr){8-10}
Model & Vol. & Log & \% $<1$ & Vol. & Log & \% $<1$ & Vol. & Log & \% $<1$ \\
\midrule
Chronos-Bolt-S & 1.057 & 1.009 & 34 & 1.092 & 1.053 & 8 & 1.180 & 1.140 & 2 \\
Chronos-Bolt-B & 1.065 & 1.003 & 54 & 1.088 & 1.039 & 22 & 1.180 & 1.146 & 0 \\
Moirai-2.0-S & 1.039 & 1.001 & 54 & 1.113 & 1.073 & 0 & 1.260 & 1.157 & 0 \\
Moirai-MoE-S & 1.025 & 1.006 & 38 & 1.077 & 1.087 & 2 & 1.147 & 1.193 & 0 \\
Lag-Llama & 1.535 & 1.071 & 0 & 1.588 & 1.173 & 0 & 1.024 & 2.030 & 0 \\
TimesFM-2.5 & 0.997 & 0.989 & 80 & 1.060 & 1.055 & 2 & 1.149 & 1.148 & 2 \\
Toto & 1.169 & 1.401 & 0 & 1.241 & 1.299 & 0 & 1.132 & 1.341 & 0 \\
Sundial & 0.999 & 0.975 & 94 & 1.087 & 1.105 & 0 & 1.183 & 1.199 & 0 \\
TTM & 0.984 & 0.993 & 62 & 0.989 & 1.021 & 26 & 0.988 & 1.024 & 42 \\
\bottomrule
\end{tabular}
\end{table}

%% file: tables/table_ml_baselines.tex
\begin{table}[htbp]
\centering
\singlespacing
\caption{Supervised machine learning baselines. Mean across the 50 assets of the per-asset QLIKE loss ratio to Log-HAR, with the number of assets, of 50, with a ratio below one ($<1$). XGBoost and the feedforward network take the logarithms of the average volatility over the last 1, 5 and 22 days, and the LSTM takes the last 22 values of log volatility. Each learner forecasts log volatility $h$ days ahead directly. We fit it on the 1{,}000 days before the origin, refit it every 22 trading days and tune it every 252 trading days, and we retransform its forecast to volatility with the rule of Log-HAR (Appendix~\ref{app:ml}). LSTM, patience 100: the LSTM with early stopping at a patience of 100 epochs and at most 500 epochs. Log-HAR, refit every 22 days: Log-HAR refitted every 22 trading days rather than at every origin. TTM is as in the main comparison. Every model is scored on the dates common to every model, with forecasts clipped to the bounds of Sec.~\ref{subsec:evaluation}.}
\label{tab:ml_baselines}
\small
\setlength{\tabcolsep}{5pt}
\begin{tabular}{lrrrrrr}
\toprule
& \multicolumn{2}{c}{$h=1$} & \multicolumn{2}{c}{$h=5$} & \multicolumn{2}{c}{$h=22$} \\
\cmidrule(lr){2-3}\cmidrule(lr){4-5}\cmidrule(lr){6-7}
Model & Ratio & $<1$ & Ratio & $<1$ & Ratio & $<1$ \\
\midrule
TTM & 0.984 & 37 & 0.989 & 31 & 0.988 & 32 \\
Log-HAR, refit every 22 days & 1.011 & 2 & 1.025 & 1 & 1.005 & 16 \\
XGBoost & 1.171 & 0 & 1.367 & 0 & 1.274 & 0 \\
Feedforward network & 1.142 & 0 & 1.215 & 0 & 1.203 & 0 \\
LSTM & 1.047 & 5 & 1.224 & 0 & 1.196 & 1 \\
LSTM, patience 100 & 1.052 & 4 & 1.199 & 1 & 1.214 & 1 \\
\bottomrule
\end{tabular}
\end{table}

%% file: tables/mz_bias_corrected.tex
\begin{table}[htbp]
\centering
\small
\setlength{\tabcolsep}{5pt}
\caption{MZ bias-corrected QLIKE across horizons (cross-asset mean over 50 assets). For each model and horizon we report the original QLIKE (Orig.) on the dates after the 252-day warm-up of the correction and the QLIKE after a recursive affine MZ correction (MZ), with $\hat\alpha_t,\hat\beta_t$ estimated on an expanding window from the forecasts whose targets are observed by each origin and applied symmetrically to all 17 models. QLIKE is on the variance scale. The lowest corrected QLIKE in each horizon is in bold. $\dagger$ marks QLIKE $>1$.}
\label{tab:mz_bias_corrected}
\begin{tabular}{lcccccc}
\toprule
& \multicolumn{2}{c}{$h=1$} & \multicolumn{2}{c}{$h=5$} & \multicolumn{2}{c}{$h=22$} \\
\cmidrule(lr){2-3}\cmidrule(lr){4-5}\cmidrule(lr){6-7}
Model & Orig. & MZ & Orig. & MZ & Orig. & MZ \\
\midrule
Log-HAR & 0.196 & 0.195 & 0.298 & 0.308 & 0.545 & 0.798 \\
HAR & 0.196 & 0.194 & 0.301 & \textbf{0.305} & 0.587 & 0.864 \\
HAR-J & 0.230 & 0.233 & 0.339 & 0.350 & 0.574 & 1.032$^{\dagger}$ \\
HAR-RS & 0.238 & 0.240 & 0.344 & 0.355 & 0.575 & 1.042$^{\dagger}$ \\
HARQ & 0.248 & 0.243 & 0.360 & 0.361 & 0.591 & 0.917 \\
ARFIMA & 0.197 & 0.199 & 0.315 & 0.313 & 0.605 & 0.801 \\
ARMA & 0.197 & 0.199 & 0.315 & 0.324 & 0.626 & 0.727 \\
MEM & 0.198 & 0.198 & 0.313 & 0.318 & 0.631 & 0.925 \\
Chronos-Bolt-S & 0.208 & 0.194 & 0.327 & 0.307 & 0.652 & 0.767 \\
Chronos-Bolt-B & 0.209 & 0.193 & 0.325 & 0.308 & 0.650 & 0.769 \\
Moirai-2.0-S & 0.204 & 0.193 & 0.335 & 0.315 & 0.697 & 0.784 \\
Moirai-MoE-S & 0.200 & 0.207 & 0.319 & 0.335 & 0.622 & 0.730 \\
Lag-Llama & 0.303 & 0.326 & 0.480 & 0.528 & 0.547 & \textbf{0.661} \\
TimesFM-2.5 & 0.195 & \textbf{0.190} & 0.317 & 0.313 & 0.634 & 0.816 \\
Toto & 0.229 & 0.223 & 0.389 & 0.355 & 0.619 & 0.735 \\
Sundial & 0.195 & 0.196 & 0.326 & 0.335 & 0.659 & 0.853 \\
TTM & 0.192 & 0.198 & 0.293 & 0.317 & 0.535 & 0.750 \\
\bottomrule
\end{tabular}
\end{table}

%% file: tables/table_combination.tex
\begin{table}[t]
\centering
\singlespacing
\caption{Forecast combinations of TTM, Log-HAR and ARMA, the most accurate of ARFIMA, ARMA and MEM at $h = 1$, each also shown alone. Panel~A: average QLIKE loss ratio to Log-HAR across the 50 assets, where a ratio below one means a lower QLIKE than Log-HAR on average and bold marks the lowest ratio in each column. Panel~B: share of the 50 assets for which each row enters the MCS, computed on the 17 models and the four combinations. The combinations are an equal-weight average and a recursive Bates--Granger minimum-variance combination whose weights are estimated from the errors of the forecasts whose targets are observed by the forecast origin (expanding window, clipped to non-negative weights, equal-weight warm-up).}
\label{tab:combination}
\small
\begin{tabular}{lrrr}
\toprule
Model & $h=1$ & $h=5$ & $h=22$ \\
\midrule
\multicolumn{4}{l}{\textit{Panel A: average QLIKE loss ratio vs Log-HAR}} \\[2pt]
Log-HAR & 1.000 & 1.000 & 1.000 \\
TTM & 0.984 & 0.989 & 0.988 \\
ARMA & 1.006 & 1.051 & 1.133 \\
TTM + Log-HAR (equal weight) & \textbf{0.978} & 0.986 & \textbf{0.982} \\
TTM + Log-HAR (Bates--Granger) & 0.982 & \textbf{0.985} & 0.991 \\
TTM + Log-HAR + ARMA (equal weight) & 0.983 & 1.000 & 1.019 \\
TTM + Log-HAR + ARMA (Bates--Granger) & 0.986 & 1.003 & 1.057 \\
\addlinespace
\multicolumn{4}{l}{\textit{Panel B: MCS inclusion rate (\% of 50 assets)}} \\[2pt]
Log-HAR & 58.0 & 92.0 & 100.0 \\
TTM & 90.0 & 98.0 & 96.0 \\
ARMA & 46.0 & 48.0 & 76.0 \\
TTM + Log-HAR (equal weight) & 100.0 & 98.0 & 100.0 \\
TTM + Log-HAR (Bates--Granger) & 98.0 & 100.0 & 98.0 \\
TTM + Log-HAR + ARMA (equal weight) & 98.0 & 92.0 & 96.0 \\
TTM + Log-HAR + ARMA (Bates--Granger) & 92.0 & 82.0 & 98.0 \\
\bottomrule
\end{tabular}
\end{table}

%% file: tables/table_combination_diagnostics.tex
\begin{table}[htbp]
\centering
\singlespacing
\caption{Diagnostics of the equal-weight combination of TTM and Log-HAR. Panel~A: on the volatility scale, $k$ is the ratio of the larger to the smaller root mean squared error (RMSE) of the two models and $\rho$ the correlation of their errors, both computed around zero. Under squared error the average of the two forecasts beats the better one when $\rho < (3-k^2)/(2k)$. Panel~B: mean across the 50 assets of the per-asset QLIKE ratio to Log-HAR for TTM and for two combinations, the harmonic mean of the two variance forecasts and the square of the average of the two volatility forecasts, which is the combination of Tab.~\ref{tab:combination}. The diversity share is the part of the combination's gain over the average QLIKE of its two members that comes from the diversity term $D_t$ of Eq.~\eqref{eq:diversity}, averaged across assets. Panel~C: each model's variance forecast multiplied by its average ratio of realized to forecast variance over the previous 250 forecasts whose targets are observed at the origin, as in Appendix~\ref{app:forecast_level}, with ratios to the uncorrected Log-HAR. Panel~D: number of assets, of 50, where a Newey--West test rejects at the 5\% level that moving weight from one model toward the other leaves the expected QLIKE unchanged. Panel~E: share of 126-day blocks in which TTM has the lower QLIKE, number of changes of the model with the lower QLIKE between consecutive blocks, and the recursive Bates--Granger combination with its weight estimated from the errors observed by each origin.}
\label{tab:combination_diagnostics}
\small
\begin{tabular}{lrrr}
\toprule
 & $h=1$ & $h=5$ & $h=22$ \\
\midrule
\multicolumn{4}{l}{\textbf{Panel A: forecast errors on the volatility scale}} \\[2pt]
RMSE ratio $k$ (median) & 1.008 & 1.007 & 1.029 \\
Error correlation $\rho$ (median) & 0.973 & 0.984 & 0.978 \\
Threshold $(3-k^2)/(2k)$ (median) & 0.985 & 0.986 & 0.944 \\
Assets with $\rho$ below the threshold & 37 & 24 & 7 \\
\midrule
\multicolumn{4}{l}{\textbf{Panel B: QLIKE loss ratio to Log-HAR}} \\[2pt]
TTM & 0.984 & 0.989 & 0.988 \\
Harmonic mean of the variance forecasts & 0.980 & 0.987 & 0.985 \\
Average of the volatility forecasts & 0.978 & 0.986 & 0.982 \\
Assets where the average beats both & 32 & 21 & 11 \\
Diversity share of the gain (\%) & 92 & 88 & 87 \\
\midrule
\multicolumn{4}{l}{\textbf{Panel C: after a real-time level correction of both models}} \\[2pt]
TTM & 0.992 & 1.004 & 1.105 \\
Log-HAR & 0.976 & 0.988 & 1.125 \\
Average of the volatility forecasts & 0.975 & 0.989 & 1.108 \\
Assets where the average beats both & 27 & 10 & 10 \\
\midrule
\multicolumn{4}{l}{\textbf{Panel D: encompassing under QLIKE, assets with $t > 1.96$}} \\[2pt]
TTM adds information to Log-HAR & 47 & 26 & 8 \\
Log-HAR adds information to TTM & 27 & 12 & 8 \\
\midrule
\multicolumn{4}{l}{\textbf{Panel E: stability over time}} \\[2pt]
126-day blocks with TTM ahead (\%) & 58 & 58 & 56 \\
Changes of the leading model per asset & 7.3 & 7.0 & 6.7 \\
Bates--Granger weight on TTM (mean) & 0.48 & 0.54 & 0.82 \\
Bates--Granger combination, loss ratio & 0.982 & 0.985 & 0.991 \\
\bottomrule
\end{tabular}
\end{table}

%% file: tables/table_equity_metrics_median.tex
\begin{table}[htbp]
\centering
\singlespacing
\caption{Forecast accuracy for 40 U.S.\ equities (VOLARE), cross-sectional medians of the per-asset loss functions. MSE is on the volatility scale and QLIKE on the variance scale. Bold indicates the best value in each column.}
\label{tab:main_results_median}
\footnotesize
\setlength{\tabcolsep}{4pt}
\begin{tabular}{lrrrrrr}
\toprule
& \multicolumn{3}{c}{MSE ($\times 10^{-6}$)} & \multicolumn{3}{c}{QLIKE} \\
\cmidrule(lr){2-4}\cmidrule(lr){5-7}
Model & $h=1$ & $h=5$ & $h=22$ & $h=1$ & $h=5$ & $h=22$ \\
\midrule
Log-HAR & 23.833 & 38.347 & 56.403 & 0.183 & 0.287 & 0.530 \\
HAR & 23.403 & 38.301 & 65.920 & 0.184 & 0.290 & 0.567 \\
HAR-J & 28.501 & 46.626 & 62.817 & 0.218 & 0.338 & 0.565 \\
HAR-RS & 29.359 & 47.267 & 62.603 & 0.221 & 0.343 & 0.565 \\
HARQ & 30.921 & 50.914 & 61.994 & 0.235 & 0.351 & 0.567 \\
ARFIMA & 24.643 & 39.726 & 54.196 & 0.184 & 0.303 & 0.577 \\
ARMA & 24.212 & 37.980 & 55.366 & 0.183 & 0.297 & 0.588 \\
MEM & 23.625 & 38.385 & 57.506 & 0.184 & 0.300 & 0.594 \\
Chronos-Bolt-S & 23.318 & \textbf{36.227} & 55.971 & 0.196 & 0.310 & 0.615 \\
Chronos-Bolt-B & 23.356 & 36.964 & 55.256 & 0.195 & 0.308 & 0.633 \\
Moirai-2.0-S & 23.445 & 38.527 & 55.238 & 0.190 & 0.315 & 0.656 \\
Moirai-MoE-S & 25.409 & 41.830 & 64.679 & 0.191 & 0.305 & 0.608 \\
Lag-Llama & 40.010 & 56.457 & 60.661 & 0.280 & 0.460 & \textbf{0.513} \\
TimesFM-2.5 & \textbf{23.214} & 37.927 & 54.073 & 0.183 & 0.303 & 0.591 \\
Toto & 28.471 & 40.656 & 65.358 & 0.211 & 0.298 & 0.593 \\
Sundial & 23.940 & 40.494 & 56.500 & 0.183 & 0.314 & 0.629 \\
TTM & 23.757 & 37.398 & \textbf{52.930} & \textbf{0.180} & \textbf{0.278} & 0.514 \\
\bottomrule
\end{tabular}
\end{table}

%% file: tables/table_context_sensitivity.tex
\begin{table}[htbp]
\centering
\singlespacing
\small
\caption{Context-length sensitivity of TSFM forecasts across 50 assets (40 equities, 5 FX, 5 futures), evaluated under the point-in-time target, mean forecast, and volatility scale. Each cell is the QLIKE on the variance scale, averaged across assets, by horizon and context length. ctx$=$1{,}000 is the default for all models except TTM, whose r2.1 release supports at most 512 days (ctx$=$1{,}000 not available, marked --) and which uses 512 as its default. TTM's ctx$=$128 and ctx$=$256 columns use the 90-day and 180-day branches of the release, the latter trained with an absolute-error loss. All context lengths are scored on the common out-of-sample window of the default run. Bold marks the best available context length for each model and horizon.}
\label{tab:context_sensitivity}
\begin{tabular}{llrrrr}
\toprule
Model & $h$ & ctx=128 & ctx=256 & ctx=512 & ctx=1000 \\
\midrule
Chronos-Bolt-S & 1 & 0.204 & \textbf{0.203} & 0.204 & 0.205 \\
 & 5 & \textbf{0.310} & 0.316 & 0.313 & 0.317 \\
 & 22 & 0.573 & 0.616 & \textbf{0.569} & 0.605 \\
\addlinespace
Chronos-Bolt-B & 1 & 0.205 & \textbf{0.204} & 0.205 & 0.207 \\
 & 5 & 0.313 & 0.316 & \textbf{0.312} & 0.315 \\
 & 22 & 0.581 & 0.618 & \textbf{0.554} & 0.603 \\
\addlinespace
TimesFM-2.5 & 1 & 0.198 & 0.198 & 0.197 & \textbf{0.193} \\
 & 5 & 0.312 & 0.321 & 0.312 & \textbf{0.307} \\
 & 22 & 0.624 & 0.626 & \textbf{0.581} & 0.587 \\
\addlinespace
Moirai-2.0-S & 1 & 0.202 & \textbf{0.201} & 0.201 & 0.202 \\
 & 5 & \textbf{0.313} & 0.314 & 0.317 & 0.323 \\
 & 22 & \textbf{0.614} & 0.625 & 0.632 & 0.645 \\
\addlinespace
Moirai-MoE-S & 1 & 0.217 & 0.215 & 0.210 & \textbf{0.198} \\
 & 5 & 0.339 & 0.324 & 0.318 & \textbf{0.311} \\
 & 22 & 0.626 & 0.620 & \textbf{0.574} & 0.580 \\
\addlinespace
Lag-Llama & 1 & \textbf{0.269} & 0.296 & 0.321 & 0.292 \\
 & 5 & \textbf{0.426} & 0.439 & 0.426 & 0.452 \\
 & 22 & 0.529 & 0.525 & \textbf{0.492} & 0.513 \\
\addlinespace
Toto & 1 & 0.276 & 0.232 & \textbf{0.223} & 0.225 \\
 & 5 & 0.356 & 0.365 & \textbf{0.330} & 0.373 \\
 & 22 & 0.643 & 0.685 & 0.584 & \textbf{0.575} \\
\addlinespace
Sundial & 1 & 0.200 & 0.197 & \textbf{0.193} & 0.193 \\
 & 5 & 0.325 & 0.315 & \textbf{0.304} & 0.314 \\
 & 22 & 0.612 & 0.597 & \textbf{0.544} & 0.608 \\
\addlinespace
TTM & 1 & 0.192 & 0.227 & \textbf{0.190} & -- \\
 & 5 & 0.288 & 0.343 & \textbf{0.285} & -- \\
 & 22 & 0.552 & 0.668 & \textbf{0.499} & -- \\
\addlinespace
\bottomrule
\end{tabular}
\end{table}

%% file: tables/table_equal_context.tex
\begin{table}[htbp]
\centering
\singlespacing
\caption{Foundation models at an equal context. Mean across the 50 assets of the per-asset QLIKE loss ratio, every forecast clipped to the bounds of Sec.~\ref{subsec:evaluation}. Panel~A: each model at its main context against Log-HAR estimated on 1{,}000 days, and at 512 days against Log-HAR estimated on 512 days, on the dates of the main comparison. TTM's r2.1 release supports at most 512 days, so its two columns coincide in the context. Panel~B: TTM by context against Log-HAR on 1{,}000 days, on the dates for which the 1{,}536-day model has forecasts. The r2.1 release is the one used in the main comparison, pretrained with daily series and a frequency token. The r2 release is the earlier one, pretrained on a smaller corpus without daily series. Its 512-day checkpoint has a 192-day forecast head and the others a 96-day head.}
\label{tab:equal_context}
\small
\setlength{\tabcolsep}{4pt}
\begin{tabular}{lrrrrrrr}
\toprule
& & \multicolumn{3}{c}{Main context} & \multicolumn{3}{c}{512 days} \\
\cmidrule(lr){3-5}\cmidrule(lr){6-8}
Model & Context & $h=1$ & $h=5$ & $h=22$ & $h=1$ & $h=5$ & $h=22$ \\
\midrule
\multicolumn{8}{l}{\textbf{Panel A: every foundation model at 512 days}} \\[2pt]
Chronos-Bolt-S & 1{,}000 & 1.057 & 1.092 & 1.180 & 1.051 & 1.073 & 1.135 \\
Chronos-Bolt-B & 1{,}000 & 1.065 & 1.088 & 1.180 & 1.055 & 1.073 & 1.107 \\
Moirai-2.0-S & 1{,}000 & 1.039 & 1.113 & 1.260 & 1.035 & 1.088 & 1.253 \\
Moirai-MoE-S & 1{,}000 & 1.025 & 1.077 & 1.147 & 1.086 & 1.095 & 1.156 \\
Lag-Llama & 1{,}000 & 1.535 & 1.588 & 1.024 & 1.688 & 1.488 & 1.000 \\
TimesFM-2.5 & 1{,}000 & 0.997 & 1.060 & 1.149 & 1.014 & 1.074 & 1.157 \\
Toto & 1{,}000 & 1.169 & 1.241 & 1.132 & 1.151 & 1.132 & 1.181 \\
Sundial & 1{,}000 & 0.999 & 1.087 & 1.183 & 0.997 & 1.048 & 1.085 \\
TTM & 512 & 0.984 & 0.989 & 0.988 & 0.981 & 0.985 & 1.003 \\
\midrule
\multicolumn{8}{l}{\textbf{Panel B: TTM by context}} \\[2pt]
TTM r2.1 & 512 & 0.988 & 0.990 & 1.008 &  &  &  \\
TTM r2 & 512 & 1.021 & 1.031 & 1.083 &  &  &  \\
TTM r2 & 1{,}024 & 1.043 & 1.030 & 1.033 &  &  &  \\
TTM r2 & 1{,}536 & 1.026 & 1.017 & 1.049 &  &  &  \\
\bottomrule
\end{tabular}
\end{table}

%% file: tables/table_avg_target.tex
\begin{table}[htbp]
\centering
\singlespacing
\caption{Averaged target. Mean across the 50 assets of the per-asset QLIKE loss ratio to Log-HAR when the target is the average of daily realized variance over the next $h$ days \citep{patton2015}, with each volatility forecast squared. Every model is scored on the dates common to the 17 models of each asset. A ratio below 1 means a lower QLIKE than Log-HAR on average, and the lowest ratio at each horizon is in bold. Tab.~\ref{tab:loss_ratios} gives the same ratios for the point-in-time target.}
\label{tab:avg_target}
\small
\begin{tabular}{lrrr}
\toprule
Model & $h=1$ & $h=5$ & $h=22$ \\
\midrule
Log-HAR & 1.000 & 1.000 & 1.000 \\
HAR & 0.999 & 1.005 & 1.057 \\
HAR-J & 1.181 & 1.603 & 1.252 \\
HAR-RS & 1.216 & 1.583 & 1.240 \\
HARQ & 1.264 & 1.520 & 1.232 \\
ARFIMA & 1.016 & 1.060 & 1.148 \\
ARMA & 1.006 & 1.053 & 1.166 \\
MEM & 1.016 & 1.036 & 1.158 \\
Chronos-Bolt-S & 1.057 & 1.106 & 1.241 \\
Chronos-Bolt-B & 1.065 & 1.114 & 1.261 \\
Moirai-2.0-S & 1.039 & 1.120 & 1.341 \\
Moirai-MoE-S & 1.025 & 1.015 & 1.031 \\
Lag-Llama & 1.535 & 1.883 & 1.376 \\
TimesFM-2.5 & 0.997 & 1.045 & 1.184 \\
Toto & 1.169 & 1.978 & 3.019 \\
Sundial & 0.999 & 1.050 & 1.211 \\
TTM & \textbf{0.984} & \textbf{0.971} & \textbf{0.992} \\
\bottomrule
\end{tabular}
\end{table}

%% file: tables/table_forecast_level.tex
\begin{table}[htbp]
\centering
\singlespacing
\caption{Forecast level and level adjustments. Level: mean across the 50 assets of the per-asset average of $RV_{t+h}/\widehat{\sigma}^{2}_{t+h}$, one when the variance forecast is right on average. This column is a ratio of variances, not a loss ratio. Log-normal: QLIKE loss ratio when Log-HAR, ARFIMA and ARMA are scored on their log-normal expected variance $\exp(2\hat{m}_{t+h} + 2\hat{v}_{t+h})$ and the other models are unchanged, relative to Log-HAR scored the same way. Rescaled: QLIKE loss ratio when every variance forecast is multiplied by its average ratio of realized to forecast variance over its last 250 forecasts whose targets are observed before the origin, relative to the rescaled Log-HAR. Log-HAR therefore has a ratio of one under both adjustments. Every forecast is clipped to the window bounds of Sec.~\ref{subsec:evaluation}, and every model is scored on the dates common to the 17 models.}
\label{tab:forecast_level}
\small
\setlength{\tabcolsep}{4pt}
\begin{tabular}{lrrrrrrrrr}
\toprule
& \multicolumn{3}{c}{Level} & \multicolumn{3}{c}{Log-normal} & \multicolumn{3}{c}{Rescaled} \\
\cmidrule(lr){2-4}\cmidrule(lr){5-7}\cmidrule(lr){8-10}
Model & $h=1$ & $h=5$ & $h=22$ & $h=1$ & $h=5$ & $h=22$ & $h=1$ & $h=5$ & $h=22$ \\
\midrule
Log-HAR & 1.125 & 1.165 & 1.307 & 1.000 & 1.000 & 1.000 & 1.000 & 1.000 & 1.000 \\
HAR & 1.120 & 1.192 & 1.415 & 1.030 & 1.029 & 1.024 & 1.001 & 0.990 & 1.005 \\
HAR-J & 1.030 & 1.048 & 1.171 & 1.218 & 1.153 & 1.007 & 1.200 & 1.144 & 1.089 \\
HAR-RS & 1.034 & 1.052 & 1.171 & 1.253 & 1.171 & 1.008 & 1.227 & 1.160 & 1.092 \\
HARQ & 1.075 & 1.087 & 1.190 & 1.303 & 1.219 & 1.035 & 1.268 & 1.198 & 1.106 \\
ARFIMA & 1.119 & 1.231 & 1.493 & 1.020 & 1.018 & 0.952 & 1.006 & 1.019 & 1.016 \\
ARMA & 1.130 & 1.249 & 1.549 & 1.005 & 0.998 & 0.956 & 1.003 & 1.001 & 1.000 \\
MEM & 1.126 & 1.205 & 1.469 & 1.048 & 1.065 & 1.088 & 1.014 & 1.014 & 1.043 \\
Chronos-Bolt-S & 1.209 & 1.329 & 1.599 & 1.090 & 1.115 & 1.133 & 0.996 & 0.986 & 1.044 \\
Chronos-Bolt-B & 1.223 & 1.323 & 1.607 & 1.098 & 1.110 & 1.133 & 0.991 & 0.985 & 1.031 \\
Moirai-2.0-S & 1.203 & 1.344 & 1.723 & 1.072 & 1.136 & 1.209 & 0.983 & 0.996 & 1.034 \\
Moirai-MoE-S & 1.060 & 1.210 & 1.481 & 1.057 & 1.099 & 1.101 & 1.054 & 1.059 & 1.085 \\
Lag-Llama & 1.101 & 1.154 & 1.191 & 1.582 & 1.621 & 0.983 & 1.582 & 1.626 & 1.096 \\
TimesFM-2.5 & 1.151 & 1.268 & 1.558 & 1.028 & 1.083 & 1.102 & 0.980 & 1.001 & 1.037 \\
Toto & 1.127 & 1.248 & 1.414 & 1.205 & 1.268 & 1.087 & 1.186 & 1.185 & 1.138 \\
Sundial & 1.125 & 1.233 & 1.541 & 1.031 & 1.110 & 1.135 & 0.999 & 1.054 & 1.095 \\
TTM & 1.052 & 1.133 & 1.386 & 1.014 & 1.010 & 0.948 & 1.016 & 1.016 & 0.982 \\
\bottomrule
\end{tabular}
\end{table}

%% file: tables/table_contamination_years.tex
\begin{table}[htbp]
\centering
\singlespacing
\caption{QLIKE loss ratio to Log-HAR by calendar year of the forecast target. Mean across the 50 assets of the per-asset ratio of mean QLIKE to Log-HAR's on the same days. The last column covers January 2026, the end of the sample.}
\label{tab:contamination_years}
\footnotesize
\setlength{\tabcolsep}{4pt}
\begin{tabular}{lrrrrrrrr}
\toprule
Model & 2019 & 2020 & 2021 & 2022 & 2023 & 2024 & 2025 & Jan.~2026 \\
\midrule
\multicolumn{9}{l}{\textbf{Panel A: $h=1$}} \\[2pt]
HAR & 0.999 & 0.954 & 1.023 & 1.029 & 1.013 & 1.000 & 1.000 & 0.943 \\
ARFIMA & 1.078 & 1.000 & 0.991 & 0.985 & 1.010 & 0.993 & 1.037 & 1.265 \\
ARMA & 1.009 & 1.002 & 1.005 & 1.003 & 1.029 & 0.997 & 1.013 & 1.010 \\
MEM & 1.021 & 1.004 & 1.018 & 1.042 & 1.009 & 1.006 & 1.020 & 1.048 \\
HAR-J & 1.020 & 1.840 & 1.125 & 1.093 & 1.140 & 1.056 & 0.982 & 0.861 \\
HAR-RS & 1.044 & 1.859 & 1.159 & 1.118 & 1.168 & 1.086 & 1.009 & 0.802 \\
HARQ & 1.015 & 2.113 & 1.171 & 1.122 & 1.218 & 1.077 & 1.019 & 0.883 \\
Chronos-Bolt-S & 1.050 & 1.006 & 1.063 & 1.064 & 1.032 & 1.092 & 1.069 & 1.112 \\
Chronos-Bolt-B & 1.065 & 1.008 & 1.070 & 1.096 & 1.017 & 1.112 & 1.075 & 1.161 \\
Moirai-2.0-S & 1.034 & 1.003 & 1.040 & 1.041 & 1.006 & 1.060 & 1.065 & 1.083 \\
Moirai-MoE-S & 1.055 & 0.961 & 1.078 & 1.037 & 1.036 & 1.033 & 1.048 & 0.997 \\
Lag-Llama & 1.270 & 2.750 & 1.450 & 1.673 & 1.211 & 1.163 & 1.329 & 1.204 \\
TimesFM-2.5 & 1.003 & 0.975 & 1.003 & 1.017 & 0.978 & 0.996 & 1.007 & 1.045 \\
Toto & 1.120 & 1.369 & 1.158 & 1.116 & 1.116 & 1.105 & 1.114 & 1.038 \\
Sundial & 1.007 & 0.999 & 0.997 & 1.003 & 0.986 & 0.991 & 1.017 & 0.968 \\
TTM & 0.997 & 0.975 & 0.998 & 0.974 & 0.988 & 0.981 & 1.010 & 0.869 \\
\midrule
\multicolumn{9}{l}{\textbf{Panel B: $h=5$}} \\[2pt]
HAR & 1.003 & 0.963 & 1.054 & 1.079 & 1.024 & 1.005 & 1.025 & 0.992 \\
ARFIMA & 1.084 & 1.136 & 0.986 & 1.009 & 0.976 & 1.009 & 1.049 & 1.405 \\
ARMA & 1.019 & 1.077 & 1.045 & 1.075 & 1.020 & 1.020 & 1.064 & 1.081 \\
MEM & 0.999 & 1.052 & 1.039 & 1.084 & 1.025 & 1.006 & 1.082 & 1.000 \\
HAR-J & 1.026 & 1.259 & 1.176 & 1.051 & 1.257 & 1.095 & 1.020 & 0.836 \\
HAR-RS & 1.026 & 1.272 & 1.174 & 1.069 & 1.299 & 1.106 & 1.031 & 0.826 \\
HARQ & 1.025 & 1.444 & 1.229 & 1.127 & 1.239 & 1.096 & 1.009 & 0.835 \\
Chronos-Bolt-S & 1.066 & 1.122 & 1.034 & 1.149 & 0.996 & 1.145 & 1.064 & 1.276 \\
Chronos-Bolt-B & 1.060 & 1.107 & 1.048 & 1.165 & 0.981 & 1.145 & 1.060 & 1.258 \\
Moirai-2.0-S & 1.054 & 1.191 & 1.063 & 1.119 & 0.978 & 1.103 & 1.116 & 1.205 \\
Moirai-MoE-S & 1.119 & 1.002 & 1.142 & 1.146 & 1.057 & 1.141 & 1.086 & 1.244 \\
Lag-Llama & 1.225 & 2.534 & 1.398 & 1.358 & 1.313 & 1.139 & 1.064 & 1.068 \\
TimesFM-2.5 & 1.017 & 1.159 & 0.998 & 1.062 & 0.952 & 1.023 & 1.039 & 1.202 \\
Toto & 1.078 & 1.888 & 1.147 & 1.227 & 1.110 & 1.075 & 1.081 & 1.074 \\
Sundial & 1.025 & 1.230 & 1.005 & 1.050 & 0.980 & 1.028 & 1.044 & 1.054 \\
TTM & 1.011 & 1.000 & 0.995 & 0.977 & 0.961 & 0.990 & 1.001 & 0.901 \\
\midrule
\multicolumn{9}{l}{\textbf{Panel C: $h=22$}} \\[2pt]
HAR & 0.993 & 1.095 & 1.000 & 1.162 & 0.993 & 1.005 & 1.041 & 1.021 \\
ARFIMA & 1.065 & 1.188 & 0.951 & 1.108 & 0.882 & 0.993 & 1.048 & 1.270 \\
ARMA & 1.005 & 1.198 & 1.028 & 1.321 & 0.925 & 1.015 & 1.092 & 1.069 \\
MEM & 0.974 & 1.180 & 1.012 & 1.238 & 1.017 & 1.053 & 1.099 & 1.017 \\
HAR-J & 1.014 & 0.954 & 1.315 & 0.985 & 1.453 & 1.254 & 1.067 & 0.914 \\
HAR-RS & 1.015 & 0.959 & 1.309 & 0.977 & 1.455 & 1.257 & 1.070 & 0.920 \\
HARQ & 1.017 & 1.013 & 1.305 & 0.983 & 1.421 & 1.241 & 1.062 & 0.926 \\
Chronos-Bolt-S & 1.034 & 1.297 & 0.999 & 1.365 & 0.889 & 1.165 & 0.995 & 1.261 \\
Chronos-Bolt-B & 1.028 & 1.283 & 1.005 & 1.420 & 0.896 & 1.179 & 0.994 & 1.255 \\
Moirai-2.0-S & 1.080 & 1.408 & 1.079 & 1.393 & 0.869 & 1.153 & 1.155 & 1.204 \\
Moirai-MoE-S & 1.123 & 1.171 & 1.200 & 1.286 & 1.021 & 1.178 & 1.140 & 1.200 \\
Lag-Llama & 1.131 & 0.921 & 1.325 & 1.192 & 1.189 & 1.110 & 1.009 & 1.047 \\
TimesFM-2.5 & 0.993 & 1.263 & 0.969 & 1.233 & 0.878 & 1.050 & 1.048 & 1.161 \\
Toto & 1.093 & 1.158 & 1.189 & 1.309 & 1.012 & 1.120 & 1.090 & 1.109 \\
Sundial & 1.016 & 1.296 & 1.015 & 1.286 & 0.969 & 1.041 & 1.046 & 1.091 \\
TTM & 0.990 & 0.979 & 1.006 & 1.063 & 0.888 & 0.992 & 1.042 & 0.965 \\
\bottomrule
\end{tabular}
\end{table}

%% file: tables/table_combination_pairs.tex
\begin{table}[htbp]
\centering
\singlespacing
\caption{Equal-weight pairs. Mean across the 50 assets of the per-asset QLIKE ratio to Log-HAR for the partner model alone, for the square of the average of the two volatility forecasts (Avg.), and for the harmonic mean of the two variance forecasts (Harm.), with the median correlation $\rho$ of the two models' errors on the volatility scale. Panel~A: Log-HAR with every other model. Panel~B: TTM with every econometric model. Panel~C: HAR with every model other than Log-HAR, whose pair with HAR is in Panel~A.}
\label{tab:combination_pairs}
\small
\setlength{\tabcolsep}{3.5pt}
\resizebox{\textwidth}{!}{%
\begin{tabular}{lrrrrrrrrrrrr}
\toprule
& \multicolumn{4}{c}{$h=1$} & \multicolumn{4}{c}{$h=5$} & \multicolumn{4}{c}{$h=22$} \\
\cmidrule(lr){2-5}\cmidrule(lr){6-9}\cmidrule(lr){10-13}
Partner & Alone & Avg. & Harm. & $\rho$ & Alone & Avg. & Harm. & $\rho$ & Alone & Avg. & Harm. & $\rho$ \\
\midrule
\multicolumn{13}{l}{\textbf{Panel A: averaged with Log-HAR}} \\[2pt]
TTM & 0.984 & 0.978 & 0.980 & 0.973 & 0.989 & 0.986 & 0.987 & 0.984 & 0.988 & 0.982 & 0.985 & 0.978 \\
Chronos-Bolt-S & 1.057 & 1.009 & 1.013 & 0.963 & 1.092 & 1.022 & 1.028 & 0.968 & 1.180 & 1.056 & 1.068 & 0.952 \\
Chronos-Bolt-B & 1.065 & 1.012 & 1.016 & 0.961 & 1.088 & 1.020 & 1.026 & 0.966 & 1.180 & 1.055 & 1.067 & 0.951 \\
TimesFM-2.5 & 0.997 & 0.989 & 0.991 & 0.982 & 1.060 & 1.018 & 1.020 & 0.977 & 1.149 & 1.052 & 1.059 & 0.963 \\
Moirai-2.0-S & 1.039 & 1.007 & 1.010 & 0.980 & 1.113 & 1.034 & 1.040 & 0.969 & 1.260 & 1.081 & 1.100 & 0.935 \\
Moirai-MoE-S & 1.025 & 0.985 & 0.985 & 0.900 & 1.077 & 1.003 & 1.006 & 0.908 & 1.147 & 1.035 & 1.043 & 0.925 \\
Lag-Llama & 1.535 & 1.088 & 1.159 & 0.823 & 1.588 & 1.093 & 1.196 & 0.826 & 1.024 & 0.976 & 0.979 & 0.920 \\
Toto & 1.169 & 0.999 & 1.050 & 0.850 & 1.241 & 0.993 & 1.092 & 0.902 & 1.132 & 1.017 & 1.038 & 0.909 \\
Sundial & 0.999 & 0.989 & 0.991 & 0.981 & 1.087 & 1.025 & 1.032 & 0.970 & 1.183 & 1.063 & 1.075 & 0.955 \\
HAR & 0.999 & 0.996 & 0.996 & 0.980 & 1.008 & 0.998 & 1.000 & 0.967 & 1.068 & 1.028 & 1.029 & 0.954 \\
ARFIMA & 1.016 & 0.996 & 0.997 & 0.984 & 1.060 & 1.017 & 1.021 & 0.983 & 1.104 & 1.034 & 1.040 & 0.977 \\
ARMA & 1.006 & 1.001 & 0.998 & 0.994 & 1.051 & 1.019 & 1.019 & 0.988 & 1.133 & 1.051 & 1.055 & 0.967 \\
MEM & 1.016 & 0.996 & 0.998 & 0.975 & 1.043 & 1.010 & 1.012 & 0.975 & 1.134 & 1.044 & 1.052 & 0.956 \\
HAR-J & 1.181 & 0.984 & 1.047 & 0.894 & 1.130 & 0.991 & 1.026 & 0.890 & 1.049 & 1.006 & 1.003 & 0.949 \\
HAR-RS & 1.216 & 1.000 & 1.061 & 0.900 & 1.147 & 0.996 & 1.033 & 0.889 & 1.050 & 1.008 & 1.003 & 0.949 \\
HARQ & 1.264 & 1.008 & 1.091 & 0.895 & 1.194 & 1.005 & 1.057 & 0.897 & 1.078 & 1.009 & 1.017 & 0.940 \\
\midrule
\multicolumn{13}{l}{\textbf{Panel B: averaged with TTM}} \\[2pt]
HAR & 0.999 & 0.975 & 0.976 & 0.969 & 1.008 & 0.982 & 0.985 & 0.958 & 1.068 & 1.010 & 1.014 & 0.926 \\
ARFIMA & 1.016 & 0.979 & 0.982 & 0.967 & 1.060 & 1.009 & 1.013 & 0.976 & 1.104 & 1.034 & 1.040 & 0.987 \\
ARMA & 1.006 & 0.981 & 0.980 & 0.970 & 1.051 & 1.004 & 1.006 & 0.976 & 1.133 & 1.040 & 1.048 & 0.971 \\
MEM & 1.016 & 0.977 & 0.979 & 0.956 & 1.043 & 0.994 & 0.999 & 0.963 & 1.134 & 1.026 & 1.039 & 0.952 \\
HAR-J & 1.181 & 0.978 & 1.042 & 0.891 & 1.130 & 0.983 & 1.018 & 0.879 & 1.049 & 0.983 & 0.979 & 0.920 \\
HAR-RS & 1.216 & 0.992 & 1.055 & 0.913 & 1.147 & 0.988 & 1.025 & 0.870 & 1.050 & 0.985 & 0.980 & 0.923 \\
HARQ & 1.264 & 0.991 & 1.077 & 0.905 & 1.194 & 0.995 & 1.047 & 0.882 & 1.078 & 0.990 & 0.995 & 0.923 \\
\midrule
\multicolumn{13}{l}{\textbf{Panel C: averaged with HAR}} \\[2pt]
TTM & 0.984 & 0.975 & 0.976 & 0.969 & 0.989 & 0.982 & 0.985 & 0.958 & 0.988 & 1.010 & 1.014 & 0.926 \\
Chronos-Bolt-S & 1.057 & 1.006 & 1.009 & 0.962 & 1.092 & 1.025 & 1.030 & 0.960 & 1.180 & 1.095 & 1.101 & 0.917 \\
Chronos-Bolt-B & 1.065 & 1.010 & 1.013 & 0.966 & 1.088 & 1.023 & 1.028 & 0.957 & 1.180 & 1.094 & 1.101 & 0.907 \\
TimesFM-2.5 & 0.997 & 0.986 & 0.988 & 0.974 & 1.060 & 1.015 & 1.021 & 0.936 & 1.149 & 1.088 & 1.091 & 0.902 \\
Moirai-2.0-S & 1.039 & 1.003 & 1.005 & 0.965 & 1.113 & 1.033 & 1.041 & 0.930 & 1.260 & 1.122 & 1.136 & 0.887 \\
Moirai-MoE-S & 1.025 & 0.984 & 0.983 & 0.928 & 1.077 & 1.005 & 1.008 & 0.933 & 1.147 & 1.068 & 1.075 & 0.924 \\
Lag-Llama & 1.535 & 1.069 & 1.150 & 0.760 & 1.588 & 1.060 & 1.186 & 0.746 & 1.024 & 0.998 & 1.003 & 0.840 \\
Toto & 1.169 & 0.999 & 1.050 & 0.901 & 1.241 & 0.997 & 1.095 & 0.919 & 1.132 & 1.046 & 1.068 & 0.893 \\
Sundial & 0.999 & 0.985 & 0.987 & 0.958 & 1.087 & 1.017 & 1.030 & 0.921 & 1.183 & 1.098 & 1.106 & 0.885 \\
ARFIMA & 1.016 & 0.990 & 0.993 & 0.956 & 1.060 & 1.011 & 1.019 & 0.935 & 1.104 & 1.066 & 1.070 & 0.920 \\
ARMA & 1.006 & 0.997 & 0.995 & 0.965 & 1.051 & 1.020 & 1.022 & 0.948 & 1.133 & 1.091 & 1.091 & 0.916 \\
MEM & 1.016 & 0.996 & 0.998 & 0.972 & 1.043 & 1.015 & 1.018 & 0.964 & 1.134 & 1.083 & 1.088 & 0.900 \\
HAR-J & 1.181 & 0.987 & 1.052 & 0.914 & 1.130 & 0.993 & 1.030 & 0.900 & 1.049 & 1.034 & 1.031 & 0.902 \\
HAR-RS & 1.216 & 1.005 & 1.068 & 0.938 & 1.147 & 0.998 & 1.038 & 0.898 & 1.050 & 1.036 & 1.031 & 0.903 \\
HARQ & 1.264 & 1.010 & 1.096 & 0.917 & 1.194 & 1.005 & 1.064 & 0.894 & 1.078 & 1.033 & 1.044 & 0.864 \\
\bottomrule
\end{tabular}%
}
\end{table}

%% file: tables/table_log_input_mcs_dm.tex
\begin{table}[htbp]
\centering
\singlespacing
\caption{MCS inclusion rates and DM pairwise win rates for 50 assets with the log version of every foundation model, retransformed to volatility as in Sec.~\ref{sec:robustness}, and the econometric models as in Tab.~\ref{tab:mcs}. MCS columns report the percentage of assets for which the model is included in the MCS at the 10\% significance level (QLIKE loss, range statistic $T_R$ and its elimination rule, block bootstrap with $B = 10{,}000$). DM columns report the percentage of pairwise comparisons (across 50 assets $\times$ 16 opponents = 800 tests) in which the row model achieves significantly lower QLIKE at the 5\% level.}
\label{tab:log_input_mcs_dm}
\small
\begin{tabular}{lrrrrrr}
\toprule
& \multicolumn{3}{c}{MCS inclusion (\%)} & \multicolumn{3}{c}{DM win rate (\%)} \\
\cmidrule(lr){2-4}\cmidrule(lr){5-7}
Model & $h=1$ & $h=5$ & $h=22$ & $h=1$ & $h=5$ & $h=22$ \\
\midrule
Log-HAR & 88.0 & 100.0 & 100.0 & 19.1 & 43.1 & 36.4 \\
HAR & 86.0 & 92.0 & 96.0 & 20.6 & 24.1 & 19.1 \\
HAR-J & 90.0 & 92.0 & 72.0 & 9.1 & 8.8 & 11.2 \\
HAR-RS & 78.0 & 84.0 & 76.0 & 6.0 & 6.0 & 10.6 \\
HARQ & 78.0 & 94.0 & 86.0 & 4.5 & 8.5 & 12.0 \\
ARFIMA & 68.0 & 96.0 & 100.0 & 13.6 & 11.4 & 22.6 \\
ARMA & 76.0 & 80.0 & 92.0 & 17.5 & 12.4 & 11.9 \\
MEM & 68.0 & 82.0 & 84.0 & 12.4 & 13.1 & 10.9 \\
Log-Chronos-Bolt-S & 76.0 & 92.0 & 94.0 & 14.6 & 8.4 & 10.4 \\
Log-Chronos-Bolt-B & 80.0 & 92.0 & 98.0 & 16.9 & 11.5 & 10.6 \\
Log-Moirai-2.0-S & 84.0 & 90.0 & 88.0 & 17.9 & 6.9 & 9.4 \\
Log-Moirai-MoE-S & 84.0 & 52.0 & 30.0 & 14.4 & 4.6 & 6.1 \\
Log-Lag-Llama & 64.0 & 58.0 & 14.0 & 5.2 & 0.8 & 0.0 \\
Log-TimesFM-2.5 & 88.0 & 98.0 & 98.0 & 28.9 & 10.0 & 11.2 \\
Log-Toto & 52.0 & 48.0 & 38.0 & 0.1 & 0.2 & 4.2 \\
Log-Sundial & 100.0 & 98.0 & 98.0 & 52.0 & 5.5 & 8.1 \\
Log-TTM & 98.0 & 100.0 & 96.0 & 22.6 & 32.0 & 19.1 \\
\bottomrule
\end{tabular}
\end{table}

%% file: tables/table_log_vs_vol_dm.tex
\begin{table}[htbp]
\centering
\singlespacing
\caption{DM tests of the log version of each foundation model against the same model with volatility as input. Percentage of the 50 assets on which the log version has a significantly lower QLIKE (Lower) or a significantly higher QLIKE (Higher) at the 5\% level, with the two-sided test of Sec.~\ref{sec:significance}, on the dates and forecasts of Tab.~\ref{tab:log_input}.}
\label{tab:log_vs_vol_dm}
\small
\begin{tabular}{lrrrrrr}
\toprule
& \multicolumn{2}{c}{$h=1$} & \multicolumn{2}{c}{$h=5$} & \multicolumn{2}{c}{$h=22$} \\
\cmidrule(lr){2-3}\cmidrule(lr){4-5}\cmidrule(lr){6-7}
Model & Lower & Higher & Lower & Higher & Lower & Higher \\
\midrule
Log-Chronos-Bolt-S & 96 & 0 & 56 & 0 & 8 & 0 \\
Log-Chronos-Bolt-B & 98 & 0 & 64 & 0 & 18 & 0 \\
Log-Moirai-2.0-S & 94 & 0 & 46 & 0 & 32 & 0 \\
Log-Moirai-MoE-S & 10 & 0 & 4 & 4 & 0 & 10 \\
Log-Lag-Llama & 98 & 0 & 8 & 0 & 0 & 94 \\
Log-TimesFM-2.5 & 22 & 0 & 12 & 6 & 2 & 6 \\
Log-Toto & 0 & 28 & 0 & 36 & 0 & 34 \\
Log-Sundial & 56 & 0 & 2 & 8 & 4 & 4 \\
Log-TTM & 2 & 28 & 4 & 34 & 0 & 32 \\
\bottomrule
\end{tabular}
\end{table}

%% file: tables/table_ml_diag.tex
\begin{table}[htbp]
\centering
\singlespacing
\caption{Level and tracking error of the supervised machine learning baselines. For the volatility forecast $\widehat{\sigma}_{t+h}$ and the realized volatility $\sigma_{t+h}$, the level error is the mean of $\log(\sigma_{t+h}/\widehat{\sigma}_{t+h})$, negative when the forecasts are above the realized values, and the tracking error is the standard deviation of $\log(\sigma_{t+h}/\widehat{\sigma}_{t+h})$ as a ratio to that of Log-HAR. The last three columns give the QLIKE loss ratio to Log-HAR after every model, Log-HAR included, is rescaled in real time as in Appendix~\ref{app:forecast_level}. Each measure is computed per asset and averaged across the 50 assets. LSTM, patience 100: the LSTM with early stopping at a patience of 100 epochs and at most 500 epochs.}
\label{tab:ml_diag}
\small
\setlength{\tabcolsep}{3.5pt}
\begin{tabular}{lrrrrrrrrr}
\toprule
& \multicolumn{3}{c}{Level error} & \multicolumn{3}{c}{Tracking error} & \multicolumn{3}{c}{Ratio after rescaling} \\
\cmidrule(lr){2-4}\cmidrule(lr){5-7}\cmidrule(lr){8-10}
Model & $h=1$ & $h=5$ & $h=22$ & $h=1$ & $h=5$ & $h=22$ & $h=1$ & $h=5$ & $h=22$ \\
\midrule
Log-HAR & -0.034 & -0.062 & -0.100 & 1.000 & 1.000 & 1.000 & 1.000 & 1.000 & 1.000 \\
TTM & -0.069 & -0.076 & -0.056 & 1.004 & 0.996 & 0.979 & 1.016 & 1.016 & 0.982 \\
XGBoost & -0.028 & -0.029 & -0.031 & 1.053 & 1.077 & 1.076 & 1.139 & 1.242 & 1.128 \\
Feedforward network & -0.029 & -0.036 & -0.035 & 1.055 & 1.054 & 1.057 & 1.106 & 1.140 & 1.122 \\
LSTM & -0.032 & -0.038 & -0.030 & 1.014 & 1.044 & 1.037 & 1.039 & 1.152 & 1.084 \\
LSTM, patience 100 & -0.031 & -0.039 & -0.030 & 1.017 & 1.044 & 1.054 & 1.044 & 1.142 & 1.099 \\
\bottomrule
\end{tabular}
\end{table}

%% file: tables/table_finetune.tex
\begin{table}[htbp]
\centering
\singlespacing
\caption{Fine-tuned TTM and Sundial. Mean across assets of the per-asset QLIKE loss ratio to Log-HAR from 3 January 2022 to 30 January 2026, for each model applied zero-shot as in the main comparison and fine-tuned in the two ways of Appendix~\ref{app:finetune}: all the weights of TTM or low-rank adapters on Sundial, and the head only, with the pretrained backbone fixed. Panel~A: the 50 assets, with the number of assets with a ratio below one ($<1$). Panel~B: the 40 stocks, which are in the fine-tuning corpus, and the 10 FX pairs and futures, which are not. Dates common to Log-HAR and the six models.}
\label{tab:finetune}
\small
\setlength{\tabcolsep}{5pt}
\begin{tabular}{lrrrrrr}
\toprule
\multicolumn{7}{l}{\textit{Panel A: all 50 assets}} \\
& \multicolumn{2}{c}{$h=1$} & \multicolumn{2}{c}{$h=5$} & \multicolumn{2}{c}{$h=22$} \\
\cmidrule(lr){2-3}\cmidrule(lr){4-5}\cmidrule(lr){6-7}
Model & Ratio & $<1$ & Ratio & $<1$ & Ratio & $<1$ \\
\midrule
TTM, zero-shot & 0.983 & 38 & 0.981 & 38 & 0.989 & 32 \\
TTM, all weights & 0.999 & 25 & 0.990 & 32 & 0.972 & 37 \\
TTM, head only & 0.987 & 32 & 0.998 & 30 & 0.989 & 31 \\
Sundial, zero-shot & 0.997 & 28 & 1.028 & 11 & 1.066 & 4 \\
Sundial, adapters & 0.997 & 28 & 1.068 & 1 & 1.062 & 5 \\
Sundial, head only & 1.045 & 14 & 1.020 & 15 & 0.971 & 33 \\
\midrule
\multicolumn{7}{l}{\textit{Panel B: by asset class}} \\
& \multicolumn{3}{c}{40 stocks} & \multicolumn{3}{c}{10 FX pairs and futures} \\
\cmidrule(lr){2-4}\cmidrule(lr){5-7}
Model & $h=1$ & $h=5$ & $h=22$ & $h=1$ & $h=5$ & $h=22$ \\
\midrule
TTM, zero-shot & 0.985 & 0.976 & 0.986 & 0.979 & 1.002 & 1.003 \\
TTM, all weights & 0.993 & 0.976 & 0.964 & 1.023 & 1.045 & 1.005 \\
TTM, head only & 0.982 & 0.995 & 0.985 & 1.004 & 1.011 & 1.004 \\
Sundial, zero-shot & 1.000 & 1.029 & 1.073 & 0.988 & 1.020 & 1.036 \\
Sundial, adapters & 1.002 & 1.070 & 1.065 & 0.979 & 1.058 & 1.050 \\
Sundial, head only & 1.046 & 1.021 & 0.964 & 1.040 & 1.016 & 0.997 \\
\bottomrule
\end{tabular}
\end{table}